\documentclass[useAMS,usenatbib]{mn2e}
\usepackage{graphicx}
\usepackage{txfonts}
\usepackage{natbib}
\usepackage{wrapfig}
\usepackage{longtable}
\usepackage[usenames]{color}

\newcommand{\bea}{\begin{eqnarray}}
\newcommand{\eea}{\end{eqnarray}}
\newcommand{\be}{\begin{equation}}
\newcommand{\ee}{\end{equation}}
\newcommand{\ave}[1]{\langle #1 \rangle}

\def\arcminf {\hbox{$.\!\!^{\prime}$}}
\def\arcdegf {\hbox{$.\!\!^{\circ}$}}

\bibpunct{(}{)}{;}{a}{}{,}

\title
{How special are Brightest Group and Cluster Galaxies?}

\author[Anja von der Linden et al.]
{Anja von der Linden$^{1}$
\thanks{E-mail:anja@mpa-garching.mpg.de}, 
Philip N. Best$^{2}$,
Guinevere Kauffmann$^{1}$,
Simon D. M. White$^{1}$
\\
$^{1}$Max Planck Institut f\"ur Astrophysik, 
              Karl-Schwarzschild-Str. 1, Postfach 1317,
              85741 Garching, Germany\\
$^{2}$SUPA, Institute for Astronomy,
              Royal Observatory Edinburgh,
              Blackford Hill, Edinburgh EH9 3HJ}
\begin{document}

\date{Submitted: 7 November 2006; accepted: 4 May 2007}
\date{Accepted 2007 May 4. Received in original form 2006 November 7.}

\pagerange{\pageref{firstpage}--\pageref{lastpage}} \pubyear{2007}

\maketitle

\label{firstpage}

\begin{abstract}
We use the Sloan Digital Sky Survey to construct a sample of 625 brightest
group and cluster galaxies (BCGs) together with control samples of
non--BCGs matched in stellar mass, redshift, and color.  We investigate
how the systematic properties of BCGs depend on stellar mass and on their
privileged location near the cluster center.  The groups and clusters that
we study are drawn from the C4 catalogue of Miller et al. (2005) but we
have developed improved algorithms for identifying the BCG and for
measuring the cluster velocity dispersion. Since the SDSS photometric
pipeline tends to underestimate the luminosities of large galaxies in
dense environments, we have developed a correction for this effect which
can be readily applied to the published catalog data.  We find that BCGs
are larger and have higher velocity dispersions than non-BCGs of the same
stellar mass, which implies that BCGs contain a larger fraction of dark
matter. In contrast to non--BCGs, the dynamical mass-to-light ratio of
BCGs does not vary as a function of galaxy luminosity.  Hence BCGs lie on
a different fundamental plane than ordinary elliptical galaxies.  BCGs
also follow a steeper Faber--Jackson relation than non-BCGs, as suggested
by models in which BCGs assemble via dissipationless mergers along
preferentially radial orbits.  We find tentative evidence that this
steepening is stronger in more massive clusters.  BCGs have similar mean
stellar ages and metallicities to non-BCGs of the same mass, but they have
somewhat higher $\alpha$/Fe ratios, indicating that star formation may
have occurred over a shorter timescale in the BCGs.  Finally, we find that
BCGs are more likely to host radio--loud active galactic nuclei than other
galaxies of the same mass, but are less likely to host an optical AGN.
The differences we find are more pronounced for the less massive
BCGs, i.e. they are stronger at the galaxy group level.

\end{abstract}


%________________________________________________________________
\section{Introduction}
\label{sect:intro}

The central galaxies in galaxy clusters seem to be special - in many cases,
the differences are visually obvious, because central cluster galaxies often
have extended envelopes (i.e. they are cD galaxies) and they are usually the
brightest (and most massive) galaxies in their clusters. The term
\textit{brightest cluster galaxy} (BCG) has thus become synonymous with the
term \textit{central galaxy}.

At first glance, it might seem evident that the location of the BCG at the
bottom of the potential well of a cluster must be the cause for any property
which distinguishes it from other (cluster) galaxies.  However, BCGs are also
the dominant population at the massive end of the galaxy luminosity
function, and thus, their properties are influenced both by their large
masses and by the cluster environment. It is very difficult to disentangle
these two influences, because it is difficult to find equally massive non-BCGs
for comparison. Since most BCGs are early-type galaxies, their properties
are often compared with the known scaling relations for elliptical galaxies.
It has been claimed that BCGs lie on the same Fundamental Plane as other
ellipticals \citep{oeh91}, but that they lie off its projections (e.g. the
Faber--Jackson and Kormendy relations) in that they have lower velocity
dispersions and larger radii than predicted by these relations
\citep{thr81,hos87,sch87,oeh91}. More
recently, it has been claimed that the slopes of the Faber--Jackson and
Kormendy relations change as a function of of galaxy luminosity for all
elliptical galaxies \citep{lfr06,dqm06}.
On the other hand, \citet{bcb05} find that the surface brightness profiles
(and thus the radii) of BCGs depend on the host cluster properties.\\

The formation mechanism of BCGs is also a subject of much debate. Early on, it
was suggested that BCGs form when galaxies sink to the bottom of the potential
well of a cluster and merge \citep[termed \textit{galactic cannibalism};
][]{ost75,whi76}.  However, \citet{mer85} argued that tidal stripping would
reduce the masses of cluster galaxies to the point where dynamical friction is
too slow for this to be a viable mechanism.  These analyses assumed
that clusters are {\it static entities}.  A further mechanism to form BCGs
{\it in situ} in the cluster is star formation in cluster \textit{cooling
flows}.  At the centers of clusters, gas reaches high enough density to cool
and condense into the cluster core \citep{sil76,fab94}. 
But while the mass
deposition rates inferred from the X-ray luminosities of cooling flow
clusters are of the order of several hundreds to $>1000 M_{\odot}$yr$^{-1}$
\citep[e.g.][]{afe96}, observed star formation rates are at most $\sim 100
M_{\odot}$yr$^{-1}$  \citep{cae99}. Recent X-ray studies have furthermore
demonstrated that the cluster gas does not cool below $\sim 2$ keV \citep[e.g. ][]{pef05}.
%Recent X-ray studies
%have demonstrated, however, that the actual 
%cooling rates in clusters are far below those necessary to form massive BCGs
%\citep[e.g. ][]{pef05}. 
Moreover, this scenario predicts that the stellar
populations of BCGs should be young and blue, which is clearly inconsistent
with observations.

These scenarios were proposed before hierarchical structure formation was
fully established as the standard cosmological paradigm, and for simplicity
they neglected many of
the processes that take place when clusters assemble through mergers.
\citet{dub98} used N-body simulations to show that a dominant galaxy forms
naturally by merging of massive galaxies when a cluster collapses along
filaments. 
Recently, \citet{deb06} investigated the formation of BCGs in the context of
the Millenium Run simulation \citep{swj05}. In their model, the stars that
make up BCGs today are formed in a number of galaxies at high redshifts,
which subsequently merge to form larger systems. The final BCGs assemble
rather late: by a redshift of $z\sim0.5$, on average about half of the final
stellar mass lies in the largest progenitor galaxy. Since many of these
mergers take place very late when most galaxies have converted the bulk of
their gas into stars, the merging events are very nearly dissipationless and
are not associated with significant star formation.  This scenario is
supported by observations that demonstrate that BCGs exhibit little scatter
in luminosity over a wide range of redshifts
\citep{san72,sgh83,pol95,abk98}, and that their color evolution is
consistent 
with a passively evolving stellar population that formed at high redshifts
($z_{\rm form} \sim 2 - 5$).

\citet{bmq06} used two--component N--body simulations of galaxy mergers to
show that the remnants of dissipationless mergers remain in the fundamental
plane. They showed, however, that the locations of the remnants {\it within}
the fundamental plane, and thus on projected relations such as the
Faber--Jackson 
and size--luminosity relations, depend on the orbits of the merging galaxies.
During cluster assembly, infall occurs primarily along filaments, suggesting
that mergers onto the BCG may take place preferentially on radial orbits.
\citet{bmq06} show that  BCGs would then be predicted to lie on steeper
Faber--Jackson and size--luminosity relations than field galaxies.  \\

Although BCGs are probably not formed in cooling flows, they are believed to
play an important role in regulating the rate at which gas cools at the
centers of groups and clusters. The central cluster galaxies often harbor
radio--loud active galactic nuclei (AGN), which may provide the necessary
heating to counteract radiative cooling. \citet{bur90} find that 10 out of 14
cD galaxies in cooling flow clusters are radio--loud, compared to 3 out of 13
in clusters without cooling cores. However, \citet{bkh05b} show that
radio--loudness also depends strongly on other galaxy parameters such as
stellar mass.  \\

This work aims at comparing the properties of BCGs to those of ``normal''
galaxies of similar stellar mass, thereby disentangling the influences of
mass and environment. Our work is based on data from the Sloan Digital Sky
Survey (SDSS; Sect.~\ref{sect:data}).  It has recently been noted that the
SDSS photometric pipeline tends to underestimate the luminosities of BCGs --
we present a method to retrieve more accurate magnitudes from the catalogued
data in Sect.~\ref{sect:data} and Appendix~\ref{sect:magnitudes}.  To
construct a sample of BCGs, we refine the C4 sample of galaxy clusters
\citep{mnr05} in the SDSS by improving the selection of the central galaxy
(Sect.~\ref{sect:selection}).  We also improve the C4 cluster velocity
dispersion measurement by developing an iterative algorithm to
simultaneously determine the redshift, the velocity dispersion, and the
virial radius of each cluster (Sect.~\ref{sect:selection}). A detailed
comparison of the measured BCG positions, redshifts, and velocity
dispersions to the original C4 values is given in Sect.~\ref{sect:compare}.
In Sect.~\ref{sect:radio} we investigate the occurence of radio--loud AGN in
BCGs. The results demonstrate that our choice of BCG is superior to the
original choice of BCG in the C4 catalogue. 
The radio properties of the BCGs
are investigated in more detail in an accompanying paper \citep{bes06b}, as
are the implications of these results for models of cluster cooling flows.
We compare BCGs to control samples of non--BCGs that are closely matched in
stellar mass; we analyze their structural properties, their positions with
respect to the fundamental plane and its projections, their stellar
populations, and their line emission in Sect.~\ref{sect:optical}.  A summary
of our paper is given in Sect.~\ref{sect:summary}.

Our refined BCG/cluster sample, together with the wealth of information 
available from the SDSS, also provides a local comparison sample for optically
selected high-redshift clusters such as the ESO Distant Cluster Survey,
EDisCS \citep{wcs05}. While this work focuses on the BCG selection and on
properties of the BCGs, another paper (von der Linden et al., in prep) will
concentrate on cluster properties such as substructure, mass estimates from
velocity dispersion as compared to total light, and the properties of
cluster galaxies as a function of their distance from the BCG.

Unless otherwise noted, we assume a concordance cosmology with $\Omega_{\rm m}
= 0.3$, $\Omega_{\Lambda} = 0.7$ and $H_0 = 100 \,h\, \mbox{km/s/Mpc}$, where
$h=0.7$.

\section{Data}
\label{sect:data}

The Sloan Digital Sky Survey \citep{slb02,yaa00} is a survey of about a
quarter of the  
extragalactic sky, obtaining photometry in five bands ($ugriz$) of more than
200 million objects and spectra of up to a million objects. The observations
are carried out in drift-scan mode on a dedicated 2.5m telescope at Apache
Point Observatory, with a large-array CCD camera that allows
near-simultaneous photometry. The imaging data is reduced by an automatic
pipeline, {\sc photo} \citep{lgi01}, and various classes of objects are then
classified for subsequent spectroscopy; those galaxies with
$14.5 < m_r < 17.7$ and $\mu < 24.5 \mbox{ mag arcsec}^{-2}$ comprise the `main
galaxy sample' \citep{swl02}.
The spectra are obtained using a
fiber-fed spectrograph on the same telescope. On each spectroscopic plate,
which has a circular field of view of radius $1\arcdegf49$, 592 object
fibers can be placed. Due to the finite fiber size, any two fibers on the
same plate need to be spaced at least $55\arcsec$ apart. The fiber
allocation is performed by a tiling algorithm, which maximizes the number of
objects that can be observed \citep{bll03}. In the case of a 
``fiber collision'' (i.e. two objects that are closer than $55\arcsec$), no
preference of objects is given within the usual constraints.

Our analysis is based on the fourth data release (DR4) of SDSS, whose main
galaxy sample provides spectra for more than $500\,000$
galaxies.

\subsection{Spectral analysis}
\label{sect:galaxy-cat}

A multitude of physical properties have been derived for galaxies in the
spectroscopic database via stellar population synthesis fitting and are
publicly available\footnote{\tt http://www.mpa-garching.mpg.de/SDSS/}.  The
stellar continuum of each galaxy is modelled as a sum of template spectra
generated from population synthesis models \citep{thk04,khw03}. These fits
also lead to measures of the stellar mass--to--light ratio, star formation
histories, and mean stellar ages \citep[][b\nocite{khw03b}]{khw03}. After
subtracting the stellar continuum, emission line fluxes can be accurately
measured, allowing studies of the star formation rates \citep{bcw04} and AGN
activity \citep{hkb04,kht03}.

\subsection{Photometry and Stellar Masses}
\label{sect:masses}

It has recently been noted that the SDSS photometry systematically
underestimates the luminosities of nearby BCGs \citep{bhs06,lfr06}. The
problem arises because the level of sky 
background is overestimated both for large objects and in crowded fields.
This is not only a problem for BCGs, but for all large galaxies, and the
problem is worse in dense cluster environments. Since any estimate of the
stellar mass of a galaxy is derived from its luminosity, it is crucial for
our analysis to avoid biases in the luminosity measurement. 

Apart from the {\tt local} {\tt sky} background measurement, which is
estimated and applied for each galaxy, {\sc photo} provides a {\tt global}
{\tt sky} measurement estimated over a whole field. In
Appendix~\ref{sect:magnitudes}, we argue that by adding up to 70\% of the
difference between {\tt local} and {\tt global} {\tt sky} to the radial
surface brightness profiles (as provided by {\sc photo}), more accurate
photometry can be achieved. We test this procedure using aperture photometry
of 35 BCGs from the survey of \citet{pol95} that are contained in the area
of sky covered by the SDSS DR5.  We also test our corrected magnitudes by
comparing with the photometry provided by 2MASS \citep{scs06}. Our
correction method (described in detail in Appendix~\ref{sect:magnitudes})
has been applied to the BCGs as well as to about 200,000 unique galaxies at
$z \lesssim 0.1$, which form the basis of our comparison samples for the
BCGs.

Since many BCGs are observed to have surface brightness profiles which do
not follow a simple de Vaucouleurs profile \citep{gzz05,bhs06}, we choose
not to use a magnitude measurement 
that assumes a certain profile, or is sensitive to the profile shape (this
includes Petrosian magnitudes, which include about 80\% of
the light from  galaxies with  de Vaucouleurs brightness
profiles, but almost 100\% of the light from galaxies with  
exponential light profiles).
Instead, we measure isophotal magnitudes, defined as the light within the
radius $r_{\rm iso23}$ where the 1D surface brightness profile reaches a
surface brightness of $(23 + 10 \log(1+z))\,\mbox{mag}/\square\arcsec$ in the
$r$-band (the redshift term accounts for cosmological surface brightness
dimming). This is a relatively bright isophote limit, chosen both to avoid
residual uncertainties in the sky background subtraction
(cf. Fig.~\ref{fig:mags_bcgs_pol95}) as well as to exclude light from the cD
envelope 
present in some BCGs \citep[which is noticable typically at surface
brightness levels one or two magnitudes fainter, cf. ][]{gzz05}.
We refer to these 
magnitudes as \textit{iso23} magnitudes.

Other studies of BCGs have quantified their luminosities within metric
apertures \citep[e.g.][]{pol95}. These studies were focused on cD galaxies,
whereas our study includes BCGs of a much broader range in mass and
size. Hence metric apertures would include very different fractions of light
for BCGs at the extreme ends of the mass range,
c.f. Fig.~\ref{fig:bcg_gallery}, and any result based on them would be very
difficult to interpret.
It can be argued that the \textit{iso23} magnitudes are also dependent on
the profile shape: For shallower brightness profiles, a larger fraction of
the total light / mass is missed. But an estimate of the total light is 
unfortunately not feasible, due to residual uncertainties in the sky
subtraction. We therefore adopt the \textit{iso23} magnitudes as the least
biased and least problematic luminosity measurements. However, we have
verified that our (qualitative)
results do not change if we use Petrosian magnitudes as an attempt to
measure total magnitudes.

Stellar mass estimates for the BCGs and the comparison galaxies are derived
from these luminosity measurements using the {\sc kcorrect} algorithm
\citep{blr06}, which is also used to  determine the k-corrections for our
galaxies. BCGs that were not observed spectroscopically
(see Sect. \ref{sect:bcg-sel}) are assumed to have a redshift  identical to
the cluster redshift. Just as the luminosities should not be taken
  as an 
estimate of the total light, the quoted stellar masses are not an attempt to
measure the total stellar mass. But since our analysis is based on comparing
objects with similar masses and colors, this is not an issue.

\subsection{Radio catalog}

\citet{bkh05a} identified the radio--emitting galaxies
within the main spectroscopic sample of the SDSS DR2, by cross--comparing
these galaxies with a combination of the National Radio Astronomy
Observatory (NRAO) Very Large Array (VLA) Sky Survey \citep[NVSS;][]{con98}
and the Faint Images of the Radio Sky at Twenty
centimetres (FIRST) survey \citep{bec95}. They then used the optical
properties of the galaxies to separate the radio--loud AGN from the
radio--detected star--forming galaxies.  This work has now been extended
to include the DR3 and DR4 data (Best et~al. in prep), and these results
were used to identify those 
galaxies that are radio-loud AGN.

\section{Selection of clusters and brightest cluster galaxies}
\label{sect:selection}

The basis of our cluster sample is the C4 cluster catalog \citep{mnr05}. The C4
catalog is derived using the SDSS spectroscopic sample  and is currently
available for Data Release 3. It identifies clusters in a parameter space of
position, redshift, and color. The algorithm  assumes that at least a
fraction of the cluster galaxies form a color--magnitude--relation.
It identifies galaxies in clustered regions, with neighbors of similar
colors, as ``C4 galaxies'' \citep[see][for a detailed discussion of this
selection]{mnr05}. From these C4 galaxies, it reconstructs the local
density field, and identifies the C4 galaxies at the peaks of this density
field as cluster centers (coined the {\it mean} galaxies).

The C4 catalog identifies 1106 clusters within $0.02 \le z \le 0.16$. In
order to ensure that our clusters span a large angular extent compared to
the minimum distance between fibers (55 arcsec), we limit our cluster sample
to $z \leq 0.1$. At this redshift, the magnitude limit of the spectroscopic
sample corresponds to $M_r \sim -20$, i.e.  slightly fainter than an $L^*$
galaxy.  This cut results in a starting sample of 833 clusters.

\subsection{Selection of the brightest cluster galaxy}
\label{sect:bcg-sel}

Our aim is to find the galaxy closest to the deepest point of the potential
well of the cluster. In many rich clusters, this choice is obvious, and
the central galaxy can easily be recognized as a cD elliptical
galaxy by its extended envelope. Typically, this is also the Brightest
Cluster Galaxy (BCG). However, in some clusters the central, dominant
galaxy may not be the brightest galaxy. 
An example is the cluster C4\_2003
\footnote{A note on the cluster IDs: the IDs used in the DR3 version
of C4 
are not identical to those in the DR2 version. Since the DR3 catalog was
released only within the SDSS collaboration, the DR2 IDs are the
``official'' IDs, so whenever possible, we identify clusters by their DR2
ID (e.g. C4\_2003). For those clusters without a DR2 ID, we use the DR3 ID,
and denote these as e.g. C4\_DR3\_2004.}
, shown in
Fig.~\ref{fig:c4_2004}. The obvious central galaxy is
SDSS~J215729.42-074744.5 at the center of the image, but it is 0.3~mag
fainter than SDSS~J215701.71-075022.5, about 6\arcmin west-south-west of the
former. We identify the former as the BCG (but concede that the term is a
misleading nomenclature in this case).

The C4 catalogue lists two  galaxies for
 each cluster that could be considered the BCG:
the {\it mean} galaxy (described above) and the brightest galaxy from the
spectroscopic catalog within
500~$h^{-1}$~kpc of the position of the mean galaxy, four times the velocity
dispersion, and without strong H$\alpha$ emission. However, due to the
problem of fiber collisions, the true BCG is not included 
in the SDSS spectroscopic data for about 30\% of the clusters and is thus
missed by the C4 algorithm.

An earlier version of the C4 catalog tried to correct for this by selecting
a brightest cluster galaxy based on the photometric 
catalog. This object was selected to lie within 1~$h^{-1}$~Mpc of the
cluster center, and have a color compatible with the color-magnitude
relation of that cluster. 
Visual checks revealed, however,
that this did not provide a reliable BCG: out of a subsamble of 128
clusters, 17 of 
the photometric BCGs identified by C4 were stars misclassified as galaxies, and 36 were
spiral galaxies (some of these located at the edge of the cluster).

To identify the BCG for each cluster, we use the following procedure: 
\setlength{\leftmargini}{0.6cm}
\begin{enumerate}
\setlength{\itemindent}{0cm}
\setlength{\labelsep}{0.1cm}
\setlength{\labelwidth}{0.5cm}
\item Based on the cluster redshift and velocity dispersion 
%$\sigma_{\rm v, cl}^{\rm C4}$
  given by C4, we estimate the virial radius of the cluster:
  \be
  R_{\rm 200} = 1.73 \frac{\sigma_{\rm v, cl}}{1000 \mbox{ km  s}^{-1}}
  \frac{1}{\sqrt{\Omega_{\Lambda}+\Omega_0(1+z_{\rm cl})^3}} h^{-1}
  \mbox{ Mpc} 
  \label{eq:r200}
  \ee 
  \citep[see][]{fzm05}. As C4 lists velocity dispersions within fixed physical
  radii (0.5, 1, 1.5, 2, 2.5 $h^{-1}$ Mpc), we use the minimum non--zero
  value of 
  these different values in this step.
\item Within the projection of the larger of $R_{200}$ and 0.5~Mpc around
  the {\it mean} galaxy, we 
  select the two brightest galaxies 
  that meet the following criteria: 
\begin{itemize}
\setlength{\itemindent}{0cm}
\item[--]
  The concentration index $c =
  R_{90}/R_{50}$ is larger than 2.5 (where $R_{90}$ is the radius containing
  $90$\% of the
  petrosian flux measured in the $i$-band amd $R_{50}$
   is the radius containing $50$\% of this flux), and {\tt fracDeV\_r}$ >0.5$
  (this is a measure of the contribution of the de Vaucouleurs profile to the
  SDSS model $r$--magnitude). These cuts select galaxies likely to
  be early--types. 
\item[--] The color is compatible with that of the C4
  {\it mean}
  galaxy to within $\Delta(u-g)\le 0.6$ (unless one of the $u$ magnitudes has a
  large error estimate), $\Delta(g-r)\le 0.5$, $\Delta(r-i)\le
  0.4$, and $\Delta(r-i)\le 0.4$ (these are the dimensions of the color
  criteria 
  originally used in the C4 algorithm to identify clustering in color
  space).
\item[--] The flag {\tt TARGET\_GALAXY} has been set and the flag {\tt
  SATURATED} is not set (these criteria allow us to identify
  stars that have been misclassified as galaxies, but they also apply to some
  low-redshift, bright galaxies).
\item[--] If the object has spectroscopic data, we require that the redshift is
  within $\Delta z < 0.01$ from the cluster redshift.
\end{itemize}
  The brightest galaxy that meets these criteria is our initial BCG
  candidate. However, it is possible that these criteria select a foreground
  elliptical. Thus, if the second brightest galaxy is more than
  one magnitude fainter than the brightest, we also  consider it as  a BCG
  candidate. 
\item We then loosen some of these critieria:
\begin{itemize}
\item[--] 
  we require only $c>2.3$,
\item[--] there is no constraint on
  $\Delta(u-g)$,
\item[--] {\tt TARGET\_GALAXY} does not need to be set, and {\tt
  SATURATED} can be set. 
\end{itemize}
  Galaxies which meet this second set of criteria and
  are brighter than (both) the candidate(s) selected in the previous step
  enter the list of candidates. Unfortunately, misclassified stars enter our
  candidate lists at this stage.
\item
\setlength{\leftmarginii}{0.5cm}
\begin{enumerate}
\setlength{\itemindent}{0cm}
\setlength{\labelsep}{0.1cm}
\setlength{\labelwidth}{0.4cm}
\item If this procedure returns only one candidate, which agrees with
  the spectroscopic BCG given by C4, then this is automatically considered the
  correct choice. This is the case for 242 clusters.
\item If this is not the case, then the BCG candidates (those given by C4 and
  those identified by our criteria) are inspected visually. For this
  purpose, we examine $2\arcminf5 \times 2\arcminf5$ color images of the BCG
  candidates (provided by the  DR4 Catalog Archive Server [CAS]\footnote{{\tt
  http://cas.sdss.org/astro/en/}}). These thumbnail images allow us to
\begin{itemize}
\item[--] identify cD galaxies by their extended envelope
\item[--] identify stars misclassified as galaxies
\item[--] identify obvious foreground ellipticals
\end{itemize}
  In the same step, a color image of the cluster (encompassing a field
  slightly larger than $R_{200}$) can be viewed, with the BCG candidates
  marked. An impression can thus be gained of how the positions of the
  candidates relate to the C4 cluster center(s) (as given by the {\it mean}
  galaxy and the geometric cluster center, based on the mean of the
  positions of the C4 cluster galaxies),  visible galaxy overdensities and 
  other structures in 
  the field. If there is more than one galaxy left in the candidate list,
  this color image allows us to choose the (brightest) one at the center of the
  galaxy overdensity identified by C4. In case such an overdensity is not
  apparent, we choose the brightest elliptical galaxy in the vicinity of the 
  the {\it mean} galaxy.

  For 472 clusters, the BCG can be identified fairly easily by means of
  these thumbnail images and the cluster image.
\item In case the previous step does not allow an unambiguous choice of the
  BCG, 
  we enlarge the candidate list by adding galaxies within
  2~Mpc that meet the criteria cited above. Along with this set of thumbnail
  images, we also inspect a color image
  where galaxies within $\Delta z < 0.01$ from the cluster redshift and
  2~Mpc from the {\it mean} galaxy have
  been marked  (see Fig.~\ref{fig:c4_2004}). It
  is this image  that allows us to visualize the clustering and 
  to follow the galaxy distribution at the redshift in
  question. BCG candidates belonging to neighboring clusters can be
  identified. Conversely, if the C4 cluster corresponds to a weak number
  density fluctuation within another cluster, the BCG of this larger cluster is
  chosen.
  We  identify
  the BCG of 54 clusters in this step. 
\item 
  The remaining 65 clusters require further scrutiny. For some of these
  clusters, this is necessary because the galaxy we identify as the central,
  dominant galaxy is not contained in the list of candidates. In other
  cases, it is evident from the cluster images that the cluster is in fact
  part of a larger cluster.  In this step, we use the {\tt Finding Chart},
  {\tt Navigate}, and {\tt Explore} tools of the CAS website interactively.
  We start by marking galaxies within 2~Mpc and $\Delta z < 0.01$ from the
  {\it mean} galaxy (as before), but then altering the radius, center, and
  redshift range to gain an impression of the clustering. We identify the
  center of the nearest galaxy overdensity and choose the most likely BCG
  within it. The color and brightness criteria that were previously used to
  determine the BCG candidate(s) are not applied in this step.
\end{enumerate}
\end{enumerate}
At this stage, a given galaxy may have been identified as the BCG for more
than one cluster. We then keep only the cluster whose {\it mean} galaxy is
closest to the BCG. This rejects 101 clusters as being substructures of
other clusters.

\subsection{Determination of the velocity dispersion and the virial radius}
\label{sect:veldisp}

The C4 algorithm measured velocity dispersions within fixed radii
(see Sect. \ref{sect:veldisp_comp}
for a more 
detailed discussion of the C4 velocity dispersions). We prefer to measure
the velocity dispersion $\sigma_{\rm v, cl}$
within the virial radius $R_{200}$, which can be related to the velocity
dispersion  and cluster redshift $z_{\rm cl}$ via
Eq.~(\ref{eq:r200}). We thus developed an iterative algorithm to determine
these three quantities:
\setlength{\leftmarginii}{0.5cm}
\begin{enumerate}
\setlength{\itemindent}{0cm}
\setlength{\labelsep}{0.1cm}
\setlength{\labelwidth}{0.4cm}
\item From the catalog of galaxies described in Sect.~\ref{sect:galaxy-cat}, 
  we select those that lie within $2 R_{\rm start}$ from the  BCG,
   where $R_{\rm start}$ is determined using   
  Eq.~(\ref{eq:r200}) with $\sigma_{\rm v, cl}$  given by the average value of  
    the cluster velocity dispersions measured by 
  C4. We also require the galaxies to lie within $\Delta z < 0.025$ of the
  cluster redshift.
\item For our first estimate of $z_{\rm cl}$, we take the redshift given by
  C4, $z_{\rm cl}^{\rm C4}$. For the first estimate of $\sigma_{\rm v, cl}$ (and
  $R_{200}$), we
  use the median absolute deviation of the starting sample of
  galaxies with respect to $z_{\rm cl}^{\rm C4}$. We also limit this
  first estimate to be less than 
  $500$~km/s, a step that is necessary to exclude surrounding large--scale
  structure.
\item \label{step:iterate1}
  For each galaxy, we calculate its velocity within the cluster rest-frame:
\be
\frac{v_i}{c} = \frac{z_i - z_{\rm cl}}{1 + z_{\rm cl}}
\ee
\item \label{step:iterate2} 
  From those galaxies within $\pm 3\sigma_{\rm v, cl}$ of $z_{\rm cl}$, and
  within $R_{200}$ 
  from the BCG, we re-determine $z_{\rm cl}$, $\sigma_{\rm v, cl}$ (and thus also
  $R_{200}$) using the 
  biweight estimator from \citet{bfg90}.
\item Steps \ref{step:iterate1} and \ref{step:iterate2} are repeated until
  convergence is reached, 
  i.e. subsequent iterations differ by less than 0.1\% in $z_{\rm cl}$ and
  $\sigma_{\rm v, cl}$. Galaxies are allowed 
  to re-enter the sample. The galaxies contained in the final sample are
  considered the cluster galaxies.

  It may happen that the iteration finds an oscillating solution, i.e.
  iteration $n+2$ yields the same solution (and the same galaxies) as
  iteration $n$. In this case, the algorithm modifies the input $\sigma_{\rm
    v, cl}$ to a random value between the two solutions and continues. Also, the
  convergence criterium is relaxed by factor of two every 200 iterations. A
  maximum of 1000 iterations is allowed.
\item To estimate the error on the velocity dispersion, we draw $10\,000$
  bootstrap realizations from the cluster galaxies and calculate the
  velocity dispersion of each. We adopt  68\% confidence intervals as 
  our estimate of the $\pm 1 \sigma$ error . 
\end{enumerate}
For 55 clusters, the algorithm does not converge or its sigma-- and
radius--clipping subsequently 
remove all (or all but one) of the galaxies from the starting sample. Since
these systems cannot be considered bound, they are removed from the cluster
list. 

At this point, all the clusters are inspected visually. We check a color image
of the cluster with the cluster galaxies marked, as well as the redshift
histogram. We mark clusters with the following indications that either the 
choice of BCG or the definition of cluster membership may be improved:
\setlength{\leftmarginii}{0.5cm}
\begin{itemize}
\setlength{\itemindent}{0cm}
\setlength{\labelsep}{0.2cm}
\setlength{\labelwidth}{0.5cm}
\item 
  The redshift histogram does not justify the value of the velocity
  dispersion. For clusters in very rich environments, the velocity 
  dispersion can be overestimated if  galaxies of neighboring groups and filaments
  are included. Clusters with a velocity dispersion exceeding 1500~km/s are
  automatically flagged, but also those with two or more spikes in the
  redshift histogram. 59 clusters are marked in this step.
\item 
  Some of the cluster images show that the selected BCG is not at the center
  of the clustering. In others, there may be another bright elliptical
  present which could be the BCG,  or 
  the true BCG did not meet the criteria used to select BCG
  candidates (i.e. mostly late--type BCGs).
  67 clusters are listed for re-investigation in this step.
\item
  Clusters in which the BCG redshift (if in the spectroscopic database)
  differs from the cluster redshift by more than $\Delta z = 0.002$ are also
  flagged (15 clusters).  
\end{itemize}

In this second round of visual checking, the choice of BCG 
or the sigma-- and radius--clipping limits can be changed. The former is
done  if the previously selected BCG is clearly associated with a
substructure of a larger cluster, but also if there is a possibly better BCG
candidate that was previously missed. 
Note that many of these alternative BCGs are in
fact foreground objects -- whenever possible, we retrieve redshift
information on these objects from CAS or the Nasa/IPAC Extragalactic
Database, NED\footnote{\tt http://nedwww.ipac.caltech.edu/}. For 21
clusters, the ``new'' BCG has not previously been selected as the BCG of
another cluster.  For 35 ``clusters'', the BCG  has been attributed to another
cluster, and so these systems are considered infall regions of these other
clusters and  discarded from the list. The sigma-- and 
radius--clipping limits are changed for 53 clusters to avoid galaxies in nearby
structures being included as cluster members
(the sigma--clipping is typically  changed from
$3 \sigma$ to   $2.5 \sigma$ or $2.0 \sigma$ -- note that this
affects only the choice of
galaxies from which $\sigma_{\rm v, cl}$ and $z_{\rm Cl}$ are determined; cluster
membership is still defined to  be within $3\sigma_{\rm v, cl}$ and $R_{200}$).
For 67 clusters, nothing is changed.

\begin{figure}
%[tbh]
\begin{center}
%\setlength{\fboxsep}{-\fboxrule}
%\fbox{
%\includegraphics*[bb=3.7cm 2cm 16.8cm 25.5cm,width=0.6\hsize]
\includegraphics[width=1.0\hsize]
{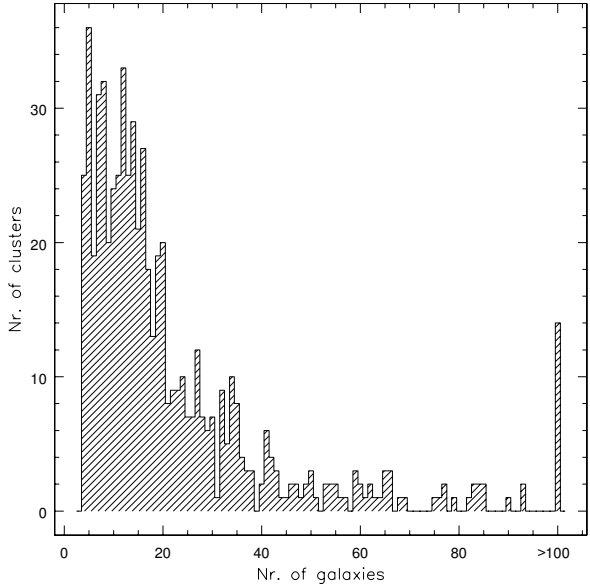}
%}
\caption{~Histogram of the number of spectroscopic members within $\pm
  3\sigma_{\rm v, cl}$
  and $1R_{200}$ for our final cluster sample. Those clusters with
  $\ge100$ members are grouped into a single bin. The cluster with the
  most members is C4\_DR3\_3031 (Abell~2199) with 263 members.}
\label{fig:ngal_histo}
\end{center}
\end{figure}

At this stage, we discard those systems that only contain 2 or 3 galaxies
within $3 \sigma_{\rm v, cl}$ and $1 R_{200}$. This leaves
625 entries in our cluster sample.
Fig.~\ref{fig:ngal_histo} shows a histogram of the number of
spectroscopic members for our cluster sample. The original C4 catalog
considered only clusters with at least 8 galaxies within 1$h^{-1}$Mpc and
$\Delta z = 0.02$. This is a much larger volume than we probe, and  it
is thus not surprising that we typically assign fewer galaxies to each cluster.

\begin{figure}
%[tbh]
\begin{center}
%\setlength{\fboxsep}{-\fboxrule}
%\fbox{
%\includegraphics*[bb=3.7cm 2cm 16.8cm 25.5cm,width=0.6\hsize]
\includegraphics[width=1.0\hsize]
{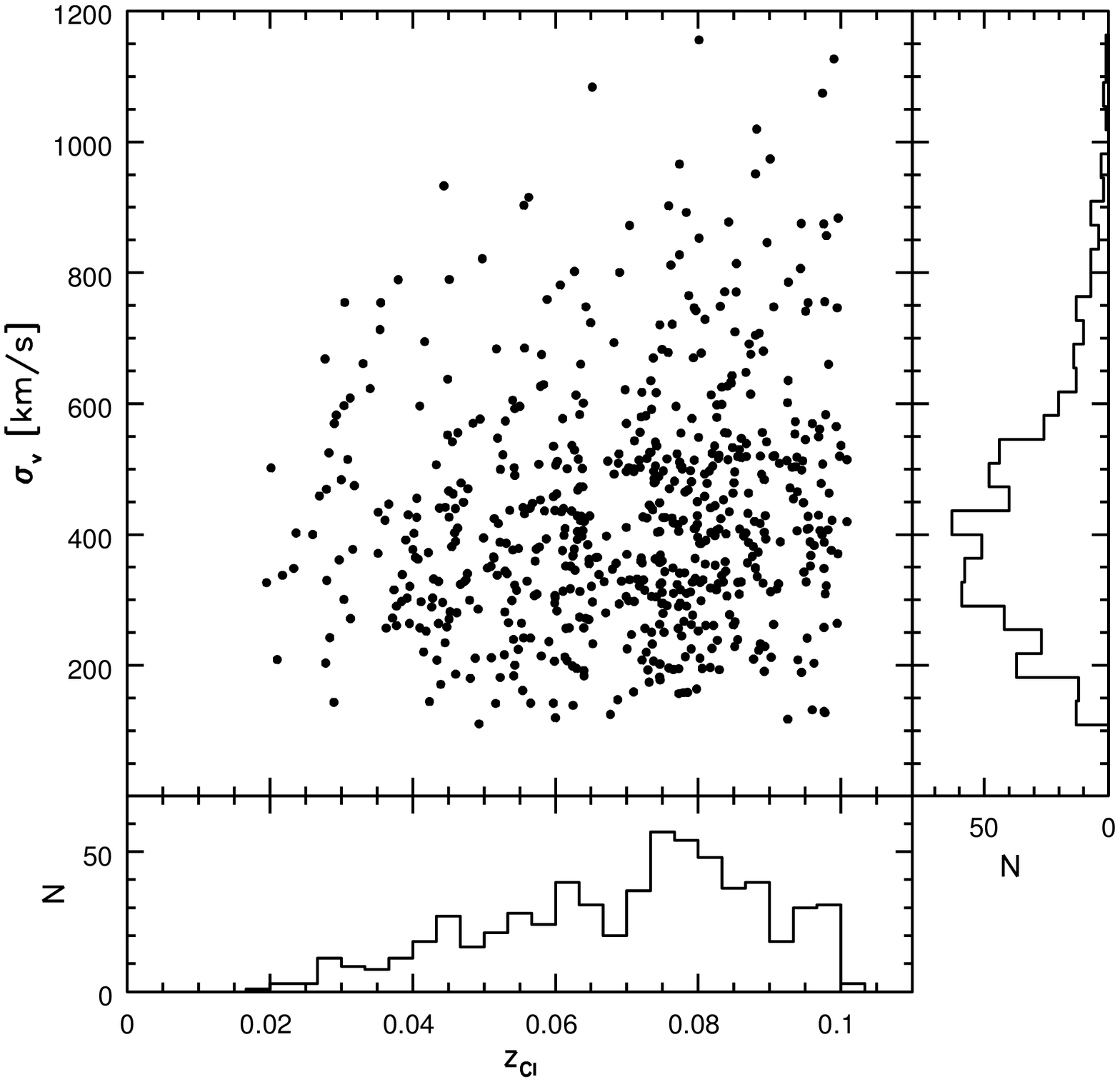}
%}
\caption{~Distribution of the redshifts and velocity dispersions we measure
  for our final cluster sample. Error bars have been omitted for
  clarity. The marginal histograms show the distribution binned in redshift
  (lower panel) and in velocity dispersion (right panel).}  
\label{fig:redshift-sigma}
\end{center}
\end{figure}

Our final cluster sample spans a large range in velocity dispersion, from
groups of $\lesssim$ 200 km/s to clusters of over 1000 km/s
(Fig.~\ref{fig:redshift-sigma}). 
The positions of our BCGs as well as the cluster properties determined by
our algorithm are listed in Table~\ref{tab:catalog}.
Those systems with velocity dispersions $\lesssim$ 300 - 400 km/s are more
likely to be galaxy groups than galaxy clusters. We use the terms `cluster'
and `brightest cluster galaxy' loosely in this paper to refer to both
clusters and groups.

\begin{table}
\begin{center}
\caption{Excerpt of our catalog of Brightest Cluster Galaxies for the C4
cluster catalog, along with the cluster properties derived by our
algorithm. The complete catalog is available electronically.            
Note that columns (1) and (2) refer to the C4 cluster ID (see footnote in
Section~\ref{sect:bcg-sel}); column (7) gives
the number of galaxies from which the cluster redshift and velocity
dispersion were determined.
}
\label{tab:catalog}
\renewcommand{\arraystretch}{1.3}
\begin{tabular}{c c r r r r r}
ID\_2&
ID\_3&
\multicolumn{1}{c}{$\alpha_{\rm BCG}$ [$^{\circ}$]} & 
\multicolumn{1}{c}{$\delta_{\rm BCG}$ [$^{\circ}$]} & 
\multicolumn{1}{c}{$z_{\rm cl}$} & 
\multicolumn{1}{c}{$\sigma_{\rm v}$ [km/s]} & 
%\multicolumn{1}{c}{$R_{200}$ [Mpc]} & 
\multicolumn{1}{c}{$N_{\rm gal}$} \\
\multicolumn{1}{c}{ (1) } &
\multicolumn{1}{c}{ (2) } &
\multicolumn{1}{c}{ (3) } &
\multicolumn{1}{c}{ (4) } &
\multicolumn{1}{c}{ (5) } &
\multicolumn{1}{c}{ (6) } &
%\multicolumn{1}{c}{ (7) } &
\multicolumn{1}{c}{ (7) } \\
\hline
\hline
1000 & 1000 & 202.5430 & -2.1050 & 0.087 & $648^{81}_{-90}$ & 35\\
1001 & 1001 & 208.2767 & 5.1497 & 0.079 & $746^{57}_{-59}$ & 82\\
1002 & 1002 & 159.7776 & 5.2098 & 0.069 & $800^{57}_{-56}$ & 90\\
1003 & 1004 & 184.4214 & 3.6558 & 0.077 & $966^{58}_{-60}$ & 127\\
1004 & 1005 & 149.7174 & 1.0592 & 0.081 & $458^{48}_{-52}$ & 21\\
1005 & 1006 & 191.3037 & 1.8048 & 0.048 & $340^{53}_{-55}$ & 24\\
1007 & 1009 & 177.4721 & 5.7008 & 0.075 & $404^{40}_{-43}$ & 27\\
1009 & 1011 & 198.0566 & -0.9745 & 0.085 & $631^{71}_{-75}$ & 36\\
1010 & 1012 & 192.0112 & -1.6528 & 0.088 & $420^{77}_{-80}$ & 14\\
1011 & 1013 & 227.1073 & -0.2663 & 0.091 & $748^{61}_{-66}$ & 42\\
- & 1014 & 220.1785 & 3.4654 & 0.027 & $459^{34}_{-35}$ & 105\\
1013 & 1015 & 203.0701 & 1.2233 & 0.079 & $327^{59}_{-74}$ & 10\\
1014 & 1016 & 175.2992 & 5.7348 & 0.098 & $660^{54}_{-56}$ & 55\\
1015 & 1017 & 182.5701 & 5.3860 & 0.077 & $596^{53}_{-59}$ & 41\\
1016 & 1018 & 154.9344 & -0.6384 & 0.093 & $455^{137}_{-153}$ & 17\\
1017 & 1019 & 227.8480 & -0.0593 & 0.091 & $509^{59}_{-64}$ & 36\\
1018 & 1020 & 214.3980 & 2.0532 & 0.054 & $605^{51}_{-53}$ & 69\\
1019 & 1021 & 195.7262 & 3.3174 & 0.071 & $496^{56}_{-59}$ & 24\\
- & 1024 & 199.8197 & -0.9954 & 0.083 & $579^{87}_{-91}$ & 37\\
1023 & 1025 & 153.4095 & -0.9254 & 0.045 & $790^{52}_{-57}$ & 66\\
1341 & 1026 & 155.6325 & 2.3608 & 0.072 & $580^{71}_{-76}$ & 26\\
- & 1027 & 191.9269 & -0.1373 & 0.088 & $1020^{87}_{-91}$ & 55\\
1027 & 1028 & 199.1357 & 0.8702 & 0.080 & $364^{54}_{-60}$ & 16\\
1030 & 1030 & 206.1357 & 2.9541 & 0.077 & $511^{66}_{-73}$ & 29\\
- & 1032 & 218.4964 & 3.7780 & 0.029 & $570^{58}_{-60}$ & 76\\
1032 & 1033 & 211.4731 & -1.2045 & 0.054 & $184^{47}_{-64}$ & 4\\
- & 1034 & 165.7398 & 7.6039 & 0.072 & $321^{41}_{-44}$ & 20\\
1034 & 1036 & 192.3087 & -1.6874 & 0.085 & $771^{63}_{-67}$ & 64\\
1036 & 1038 & 151.8861 & 0.5942 & 0.097 & $550^{61}_{-65}$ & 29\\
- & 1039 & 186.8781 & 8.8246 & 0.090 & $846^{63}_{-66}$ & 50\\
1037 & 1040 & 213.6360 & 1.7316 & 0.054 & $299^{37}_{-39}$ & 16\\
1038 & 1041 & 179.3707 & 5.0891 & 0.076 & $678^{66}_{-69}$ & 62\\
1039 & 1042 & 228.8088 & 4.3862 & 0.098 & $857^{86}_{-87}$ & 53\\
1040 & 1043 & 168.3339 & 2.5467 & 0.074 & $403^{56}_{-65}$ & 30\\
1041 & 1044 & 194.6729 & -1.7615 & 0.084 & $771^{78}_{-81}$ & 60\\
1044 & 1047 & 197.3295 & -1.6225 & 0.083 & $521^{84}_{-89}$ & 22\\
1045 & 1048 & 205.5402 & 2.2272 & 0.077 & $828^{77}_{-80}$ & 75\\
- & 1050 & 206.1075 & 2.1099 & 0.072 & $514^{84}_{-92}$ & 14\\
1176 & 1051 & 189.7348 & 6.1584 & 0.074 & $486^{74}_{-82}$ & 19\\
1048 & 1053 & 147.9551 & 1.1339 & 0.063 & $346^{36}_{-39}$ & 13\\
1049 & 1054 & 188.7581 & 1.7986 & 0.079 & $577^{61}_{-65}$ & 35\\
1051 & 1057 & 177.8878 & 5.1015 & 0.075 & $251^{22}_{-25}$ & 11\\
1052 & 1058 & 195.7191 & -2.5164 & 0.083 & $749^{62}_{-65}$ & 68\\
- & 1059 & 156.4666 & 1.1906 & 0.097 & $478^{127}_{-154}$ & 12\\
%1053 & 1061 & 228.2207 & 4.5140 & 0.038 & $789^{54}_{-60}$ & 63\\
%- & 1066 & 202.7951 & -1.7303 & 0.085 & $814^{48}_{-48}$ & 84\\
%1056 & 1067 & 212.8853 & 4.8688 & 0.094 & $208^{49}_{-60}$ & 7\\
\end{tabular}
\end{center}
\end{table}

\begin{figure*}
%[p]
\begin{center}
%\setlength{\fboxsep}{-\fboxrule}
%\fbox{
\includegraphics*[bb=4.7cm 6.9cm 17.2cm 24.5cm,width=0.8\hsize]
{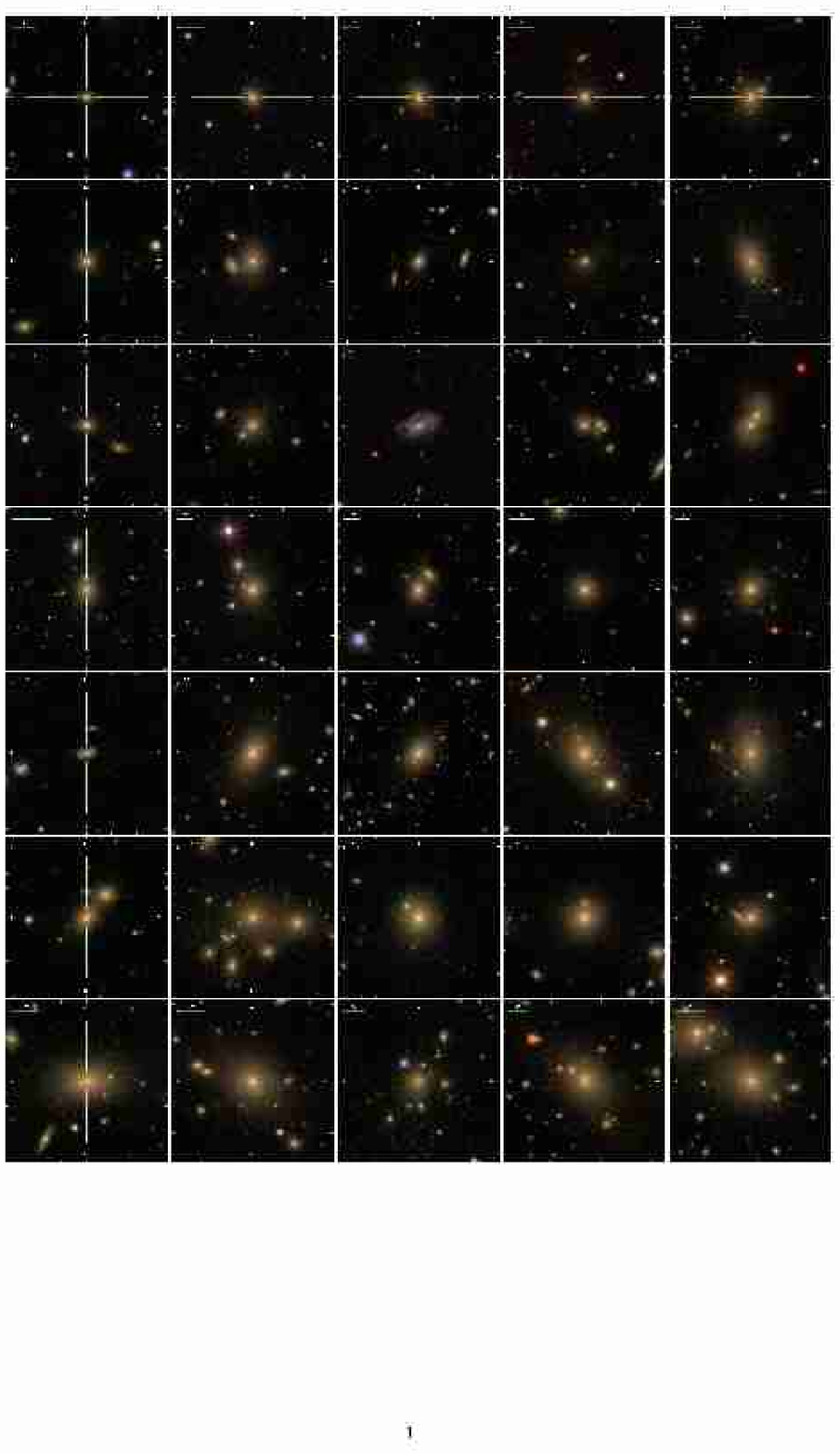}
%}
\caption{A gallery of BCGs. Each image is 200~kpc on the
  side. From left to right, top to bottom:\newline
C4\_DR3\_3351~($\sigma_{\rm v, cl}=159\,$km/s),
C4\_2042~($\sigma_{\rm v, cl}=195\,$km/s),
C4\_DR3\_1343~($\sigma_{\rm v, cl}=210\,$km/s),
C4\_3275~($\sigma_{\rm v, cl}=233\,$km/s),
C4\_3087~($\sigma_{\rm v, cl}=256\,$km/s),
C4\_DR3\_3201~($\sigma_{\rm v, cl}=264\,$km/s),
C4\_DR3\_3106~($\sigma_{\rm v, cl}=283\,$km/s),
C4\_1224~($\sigma_{\rm v, cl}=299\,$km/s),
C4\_3206~($\sigma_{\rm v, cl}=312\,$km/s),
C4\_2065~($\sigma_{\rm v, cl}=324\,$km/s),
C4\_DR3\_3272~($\sigma_{\rm v, cl}=329\,$km/s),
C4\_DR3\_1366~($\sigma_{\rm v, cl}=342\,$km/s),
C4\_DR3\_2140~($\sigma_{\rm v, cl}=355\,$km/s),
C4\_3059~($\sigma_{\rm v, cl}=365\,$km/s),
C4\_DR3\_3386~($\sigma_{\rm v, cl}=378\,$km/s),
C4\_DR3\_1355~($\sigma_{\rm v, cl}=392\,$km/s),
C4\_3068~($\sigma_{\rm v, cl}=403\,$km/s),
C4\_DR3\_3034~($\sigma_{\rm v, cl}=410\,$km/s),
C4\_1025~($\sigma_{\rm v, cl}=425\,$km/s),
C4\_1226~($\sigma_{\rm v, cl}=435\,$km/s),
C4\_DR3\_1360~($\sigma_{\rm v, cl}=448\,$km/s),
C4\_3055~($\sigma_{\rm v, cl}=467\,$km/s),
C4\_DR3\_1356~($\sigma_{\rm v, cl}=484\,$km/s),
C4\_1024~($\sigma_{\rm v, cl}=500\,$km/s),
C4\_1076~($\sigma_{\rm v, cl}=509\,$km/s),
C4\_1191~($\sigma_{\rm v, cl}=519\,$km/s),
C4\_1073~($\sigma_{\rm v, cl}=536\,$km/s),
C4\_DR3\_3105~($\sigma_{\rm v, cl}=556\,$km/s),
C4\_3009~($\sigma_{\rm v, cl}=583\,$km/s),
C4\_DR3\_1275~($\sigma_{\rm v, cl}=617\,$km/s),
C4\_DR3\_3027~($\sigma_{\rm v, cl}=670\,$km/s),
C4\_1058~($\sigma_{\rm v, cl}=721\,$km/s),
C4\_DR3\_3084~($\sigma_{\rm v, cl}=781\,$km/s),
C4\_DR3\_3349~($\sigma_{\rm v, cl}=884\,$km/s),
C4\_3002~($\sigma_{\rm v, cl}=1156\,$km/s).
}
\label{fig:bcg_gallery}
\end{center}
\end{figure*}

Fig.~\ref{fig:bcg_gallery} presents a gallery of BCGs, sorted according to
the velocity dispersion of the parent cluster (every 18th BCG is
shown). The appearance of the BCG is certainly
a function of $\sigma_{\rm v, cl}$, but it is not a monotonic one. While the BCGs of
groups are mostly  fairly isolated, rather spherical  elliptical galaxies, the
BCGs of more massive systems are in general larger and more elongated, they often
have a cD envelope and are surrounded by many satellite
galaxies.

\section{Comparison to C4}
\label{sect:compare}

\subsection{Selected BCGs}

For 31 clusters in the final sample, the BCG we identified corresponds to both the C4
{\it mean} galaxy and the spectroscopic BCG identified by C4. 19 of these were
classified automatically, as there was no other candidate in our list.

In 343 clusters, our BCG is the same as the  C4 spectroscopic BCG, but not the {\it
  mean} galaxy (183 automatically classified).

In 41 cases, the BCG is the same as the C4 {\it mean} galaxy, but not the C4
spectroscopic BCG. These are predominantly small systems, where the C4
spectroscopic BCG belongs to another system.

The BCGs of 210 clusters correspond neither to the {\it mean} galaxy, nor to
the C4 spectroscopic BCG. Of these, 141 (i.e. 23\% of the whole sample) are
not in the spectroscopic 
catalog. \\
\newline
\begin{figure}
%[tbp]
\begin{center}
%\setlength{\fboxsep}{-\fboxrule}
%\fbox{
%\includegraphics*[bb=3.7cm 2cm 16.8cm 25.5cm,width=0.6\hsize]
\includegraphics[width=1.0\hsize]
{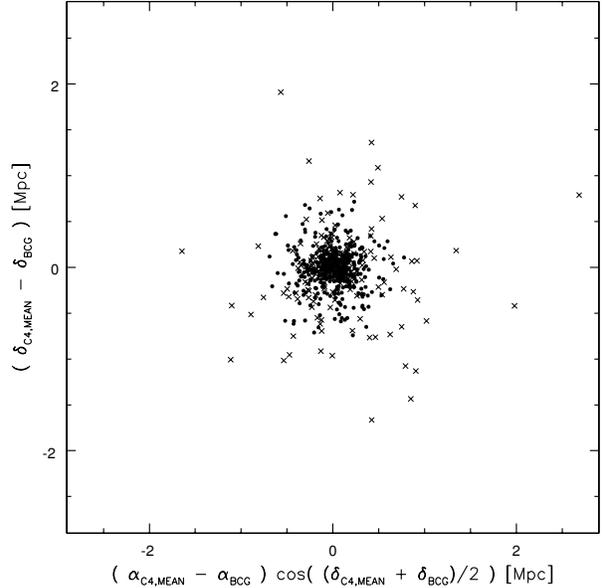}
%}
\caption{The differences in the positions of the C4 {\it mean} galaxy and
  our BCG, expressed in Mpc. Clusters in which the BCG is neither the {\it
    mean} galaxy nor the C4 spectroscopic BCG are shown as crosses, the
  other clusters are shown as filled circles.}
\label{fig:position_comp}
\end{center}
\end{figure}
In Fig. \ref{fig:position_comp}, we compare the positions of our BCGs with those of
the C4 {\it mean} galaxies.
For the majority of the clusters, these two positions fall within 500~kpc of
one another. 53 BCGs lie farther than 500~$h^{-1}$~kpc from the {\it
  mean} galaxy \footnote{500 
$h^{-1}$
kpc is 
the radius within which the C4 algorithm identifies its
spectroscopic BCG} and of these, 21 lie farther away than 1~Mpc. At these
distances, the original cluster center is well in the outskirts of the
structures we identify.

An example is C4\_DR3\_1283 (see Fig.~\ref{fig:c4_1283}), where the BCG and the
original {\it  mean} galaxy are separated by 2.8~Mpc, equivalent to $1.6 R_{200}$
according to the velocity dispersion we measure. The
{\it mean} galaxy is a 
$10^{10} M_{\odot}$, star--forming galaxy at $z=0.099$, in a rich field that
is likely to be an infall region of the cluster.
The BCG we identified (a
$4\times10^{11} 
M_{\odot}$ elongated elliptical with a cD envelope) is at the center of a
cluster of 22 other galaxies. It is curious that this cluster was not picked
up by the C4 algorithm. The comparatively high redshift of the
cluster ($z=0.095$) may possibly  play a role in this. The cases of the other clusters
with large 
separations between the BCG we identify and the C4 {\it mean} galaxy are similar, 
though less striking.

\subsection{Cluster redshift measurements}

The original C4 algorithm measures the cluster redshift using the biweight
estimator of \citet{bfg90} applied to all spectroscopic members within an
aperture of 1~$h^{-1}$~Mpc from the luminosity-weighted geometrical center
of the cluster and within $\Delta z = 0.02$ of the peak of the redshift
histogram defined by these galaxies. While 1~$h^{-1}$~Mpc is comparable to
$R_{200}$ for a cluster with a velocity dispersion of $\sim~600$~km/s, the
corresponding redshift interval from which we determine the redshift of such
a cluster would be only $\Delta z = 0.006 (1+z_{C\rm cl})$.

\begin{figure}
%[tbhp]
\begin{center}
%\setlength{\fboxsep}{-\fboxrule}
%\fbox{
%\includegraphics*[bb=3.7cm 2cm 16.8cm 25.5cm,width=0.6\hsize]
\includegraphics[width=1.0\hsize]
{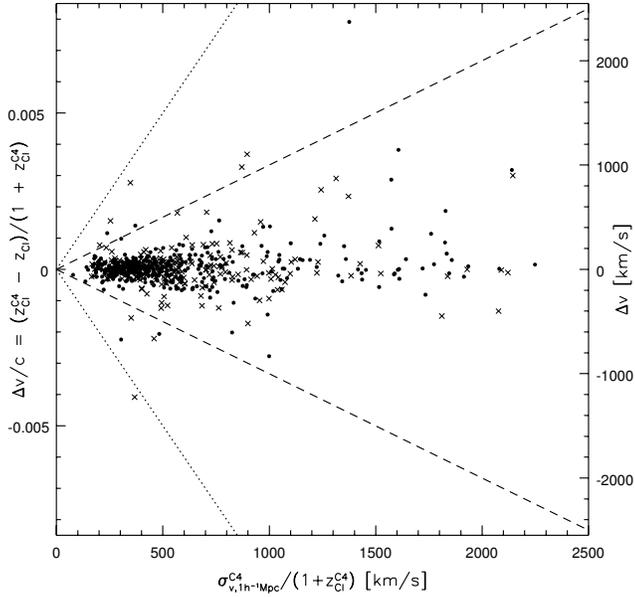}
%}
\caption{The differences between the C4 and our cluster redshift
  measurement (expressed as relative velocity) compared to the velocity
  dispersion $\sigma^{\rm C4}_{\rm v, cl}$ measured by C4 within
  1~$h^{-1}$~Mpc (expressed in the cluster rest-frame). The dashed line
  indicates $\sigma^{\rm C4}_{\rm v, cl}$, and the 
  dotted line $3 \sigma^{\rm C4}_{\rm v, cl}$. As in Fig. \ref{fig:position_comp},
  clusters in which the BCG is neither the {\it 
    mean} galaxy nor the C4 spectroscopic BCG are shown as crosses, and as
  filled circles otherwise.}
\label{fig:diff_sigma}
\end{center}
\end{figure}

In Fig.~\ref{fig:diff_sigma}, we plot the relative differences between the
two redshift measurements as a function of the velocity dispersion
$\sigma^{C4}_{\rm v, cl}$ measured by C4 within 1~$h^{-1}$~Mpc.  Our new
redshift lies outside the $1 \sigma^{C4}_{\rm v, cl}$ redshift interval for
only a few clusters , and only one lies outside the $3 \sigma^{C4}_{\rm v,
  cl}$ limit.  The most notable outlier is C4\_DR3\_2163, with a velocity offset
of about $2400$~km/s. In our cluster sample, C4\_DR3\_2163 is a group of four
galaxies at a redshift of $z_{\rm cl} = 0.070$ and a velocity dispersion of
$225^{+73}_{-101}$km/s. Its redshift histogram shows another spike of
galaxies at a redshift of $0.082$ which can be associated with C4\_2124(see
Fig.~\ref{fig:c4_2163} and Fig.~\ref{fig:c4_2101}).  We thus conclude that
the C4 algorithm considered these two structures as a single cluster,
whereas our algorithm was able to separate them.

\subsection{Velocity dispersion measurements}
\label{sect:veldisp_comp}

The C4 catalog provides five measures of a cluster's velocity dispersion,
measured within 0.5, 1, 1.5, 2, and 2.5 $h^{-1}$~Mpc from the positional
centroid measured on the sky.  A first estimate for the velocity dispersion
is made using the biweight estimator for all galaxies within $\Delta z =
0.02$ of the estimated cluster redshift. The final velocity dispersion
(expressed in the observer's frame) is recomputed from galaxies within the
redshift interval equal to $\pm 4 \sigma_{\rm v, cl}^{\rm C4}$.

\begin{figure}
%[tbhp]
\begin{center}
%\setlength{\fboxsep}{-\fboxrule}
%\fbox{
%\includegraphics*[bb=3.7cm 2cm 16.8cm 25.5cm,width=0.6\hsize]
\includegraphics[width=1.0\hsize]
{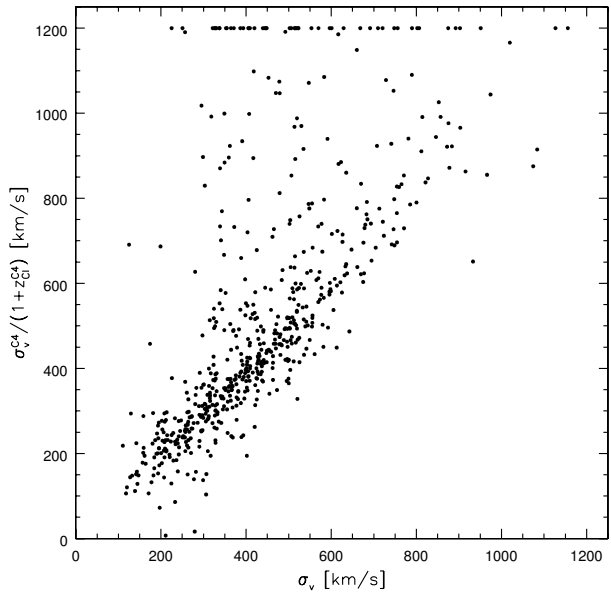}
%}
\caption{Comparison of the cluster velocity dispersions measured by C4 and
  by our algorithm. We adopt the C4 velocity dispersion measured within the radius
  that best corresponds to the virial radius measured from
  our algorithm. Those C4 values that are larger than 1200~km/s are plotted at
  this value.}  
\label{fig:sigma-comp_cut}
\end{center}
\end{figure}

Fig.~\ref{fig:sigma-comp_cut} shows a comparison of our velocity dispersions
to the C4 velocity dispersion within the radius that best corresponds to our
estimate of the virial radius.  At low velocity dispersions ($\sigma_{\rm v, cl}
\lesssim 600 \mbox {km/s}$), the measurements agree well for many clusters.
At higher velocity dispersions (as measured by C4) our algorithm obtains
lower values for the majority of the clusters.  As was previously shown for
C4\_DR3\_2163, this is mainly caused by the fact that our iterative algorithm
separates neighboring groups/clusters better than C4.

\section{Radio--loud AGN activity of BCGs}
\label{sect:radio}

\begin{figure*}
%[tbp]
\begin{center}
%\setlength{\fboxsep}{-\fboxrule}
%\fbox{
%\includegraphics*[bb=3.7cm 2cm 16.8cm 25.5cm,width=0.6\hsize]
\includegraphics[width=0.49\hsize]
{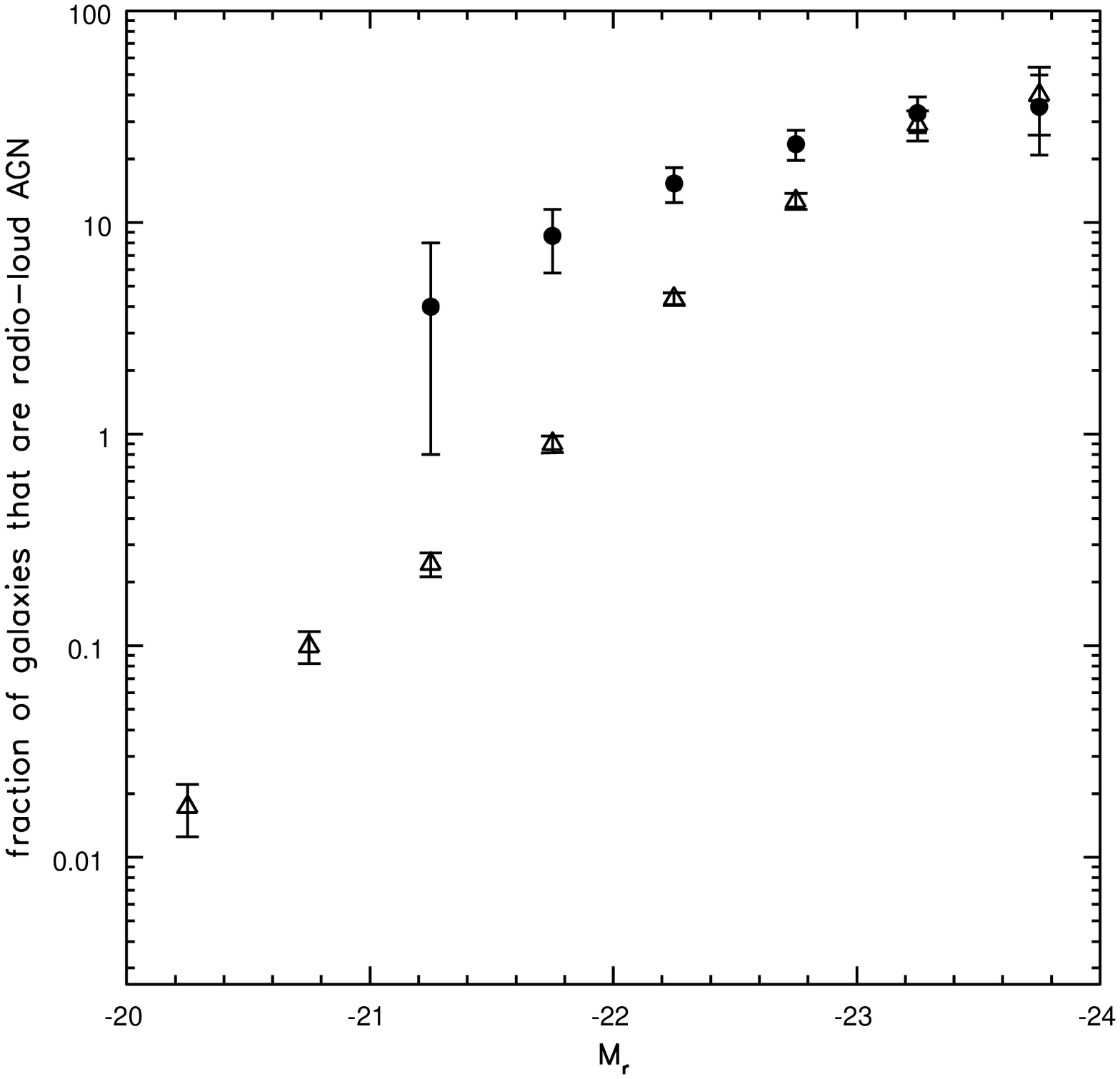}
\includegraphics[width=0.49\hsize]
{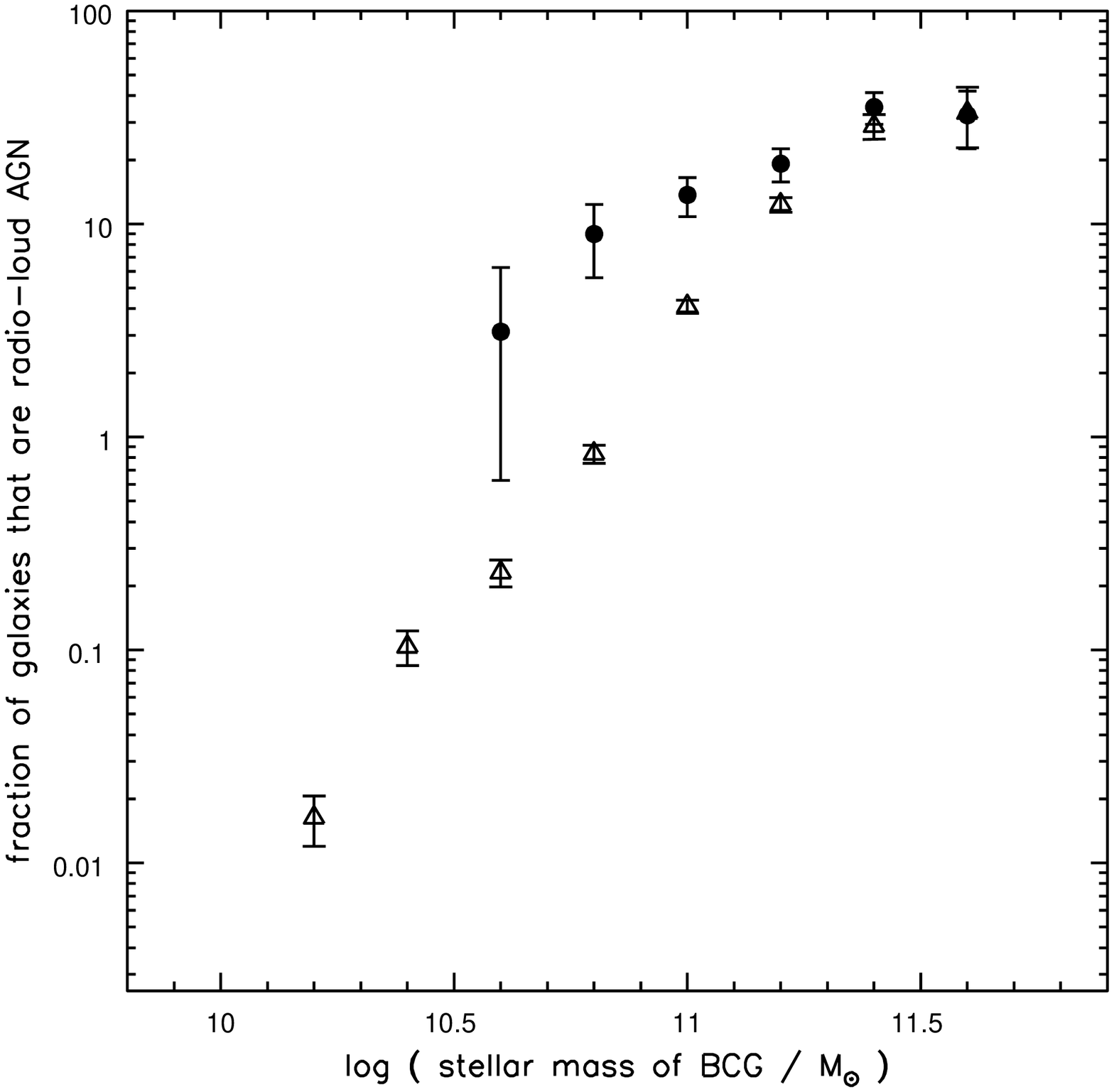}
%}
\caption{~The fraction of galaxies that are radio loud AGN, as a function of
  luminosity (left) and stellar mass (right). Results are plotted for all
  galaxies at $z<0.1$ (open
  triangles), and for the BCGs (solid circles). Galaxies are considered
  radio--loud if their 1.4~GHz radio luminosity is greater than $10^{23}
  \mbox{ W/Hz}$, and they are not classified as star-forming.}
\label{fig:radio-bcgs}
\end{center}
\end{figure*}

It is known that BCGs often host radio--loud AGNs \citep[e.g.][]{bur90}.  It
has previously not been investigated whether this is simply a consequence of
the strong dependence of radio--AGN activity on galaxy stellar mass ($f_{\rm
  radio-loud} \propto M_*^{2.5}$ \citep{bkh05b}), or whether this is a
special property of BCGs. With our large sample of BCGs, it is possible to
disentangle the mass dependence and the influence of the cluster
environment. Fig.~\ref{fig:radio-bcgs} compares the fraction of galaxies
that are radio--loud for the BCG sample with the results found for all SDSS
galaxies that overlap the NVSS and FIRST surveys.  BCGs of all luminosities / masses are
more likely to be radio--loud than other galaxies of the same luminosity / stellar mass.
This enhancement ranges from a factor of 10 at masses of $5 \times 10^{10}
M_\odot$ to less than a factor of two above $4 \times 10^{11} M_\odot$. 
\citet{bkh05b} have argued that radio--AGN activity is fuelled from
the hot gas envelopes of galaxies. In this scenario, groups and clusters
provide an additional hot gas reservoir, which boosts the radio--AGN
activity of the central galaxies.
This
result, and its implications for the cooling flow model, are investigated in
more detail in the accompanying paper \citep{bes06b}.

\begin{figure}
%[tbp]
\begin{center}
%\setlength{\fboxsep}{-\fboxrule}
%\fbox{
%\includegraphics*[bb=3.7cm 2cm 16.8cm 25.5cm,width=0.6\hsize]
\includegraphics[width=1.0\hsize]
{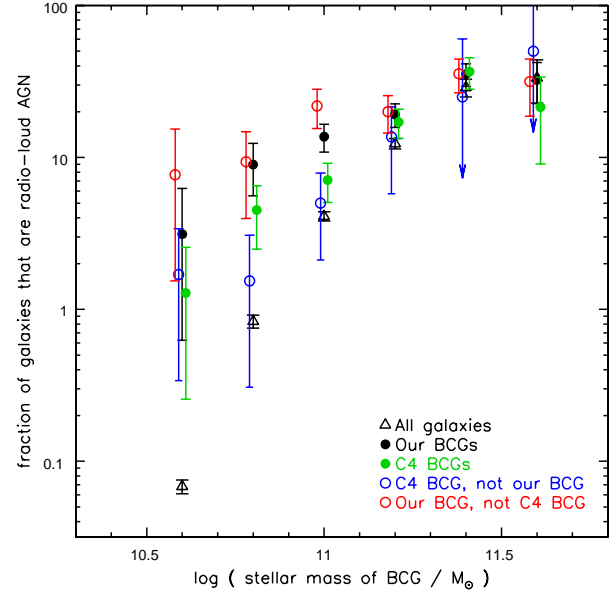}
%}
\caption{~The radio--loud fraction of BCGs identified by C4 (green, filled
  circles), compared to those identified by our method (black, filled
  circles). We also show the radio--loud fractions derived only from
  clusters where the two choices differ: C4 BCGs are shown as blue, open
  circles, our BCGs are shown as red, open circles. 
 Arrows indicate that the result is an upper limit
only. The points are slightly offset in mass for clarity.}  
\label{fig:radio-c4bcgs}
\end{center}
\end{figure}

\begin{figure}
%[tbp]
\begin{center}
%\setlength{\fboxsep}{-\fboxrule}
%\fbox{
%\includegraphics*[bb=3.7cm 2cm 16.8cm 25.5cm,width=0.6\hsize]
\includegraphics[width=1.0\hsize]
{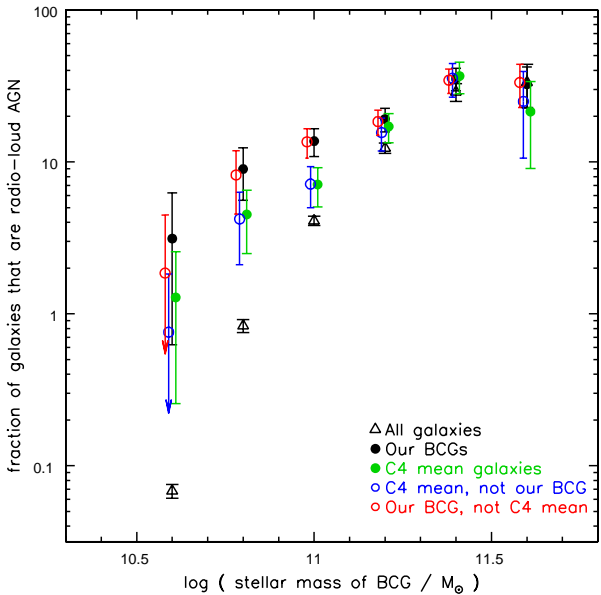}
%}
\caption{~The same as Fig.~\ref{fig:radio-c4bcgs}, but for the C4
  \textit{mean} galaxies.}  
\label{fig:radio-c4mean}
\end{center}
\end{figure}

In this paper, we use the enhanced radio--AGN fraction of BCGs as a diagnostic
for the {\it reliability} of the BCG selection, i.e. are our BCGs indeed
better tracers of the bottom of the clusters' potential wells than the
original C4 BCGs? In Fig.~\ref{fig:radio-c4bcgs}, we repeat the previous
analysis for C4 BCGs.
At the highest mass bins ($> 10^{11.1} M_{\odot}$), the C4 BCGs have
a similar radio--loud 
fraction to our BCGs, but at lower masses, the radio--loud fraction is lower
than in ours. We also investigate the radio--loud fraction  in
clusters where our method and the C4 algorithm select different BCGs  (252
clusters). Here, the difference becomes even clearer: our BCGs have a much
higher  radio--loud fraction than those identified by C4.

In Fig.~\ref{fig:radio-c4mean} we repeat this analysis on the C4
\textit{mean} galaxies. A similar trend as for the C4 BCGs is seen.

These results can easily be explained by the difference in selection
algorithm: in a cluster where the BCG is the most massive and
brightest galaxy and has spectroscopic information available, C4 will
correctly identify it, and hence the agreement
is good in the high mass bins. However, if C4 misses the `real' BCG,
for example due to fiber collisions, it classifies a less
massive galaxy  (typically not at the bottom of the potential well)
as a BCG. Since these galaxies are normal cluster galaxies, their
radio--loud fraction is lower than that of BCGs of equal mass. Hence, the C4
algorithm results in  an underestimate of the 
radio--loud fraction at low masses.

\section{Optical properties of BCGs}
\label{sect:optical}

Our large sample of BCGs and the extensive SDSS database allow us to compare
the structural properties of BCGs with those of non-BCGs in order to
distinguish the roles of mass and environment in governing their properties.
In order to perform the comparison, we construct a comparison sample of
``control'' galaxies for each BCG by finding its three closest neighbors in
a space spanned by (the logarithm) of galaxy stellar mass, redshift, and
$g-r$ color. The ``redshift axis'' of this space is scaled by a factor of
five, so that a difference of 0.1 in $\log M_{\star}$ corresponds to a
redshift difference of 0.02 , and a difference of 0.1 in $g-r$. The matching
is performed in order of decreasing BCG mass, and galaxies are not allowed
to enter the comparison sample more than once.

By matching in redshift, redshift--dependent aperture effects are avoided.
The matching in $g-r$ ensures similar stellar populations and
mass--to--light ratios in the BCGs and their controls, i.e. effectively,
early--type BCGs are matched to early--type galaxies. Without the $g-r$
matching, there are more late--type galaxies in the control sample than the
BCG sample. But since our method of selecting BCGs is somewhat biased
towards selecting early--type galaxies over late--types, we cannot
unambiguously deduce that BCGs are more likely to be early--types.

The pool of galaxies from which the control sample is drawn consists of all
galaxies in the DR4 spectroscopic catalog that have not been identified as a
BCG in our sample. Yet, at the very massive end ($\log M_{\star}/M_{\odot} >
11.5$) there are not enough non--BCGs to provide three control galaxies per
BCG. On the other hand, since we draw comparison galaxies from the full DR4
database, whereas the C4 catalog is based on DR3, the control sample is
``contaminated'' by BCGs that failed to enter our sample, particularly for
the most massive galaxies.
\footnote{We have also attempted to clean the matched sample by
applying the algorithm described in Sect.~\ref{sect:veldisp} to these
galaxies, and considering those which are the brightest in structures of
more than four galaxies to be possible BCGs. About one third of the matched
galaxies are such BCG candidates. Basing our analysis on the remaining BCGs
and the respective matched galaxies does not qualitatively alter our results.}
Hence, for a large part of the analysis, we 
restrict the sample to BCGs with $\log M_{\star}/M_{\odot} < 11.3$; this
avoids the problem of finding three suitable non--BCGs for the comparison
sample, and also purifies the comparison sample since at very high masses, a
significant fraction of the comparison galaxies may themselves be BCGs. With
these criteria, we construct two comparison samples, one for the full set of
BCGs, and the other for the subset of BCGs with spectroscopic information.

In order to study scaling relations over a larger range in mass, we
construct two more comparison samples (one drawn from all BCGs and one of
them for BCGs with SDSS spectroscopy) with only one matched galaxy.
Restricting to one comparison galaxy per BCG minimizes the problem of lack
of comparison galaxies at the
high mass end. For this matching, we restrict our analysis to only
early--type BCGs and comparison galaxies by requiring $M_g - M_r > 0.75$ and
{\tt fracDeV\_r} $>0.8$.

Our four comparison samples are summarized below:
\setlength{\leftmargini}{0.9cm}
\begin{itemize}
\setlength{\itemindent}{0cm}
\setlength{\labelsep}{0.1cm}
\setlength{\labelwidth}{1cm}
\item[CS3p:] A comparison sample of three matching galaxies for BCGs with
  $\log M_{\star}/M_{\odot} < 11.3$. The galaxies are matched in mass,
  redshift, and $g-r$ color.
\item[CS3s:] Like CS3p, but for BCGs contained in the spectroscopic
  database.
\item[CS1p:] A comparison sample of one matching galaxy for each BCG (with
   no upper mass limit).  The sample is matched in mass, redshift, and
   $g-r$ color, and restricted to only early-type BCGs and comparison
   galaxies ($M_g - M_r > 0.75$ and {\tt fracDeV\_r} $>0.8$).
\item[CS1s:] Like CS1p, but for BCGs contained in the spectroscopic
  database.
\end{itemize}

The first two samples are used to compare the {\it distributions} of physical  
parameters for BCGs and non-BCGs. The latter two samples are used to
analyze  early-type galaxy scaling relations, and to probe them to the
highest masses.\\

\begin{figure*}
%[tbp]
\begin{center}
%\setlength{\fboxsep}{-\fboxrule}
%\fbox{
%\includegraphics*[bb=3.7cm 2cm 16.8cm 25.5cm,width=0.6\hsize]
\includegraphics[trim=0 5cm 0 0,width=1.0\hsize]
{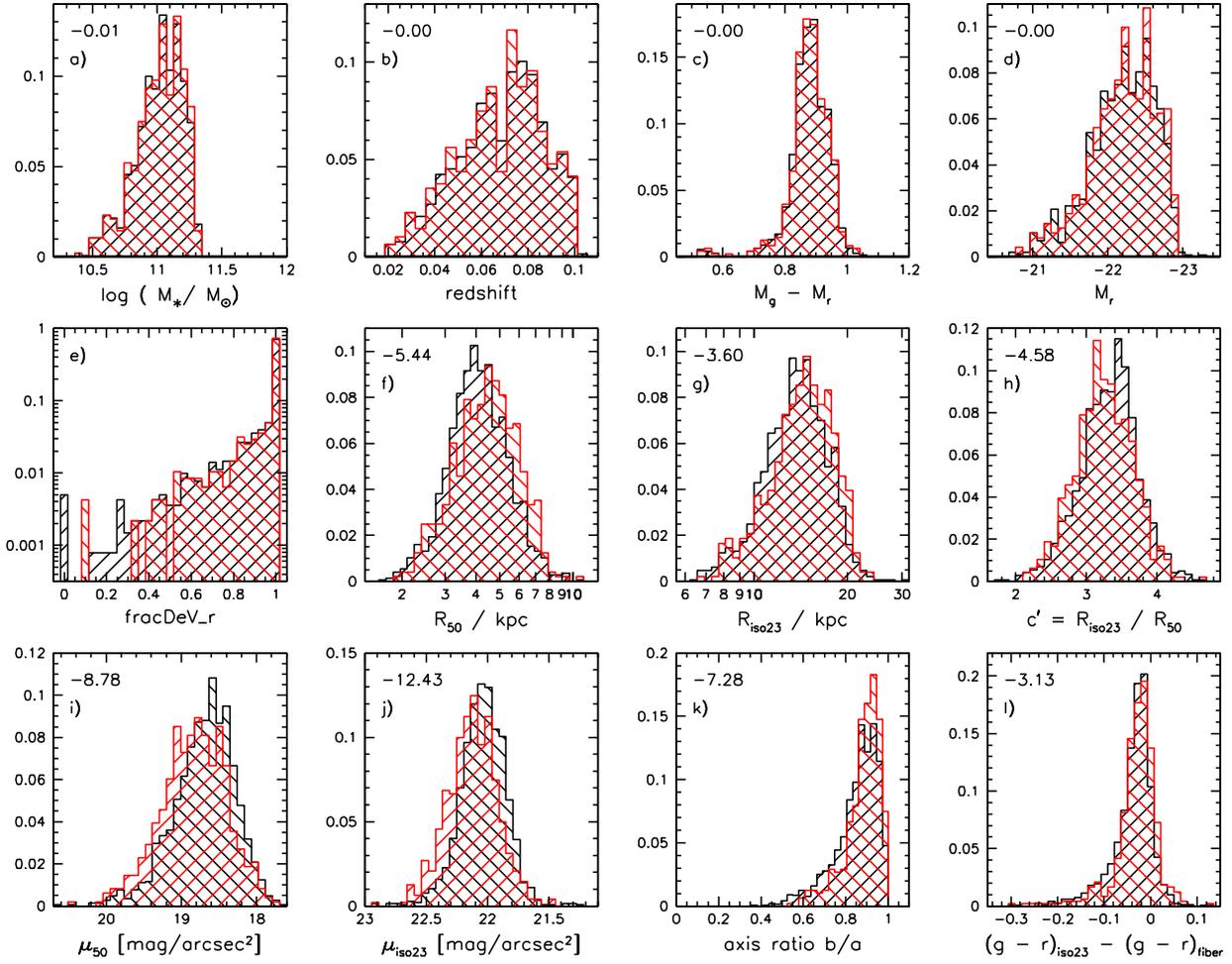}
%}
\caption{~Distributions of a variety of photometric parameters for the sample of
  BCGs (red histograms) and the comparison sample, matched in stellar mass,
  redshift and $g-r\;$ (black histograms). The ordinate of each plot shows
  the fraction of galaxies in a particular bin. In the top left corner of
  each panel we list the logarithm of the Kolmogorov-Smirnov probability
  that the two distributions are drawn from an identical parent
  population. From left to right, top to bottom, the panels show: stellar
  mass; redshift; $M_g - M_r$; $M_r$; {\tt fracDeV\_r}; the inner characteristic radius $R_{50}$,
  defined as the radius enclosing half the light measured within the isophotal
  radius $R_{\rm iso23}$;
  the $r=23\,\mbox{mag}/\square\arcsec$ isophote radius $R_{\rm iso23}$ (within
  which our magnitudes are defined); concentration parameter $c^{\prime} =
  R_{\rm iso23}/R_{50}$; average surface brightness $\mu_{50}$ within
  $R_{50}$; the average surface brightness $\mu_{\rm iso23}$ within
  $R_{\rm iso23}$; axis ratio (from the
  flux-weighted second moments measured by {\sc photo}); and the color gradient
  between the iso23 and the fiber apertures; . 
}  
\label{fig:histos_sample2}
\end{center}
\end{figure*}

\begin{figure*}
%[tbp]
\begin{center}
%\setlength{\fboxsep}{-\fboxrule}
%\fbox{
%\includegraphics*[bb=3.7cm 2cm 16.8cm 25.5cm,width=0.6\hsize]
\includegraphics[trim=0 5cm 0 0,width=1.0\hsize]
{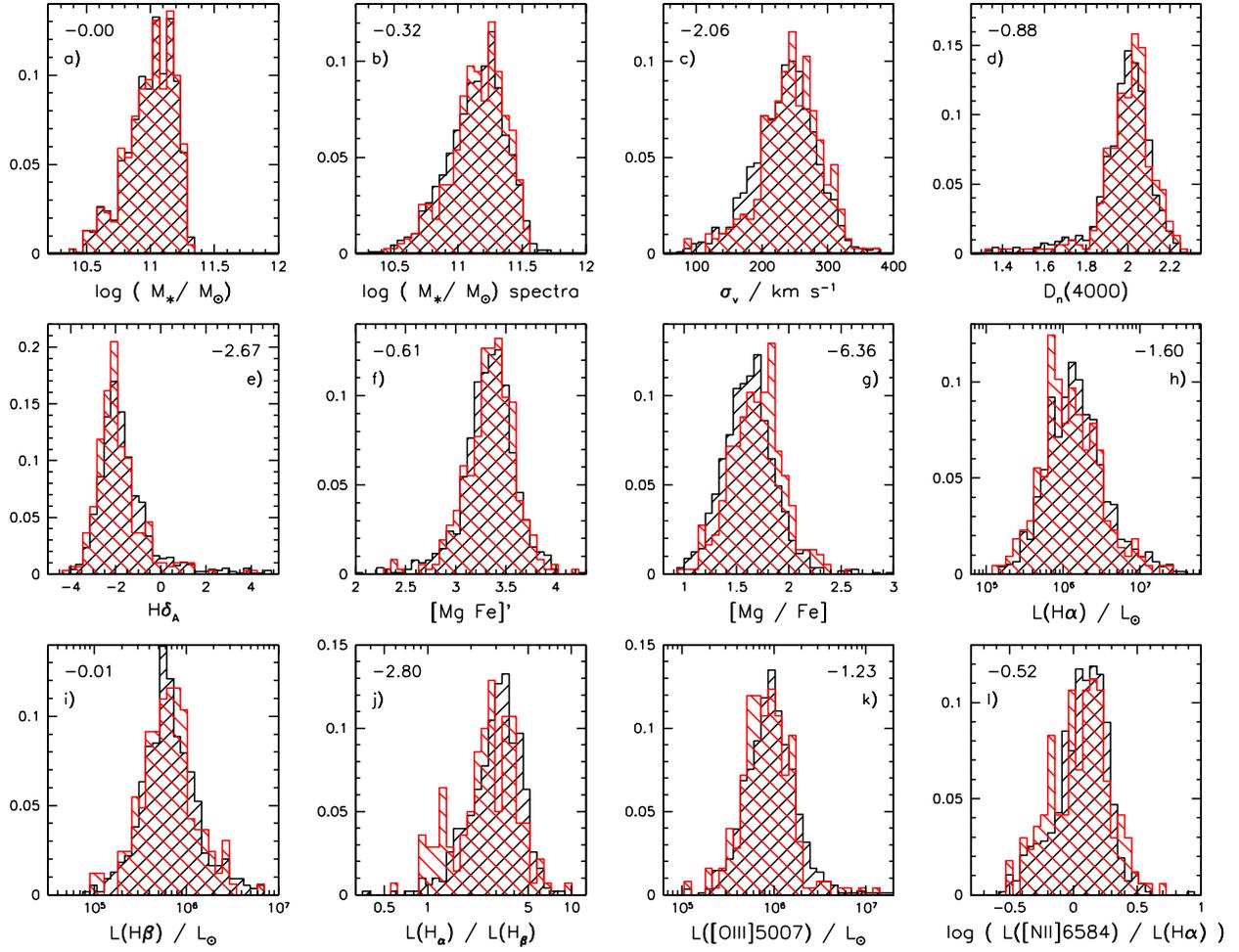}
%}
\caption{~As Fig.~\ref{fig:histos_sample2}, but for a comparison sample
  matched to the BCGs in the spectroscopic database; showing the
  distributions of various (mainly) spectroscopic parameters. The first
  panel demonstrates the match in stellar mass. The other panels show:
  stellar mass as extrapolated from the mass-to-light ratio derived from
  the continuum spectrum; velocity dispersion; strength of the
  4000\AA-break; H$\delta_{\rm A}$ 
index; the metallicity index [Mg
  Fe]'; the alpha-to-iron index Mgb/$\ave{\rm Fe}$;
  H$\alpha$ line luminosity; H$\beta$ line luminosity; the Balmer decrement
  H$\alpha$/H$\beta$; the {\sc [Oiii]}5007 line luminosity; and the line ratio {\sc [Nii]}/H$\alpha$ (a projection of
  the BPT diagram).}   
\label{fig:histos_sample2_spec}
\end{center}
\end{figure*}

\begin{table*}
\begin{center}
\caption{The 16\%, 50\%, and 84\% percentiles (left, middle, right columns
respectively) of the distributions presented in Fig.~\ref{fig:histos_sample2}.
The values for the matched sample are listed in the top rows (black), those
for the BCGs in the bottom rows (red).
}
\label{tab:blub}
\begin{tabular}{r r r @{$\qquad\qquad$} r r r  @{$\qquad\qquad$} r r r
@{$\qquad\qquad$} r r r}
\multicolumn{3}{c @{$\qquad\qquad$}}{$\log (M_{\star}/M_{\odot})$} & \multicolumn{3}{c @{$\qquad\qquad$}}{$z$} & \multicolumn{3}{c @{$\qquad\qquad$}}{$M_g-M_r$} & \multicolumn{3}{c}{$M_r$}\\\hline
10.84 & 11.04 & 11.19 & 0.05 & 0.07 & 0.09 & 0.83 & 0.88 & 0.94 & -22.62 & -22.24 & -21.76\\
\textcolor{BrickRed}{10.84} & \textcolor{BrickRed}{11.04} &
\textcolor{BrickRed}{11.20} & \textcolor{BrickRed}{0.05} &
\textcolor{BrickRed}{0.07} & \textcolor{BrickRed}{0.09} &
\textcolor{BrickRed}{0.84} & \textcolor{BrickRed}{0.88} &
\textcolor{BrickRed}{0.94} & \textcolor{BrickRed}{-22.63} &
\textcolor{BrickRed}{-22.24} & \textcolor{BrickRed}{-21.77}\\[1.8ex]
%\hline
\multicolumn{3}{c @{$\qquad\qquad$}}{fracDeV\_r} & 
\multicolumn{3}{c @{$\qquad\qquad$}}{$R_{50}$ / kpc} & 
\multicolumn{3}{c @{$\qquad\qquad$}}{$R_{\rm iso23}$ / kpc}& 
\multicolumn{3}{c}{$c^{\prime}$}\\\hline
0.86 & 1.00 & 1.00 & 3.14 & 4.08 & 5.41 & 10.76 & 13.79 & 17.20 & 2.92 & 3.34 & 3.69\\
\textcolor{BrickRed}{0.89} & \textcolor{BrickRed}{1.00} &
\textcolor{BrickRed}{1.00} & \textcolor{BrickRed}{3.24} &
\textcolor{BrickRed}{4.49} & \textcolor{BrickRed}{5.89} &
\textcolor{BrickRed}{11.14} & \textcolor{BrickRed}{14.47} &
\textcolor{BrickRed}{17.96} & \textcolor{BrickRed}{2.84} &
\textcolor{BrickRed}{3.23} & \textcolor{BrickRed}{3.65}\\[1.8ex]
%\hline
\multicolumn{3}{c @{$\qquad\qquad$}}{$\mu_{50}$} & \multicolumn{3}{c
@{$\qquad\qquad$}}{$\mu_{\rm iso23}$} & \multicolumn{3}{c @{$\qquad\qquad$}}{$b/a$} &
\multicolumn{3}{c}{$(g-r)_{\rm iso23} - (g-r)_{\rm fiber}$}\\\hline
18.31 & 18.65 & 19.09 & 21.86 & 22.04 & 22.21 & 0.76 & 0.88 & 0.95 & -0.07 & -0.03 & -0.00\\
\textcolor{BrickRed}{18.41} & \textcolor{BrickRed}{18.82} &
\textcolor{BrickRed}{19.26} & \textcolor{BrickRed}{21.93} &
\textcolor{BrickRed}{22.12} & \textcolor{BrickRed}{22.32} &
\textcolor{BrickRed}{0.81} & \textcolor{BrickRed}{0.90} &
\textcolor{BrickRed}{0.96} & \textcolor{BrickRed}{-0.06} &
\textcolor{BrickRed}{-0.02} & \textcolor{BrickRed}{0.00}\\[1.8ex]
\end{tabular}
\end{center}
\end{table*}

\begin{table*}
\begin{center}
\caption{The 16\%, 50\%, and 84\% percentiles (left, middle, right columns
respectively) of the distributions presented in Fig.~\ref{fig:histos_sample2_spec}.
The values for the matched sample are listed in the top rows (black), those
for the BCGs in the bottom rows (red).
}
\label{tab:blub2}
\begin{tabular}{r r r @{$\qquad\qquad$} r r r  @{$\qquad\qquad$} r r r
@{$\qquad\qquad$} r r r}
\multicolumn{3}{c @{$\qquad\qquad$}}{$\log (M_{\star}/M_{\odot})$} & 
\multicolumn{3}{c @{$\qquad\qquad$}}{$\log (M_{\star}/M_{\odot})_{\rm spectra}$} & 
\multicolumn{3}{c @{$\qquad\qquad$}}{$\sigma$ / km s$^{-1}$} & 
\multicolumn{3}{c}{$D_n(4000)$}\\\hline
10.82 & 11.03 & 11.18 & 10.91 & 11.17 & 11.36 & 187 & 239 & 279 & 1.89 & 2.00 & 2.09\\
\textcolor{BrickRed}{10.82} & \textcolor{BrickRed}{11.03} &
\textcolor{BrickRed}{11.19} & \textcolor{BrickRed}{10.92} &
\textcolor{BrickRed}{11.17} & \textcolor{BrickRed}{11.37} &
\textcolor{BrickRed}{203} & \textcolor{BrickRed}{246} &
\textcolor{BrickRed}{288} & \textcolor{BrickRed}{1.91} &
\textcolor{BrickRed}{2.02} & \textcolor{BrickRed}{2.10}\\[1.8ex]
\multicolumn{3}{c @{$\qquad\qquad$}}{H$\delta_{\rm A}$} & 
\multicolumn{3}{c @{$\qquad\qquad$}}{[Mg Fe]'} & 
\multicolumn{3}{c @{$\qquad\qquad$}}{Mgb/$\ave{\rm Fe}$} & 
\multicolumn{3}{c}{$\log (L_{\rm H\alpha}/L_{\odot})$}\\\hline
-2.57 & -1.88 & -0.80 & 3.11 & 3.35 & 3.56 & 1.38 & 1.62 & 1.85 & 5.81 & 6.14 & 6.53\\
\textcolor{BrickRed}{-2.72} & \textcolor{BrickRed}{-2.07} &
\textcolor{BrickRed}{-1.09} & \textcolor{BrickRed}{3.10} &
\textcolor{BrickRed}{3.37} & \textcolor{BrickRed}{3.58} &
\textcolor{BrickRed}{1.45} & \textcolor{BrickRed}{1.70} &
\textcolor{BrickRed}{1.93} & \textcolor{BrickRed}{5.75} &
\textcolor{BrickRed}{6.07} & \textcolor{BrickRed}{6.47}\\[1.8ex]
\multicolumn{3}{c @{$\qquad\qquad$}}{$\log (L_{\rm H\beta}/L_{\odot})$} & 
\multicolumn{3}{c @{$\qquad\qquad$}}{$L_{\rm H\alpha}/L_{\rm H\beta}$} & 
\multicolumn{3}{c @{$\qquad\qquad$}}{$\log (L_{\rm [OIII]}/L_{\odot})$} & 
\multicolumn{3}{c}{$L_{\rm [NII]}/L_{\rm H\alpha}$}\\\hline
5.55 & 5.80 & 6.09 & 1.96 & 3.11 & 4.32 & 5.70 & 5.96 & 6.22 & -0.13 & 0.09 & 0.25\\
\textcolor{BrickRed}{5.53} & \textcolor{BrickRed}{5.81} &
\textcolor{BrickRed}{6.07} & \textcolor{BrickRed}{1.34} &
\textcolor{BrickRed}{2.67} & \textcolor{BrickRed}{3.94} &
\textcolor{BrickRed}{5.69} & \textcolor{BrickRed}{5.92} &
\textcolor{BrickRed}{6.17} & \textcolor{BrickRed}{-0.19} &
\textcolor{BrickRed}{0.07} & \textcolor{BrickRed}{0.26}\\[1.8ex]
\end{tabular}
\end{center}
\end{table*}

In Fig.~\ref{fig:histos_sample2} and Fig.~\ref{fig:histos_sample2_spec} we
present the distributions of  a variety of photometric and spectroscopic parameters
for the BCGs and the comparison samples CS3p and CS3s. For
each parameter, we list the decimal logarithm of $1$ minus the
Kolmogorov-Smirnov confidence
level at which the null hypothesis that the distributions are drawn from 
identical parent populations  is rejected (i.e. a 99\% probability that
the distributions are different will have a value of $-2$) .

By construction, the BCGs and the comparison sample are
identical in stellar mass, redshift, and color. Because both stellar mass
and color are the same, the distributions of luminosity  are also equivalent. 

Note that our stellar masses are calculated using the 
{\tt kcorrect} algorithm applied to  isophotal magnitudes
that have been corrected for sky-subtraction errors.
These masses are not the same as the stellar masses estimated using
the methods described in 
\citet{khw03}, which we compare in Fig.~\ref{fig:histos_sample2_spec}, panel
(c). For the
latter, the mass-to-light ratio determined from the continuum spectrum is
applied to the  SDSS petrosian
magnitude. Since these magnitudes underestimate the luminosity, the galaxy
mass is underestimated as well. We also find that non-BCGs
show stronger gradients between their fiber colors and their iso23 aperture colors
(Fig.~\ref{fig:histos_sample2}, cf. Sect.~\ref{sect:stellarpop}).
Color gradients imply that the mass-to-light ratio varies across the galaxy and this
is not accounted for when deriving stellar masses from the spectra.

\subsection{Structural parameters}
\label{sect:structure}

\subsubsection{Radii and surface brightness}

In agreement with previous studies, we find that BCGs are larger
(Fig.~\ref{fig:histos_sample2}, panels [f] and [g]) and have lower surface
brightnesses than non--BCGs (Fig.~\ref{fig:histos_sample2}, panels [h] and
[i]).  The difference is more prominent for the inner characteristic radius
$R_{50}$ (defined as the radius containing half the galaxy's light measured
within the $r=23\,\mbox{mag}/\square\arcsec$ isophote) than for the outer
isophotal radius $R_{\rm iso23}$, within which we measure the luminosity of
the galaxy.  This is also evident in the distributions of the concentration
parameter $c^{\prime} = R_{\rm iso23}/R_{50}$: for a given $R_{\rm iso23}$,
a BCG has a larger $R_{50}$ than a non-BCG.  This indicates that the light
profiles of BCGs are systematically different to those of non--BCGs. To
first order, these differences can be explained by BCGs having shallower
light profiles. Indeed, \citet{gzz05} and \citet{bhs06} find comparatively
large Sersic indices (and thus shallow profiles) when fitting BCG light
profiles with Sersic profiles.

\subsubsection{Size-luminosity relation}

The sizes and luminosities of elliptical
galaxies have been shown to obey the scaling  $R_{50} \propto L^{\alpha}$, with
$\alpha \simeq 0.6$ \citep[e.g. ][]{bsa03b}. However, at the massive end,
this relation steepens \citep{lfr06}. \citet{bhs06} argue that BCGs have
larger radii and that this steepening is caused  by an increasing fraction of
BCGs. \citet{dqm06} still find a steepening after removing the  C4
BCGs from their  sample of SDSS elliptical galaxies and argue that the steepening
is not solely attributable to `contamination' from a population of galaxies
with intriniscally larger radii (BCGs).

\begin{figure*}
\begin{center}
%\setlength{\fboxsep}{-\fboxrule}
%\fbox{
%\includegraphics*[bb=3.7cm 2cm 16.8cm 25.5cm,width=0.6\hsize]
\includegraphics[width=0.49\hsize]
{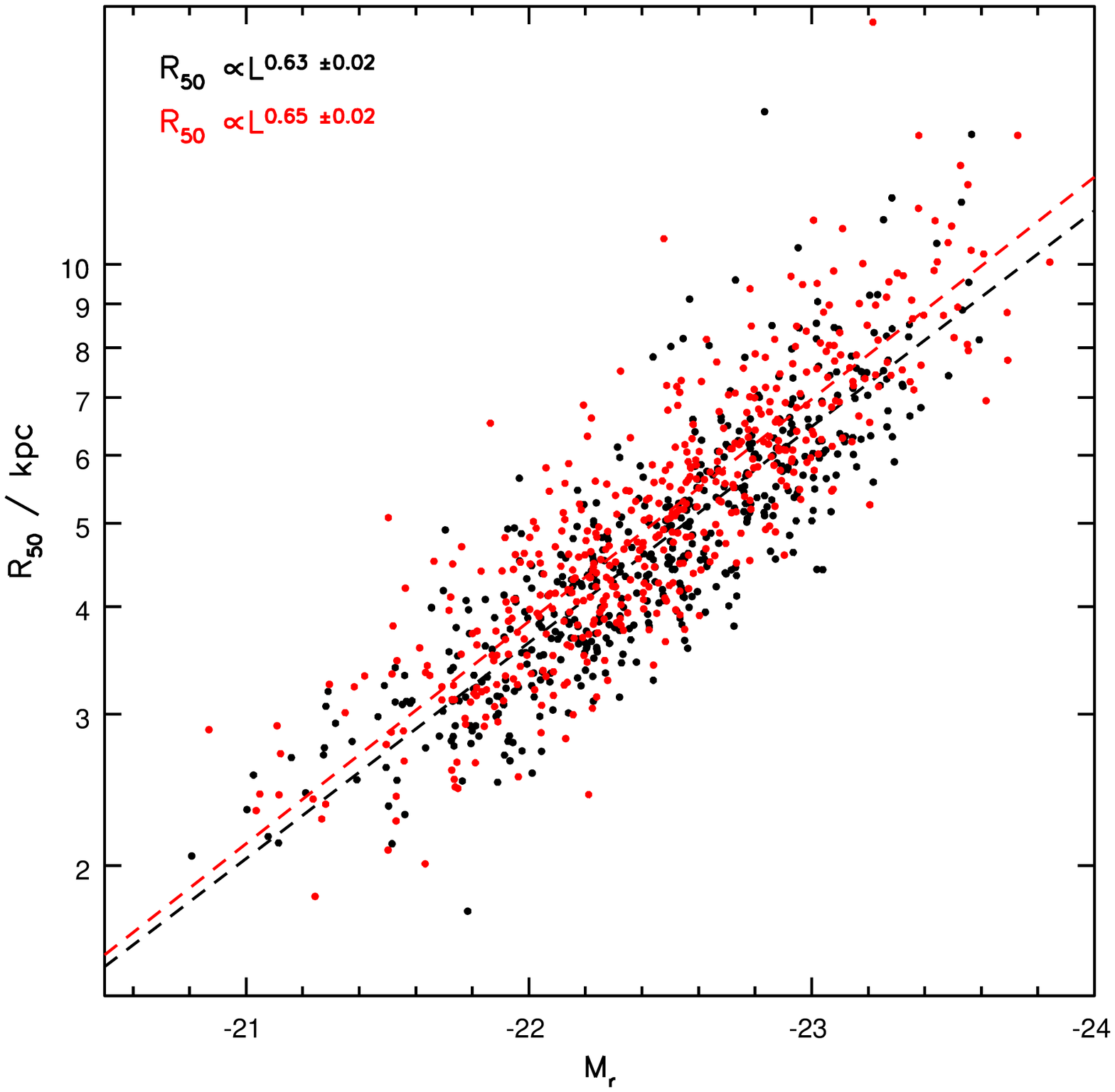}
\includegraphics[width=0.49\hsize]
{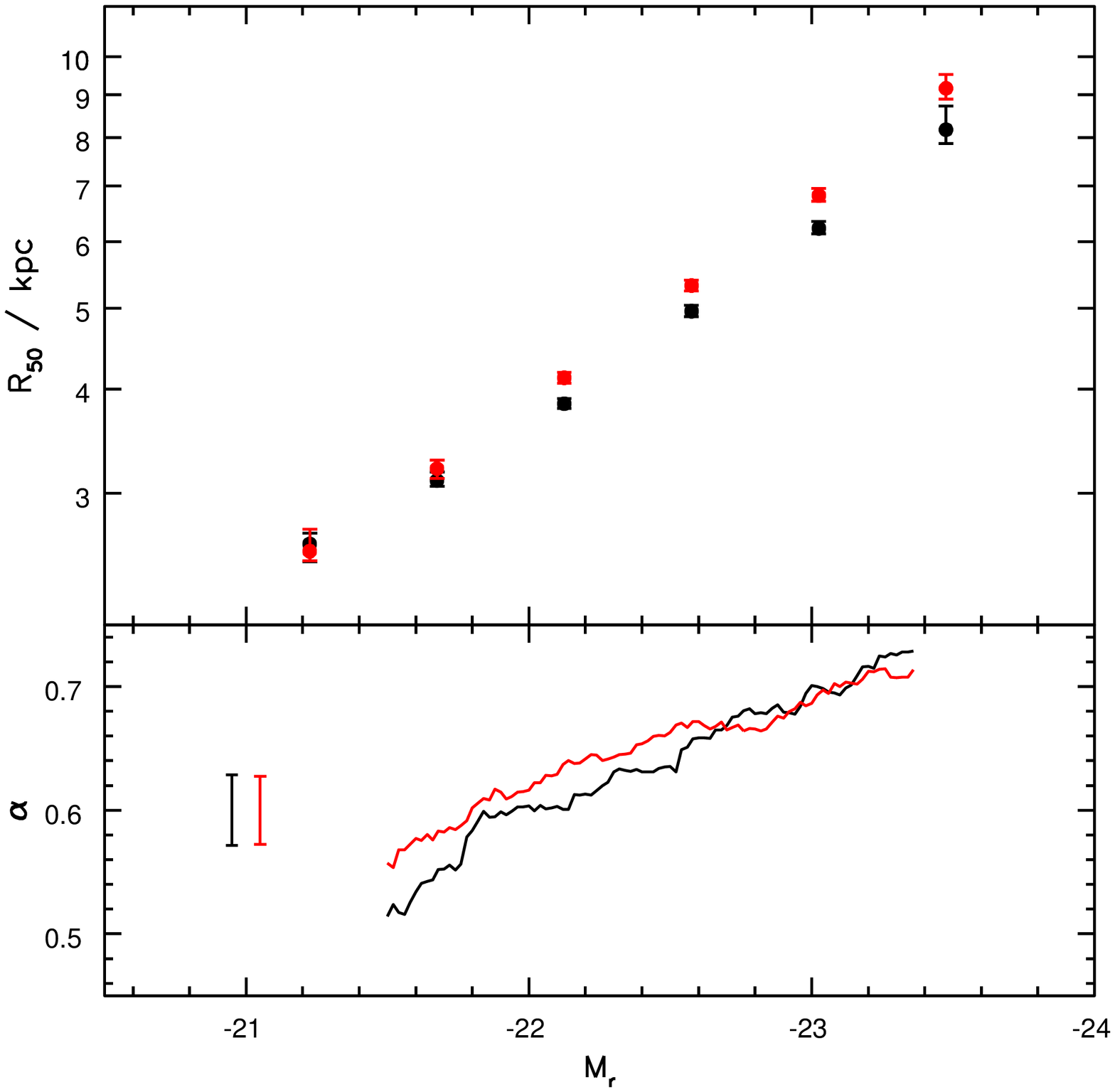}
%}
\caption{~Left panel: the size-luminosity relation for the BCGs (red) and
  for the comparison sample CS1p (black) along with the best-fit linear
  relations (dashed lines). Top right panel: the median radii in bins of
  magnitude. The error bars represent the 68\% confidence levels divided by
  the square root of the number of galaxies in the respective bin. Lower
  left panel: the variation of $\alpha$, the exponent of 
  the size-luminosity relation, as a function of magnitude $M_r$, determined
  from galaxies within $M_r \pm 1.0$ .} 
\label{fig:size_lum}
\end{center}
\end{figure*}

The top right panel of Fig.~\ref{fig:size_lum} {\it demonstrates}  that 
BCGs are larger than non-BCGs at all luminosities or stellar masses. 
Symmetric linear fits to the individual data points
yield very similar exponents for the radius-luminosity relation for
the BCG sample and the comparison sample:
\begin{eqnarray*}
R_{50, \rm BCGs} &\propto& L^{\;0.65 \pm 0.02} \; ,\\
R_{50, \rm CS1s} &\propto& L^{\;0.63 \pm 0.02} \; .
\end{eqnarray*}
However, we also find that the relation displays curvature, i.e. it steepens
with luminosity. This is shown in the lower right panel of
Fig.~\ref{fig:size_lum}. The range of exponents we find ($\alpha \sim 0.5 -
0.7$) is broadly consistent with that of \citet{dqm06}. We note that
$\alpha$ is only slightly larger for the BCGs than the non-BCGs, even at the
highest luminosities. We do not find the significantly steeper relations
claimed by \citet[$\alpha = 0.92$]{bhs06} and \citet[$\alpha =
1.18$]{lfr06}. Both of these samples include both BCGs and non-BCGs - the
\citet{bhs06} study uses the original C4 BCGs, which we have shown to be
contaminated by non-BCGs, and the \citet{lfr06} study is based on galaxies
with $M_V<-21$ and detectable core radii. Such contamination from non-BCGs
is likely to be most important at lower luminosities, and will thus mimic a
steeper slope. Even if we take this effect into account, our data do not
support very large values of $\alpha$; if we fit a relation to the BCGs at
$M_r \sim -23.5$ and non--BCGs at $M_r \sim -23$, we find a value of
$\alpha$ of only 0.93.  We speculate that one possible source for the
discrepancy could be the different definitions of $R_{50}$ used in the
different studies.

\subsubsection{Ellipticity}

We calculate the axial ratios of the galaxies in our
sample (Fig.~\ref{fig:histos_sample2}, panel [k]) from the flux-weighted 
second moments as measured
by {\sc photo}, i.e.
\be
b/a = \frac{1-\sqrt{Q^2+U^2}}{1+\sqrt{Q^2+U^2}}
\ee
where $Q$ and $U$ are the Stokes parameters listed in the {\sc photo}
database. We choose this measurement since it is not based on fitting a
particular model to the surface brightness profile of the galaxy
and it is  also not as
sensitive to the sky subtraction as  isophotal ellipticity measures.
The majority of  BCGs and non-BCGs are  round, with axis
ratios $b/a \gtrsim 0.8$.  Both samples exhibit a tail to lower axial ratios, but
this tail is more prominent for the non-BCGs. The median axial ratio is very
similar for both samples (0.90 for the BCGs and 0.88 for the
non-BCGs). This is qualtitatively consistent with the results of
\citet{rlp93}. It should be noted, however,  that \citet{psh91} find that the 
ellipticity of BCGs increases as a function of the radius at which it is measured
\citep[see also][]{gzz05}.

\subsection{Dynamical Structure}

\subsubsection{Velocity dispersion}

As in previous studies of elliptical galaxies
that used spectra taken within fixed-sized 
apertures \citep{jfk95,bsa03a},
we correct the galaxy velocity 
dispersion to its expected value at one-eighth of the effective radius:
\be
\sigma_{\rm v} = \sigma_{\rm v, meas}
\left( \frac{r_{\rm fiber}}{r_{50}/8} \right)^{0.04}
\ee
where $\sigma_{\rm v, meas}$ is the measured velocity dispersion, 
$r_{\rm fiber}$ is the radius of the SDSS fiber (1.5$\arcsec$), and $r_{50}$ is the 
inner characteristic radius, measured in arcseconds. Strictly speaking,
since the iso23 magnitudes do not attempt to measure the total galaxy light,
$r_{50}$ is not exactly the same as the effective radius, but 
because the correction does not scale very steeply with radius,
this difference is negligible. 
This correction also assumes
a universal velocity dispersion profile. While this seems to be applicable
to most elliptical galaxies \citep{jfk95}, it has not yet been demonstrated  that
it also applies to BCGs.

We find that BCGs have systematically larger velocity dispersions than
non-BCGs ((Fig.~\ref{fig:histos_sample2_spec}, panel [c]; this also holds for
the uncorrected velocity dispersions).

\subsubsection{Dynamical mass}

\begin{figure}
\begin{center}
%\setlength{\fboxsep}{-\fboxrule}
%\fbox{
%\includegraphics*[bb=3.7cm 2cm 16.8cm 25.5cm,width=0.6\hsize]
\includegraphics[width=1.0\hsize]
{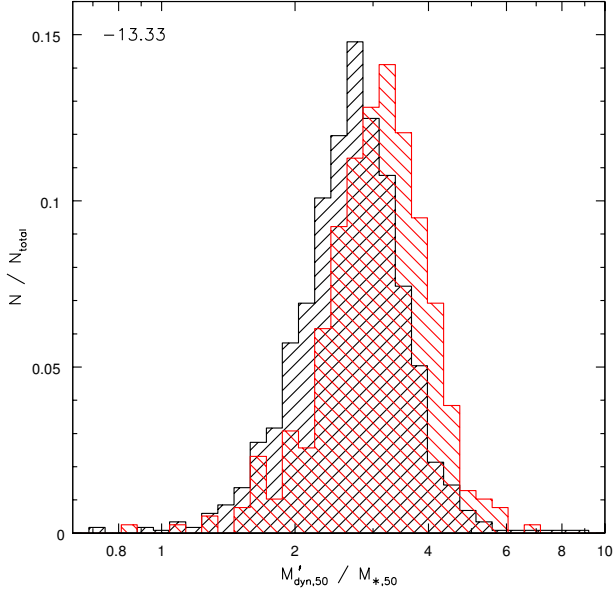}
%}
\caption{~The ratio of dynamical mass to stellar mass (within $R_{50}$) for
  the BCGs (red) and the comparison sample CS3s (black). We have assumed
  that $c_2=(1.65)^2$.}  
\label{fig:virmass}
\end{center}
\end{figure}

The larger radii and higher velocity dispersions of BCGs 
imply that they have larger dynamical-to-stellar mass ratios
 than non-BCGs. 
The dynamical mass within $R_{50}$ can be derived via a projection of the
scalar virial theorem onto observable quantities:
\be
M_{\rm dyn, 50} = c_2 \frac{\sigma_{\rm v}^2 R_{50}}{G}
\label{eq:virial}
\ee
where $c_2$ depends on the profiles of both the dark matter and the luminous
matter components. If the former follows an NFW profile \citep{nfw97}, and
the latter a \citet{her90} profile, then $c_2 = (1.65)^2$ \citep{pss04}.
For calculating the dynamical mass, we assume that $c_2 = (1.65)^2$ :
\be
M_{\rm dyn, 50}^{\prime} = M_{\rm dyn, 50} \frac{(1.65)^2}{c_2} = (1.65)^2 \frac{\sigma_{\rm v}^2 R_{50}}{G}
\label{eq:dyn_mass}
\ee
We also assume that the stellar mass within
$R_{50}$ is  50\% of the stellar mass within $R_{\rm iso23}$ \citep{pss04}.
We find that the ratio of dynamical mass to stellar mass is
indeed considerably larger for BCGs (Fig.~\ref{fig:virmass}).
This difference is likely the consequence of  the position of BCGs at or near 
the centers of galaxy clusters.
As a result, there is a greater contribution from the dark matter halo to the 
dynamical mass of the BCG.

\subsubsection{The fundamental plane}

Early-type galaxies seem to be well described by a two-parameter set of
equations, as is
evidenced by the Fundamental Plane: they lie on a plane in a coordinate
system  defined by the logarithmic values of velocity dispersion
$\sigma_{\rm v}$, 
effective radius, 
and average surface brightness within the effective radius
\citep{djd87}. The plane is typically expressed as
\be
R_{e}\; \propto\; \sigma_{\rm v}^{\;a}\, I_{e}^{\;-b} \; .
\ee 
While there is agreement that  $b \simeq 0.8$, the parameter $a$ is
dependent on filter bands and may also be sensitive to a variety of selection effects and the
precise definitions of $\sigma_{\rm v}$, $R_{e}$ and $I_{e}$. 
Typical values of $a\sim 1.2 - 1.6$ are quoted in the literature
\citep[see for example the compilation of observed FP coefficients in][]{bsa03c}.

\begin{figure}
\begin{center}
%\setlength{\fboxsep}{-\fboxrule}
%\fbox{
%\includegraphics*[bb=3.7cm 2cm 16.8cm 25.5cm,width=0.6\hsize]
\includegraphics[width=1.0\hsize]
{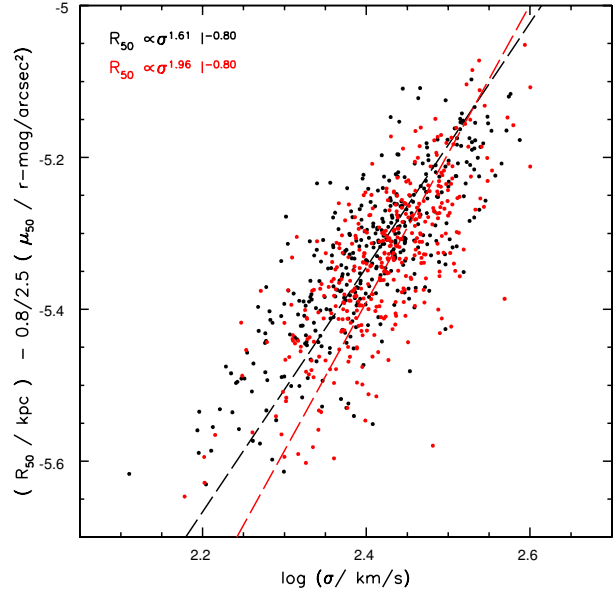}
%}
\caption{~Projection along the fundamental plane BCGs (red) and the galaxies
  of CS1s (black). The dashed lines show the respective best fit for $a$,
  keeping $b=0.8$.}  
\label{fig:f_plane_scatter}
\end{center}
\end{figure}

We assume that $b=0.8$ and plot $\log(R_{50} I_{50}^{\;0.8})$ as a function
of $\log\sigma_{\rm v}$ in Fig.~\ref{fig:f_plane_scatter} for the early--type BCGs
and the 
comparison sample CS1s. In this diagram, the BCGs and the comparison galaxies  
only diverge for galaxies with small radii and/or
high surface brightness (i.e. these are not the cD galaxies). The difference is  in the
sense that the 
velocity dispersions of the BCGs are larger.
A symmetric fit \citep[{\tt fitexy} from][]{ptv92} yields 
\begin{eqnarray*}
a_{\rm BCGs} &=& 1.96 \pm 0.10 \; ,\label{eq:bcg_fp}\\
a_{\rm CS1s} &=& 1.61 \pm 0.07 \; .
\end{eqnarray*}
For the comparison sample, the value of $a$ lies close to the values that have
been measured in the near-infrared \citep[e.g. ][$a=1.53\pm0.08$]{pdc98},
and also in the  SDSS \citep[$a=1.49\pm0.05$]{bsa03c}. For the BCGs, $a$
is significantly larger, indicating
that BCGs do not lie on the same fundamental plane as ``normal''
ellipticals. It is interesting to note that it is predominantly the small,
low velocity dispersion BCGs which deviate from the generic fundamental
plane.  

The fundamental plane relation is essentially an expression of the virial
theorem.
If we write
\be
M_{\star} = c_1 L
\quad \mbox{and} \quad
L = 2\pi I_{50} R_{50}^{\;2} \; ,
\label{eq:surface_brightness}
\ee
Eq.~(\ref{eq:virial}) can be rewritten as
\be
R_{50} = \frac{1}{2\pi G} \frac{c_2}{c_1} \frac{M_{\star}}{M_{\rm dyn,50}} 
\sigma_{\rm v}^{\;2} I_{50}^{\;-1} \; .
\label{eq:f_plane}
\ee
The deviation of the observed fundamental plane from the theoretical
one ($a = 2$ and $b = 1$) is referred to as the `tilt' of the fundamental
plane. The tilt implies that $\frac{c_2}{c_1} \frac{M_{\star}}{M_{\rm
    dyn,50}}$ varies for different elliptical galaxies.  
The  proportionality constant $c_1$ is an expression of the stellar mass-to-light
ratio and varies for different stellar populations. $c_2$
depends on the profile shapes of both the luminous and the dark 
matter components. If $c_2$ were constant, elliptical galaxies would be
structurally homologous systems.
There are contradictory results in the literature as to whether it is
predominantly non--homology or variation in $L/M_{\rm dyn,50}$
that is responsible for the tilt of the fundamental plane.

We are unable to distinguish non--homology from
variation of $L/M_{\rm dyn,50}$ with our data. When calculating dynamical mass, we assume
$c_2 = (1.65)^2$ (Eq.~[\ref{eq:dyn_mass}]), but we caution that this
approach necessarily neglects effects from non--homology.

\begin{figure}
\begin{center}
%\setlength{\fboxsep}{-\fboxrule}
%\fbox{
%\includegraphics*[bb=3.7cm 2cm 16.8cm 25.5cm,width=0.6\hsize]
\includegraphics[width=1.0\hsize]
{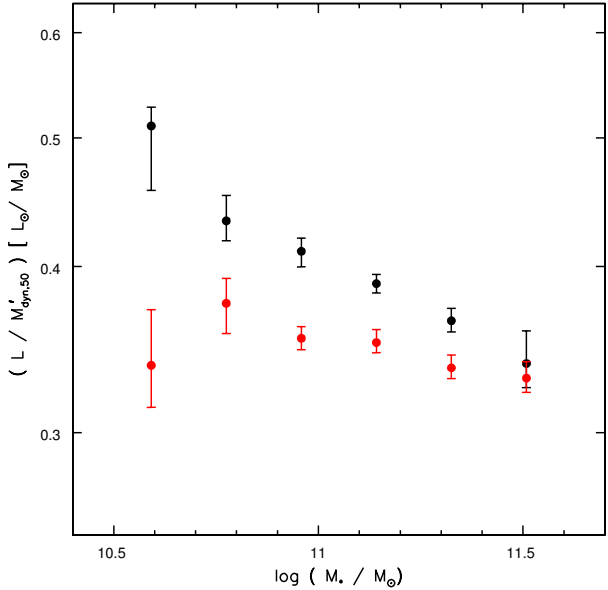}
%}
\caption{~The median luminosity to dynamical mass ratio of BCGs (red) and
  the comparison sample CS1s (black) in bins of galaxy stellar mass. It is
  the variation of this ratio which determines the tilt of the fundamental
  plane.}
\label{fig:fp_coeff}
\end{center}
\end{figure}

In Fig.~\ref{fig:fp_coeff}, we investigate how the pre-factor
$\frac{c_2}{c_1} \frac{M_{\star}}{M_{\rm dyn,50}} = \frac{L}{M_{\rm dyn, 50}
  / c_2}$ varies as a function of stellar mass for BCGs compared to
non-BCGs. The results show that the variation is much smaller for the BCGs.
This is an affirmation of our previous result that BCGs lie on a different
fundamental plane than non-BCGs. It also demonstrates that this result does
not come from a few outliers, but applies to the majority of galaxies
with $M_{\star} < 10^{11.3} M_{\odot}$. Since $\frac{L}{M_{\rm dyn, 50} /
  c_2}$ varies so little, the BCG fundamental plane is closer to the
expectations of the virial theorem ($a=2$ and $b=1$).

Again, it is for low mass galaxies that BCGs differ most from non--BCGs. The
similarity between BCGs and non--BCGs at high stellar masses implies that
the process(es) which cause this ratio to be approximately constant for
BCGs also apply to massive non--BCGs. Possibilities include assembly history
\citep[e.g. the influence of the orbital elements during dissipationless
mergers, ][]{bmq06}, and the fact that both BCGs and massive galaxies in
general 
are found in dense environments \citep{kwh04}.

\subsubsection{Faber-Jackson relation}

Several
studies suggest that BCGs follow a different relation between luminosity and
velocity dispersion than less massive elliptical galaxies
\citep{oeh91,lfr06,bhs06}.  Parametrizing this relation as $L \propto
\sigma^{\;\beta}$, the canonical value is $\beta=4$, as can be seen from
Eq.~(\ref{eq:surface_brightness}) and Eq.~(\ref{eq:f_plane}), assuming that
$\frac{c_2}{c_1} \frac{M_{\star}}{M_{\rm dyn,50}} \frac{1}{I_{50}}$ is
constant. Most measurements reported in the literature are consistent with
$\beta \simeq 4$ \citep[e.g. ][]{bsa03b}. However, for samples of BCGs
\citep{oeh91} and very massive galaxies \citep{lfr06}, it is found
that $\beta > 4$, i.e.  $\sigma$ increases less steeply with luminosity than
predicted by the standard Faber-Jackson relation.
\begin{figure}
\begin{center}
%\setlength{\fboxsep}{-\fboxrule}
%\fbox{
%\includegraphics*[bb=3.7cm 2cm 16.8cm 25.5cm,width=0.6\hsize]
\includegraphics[width=1.0\hsize]
{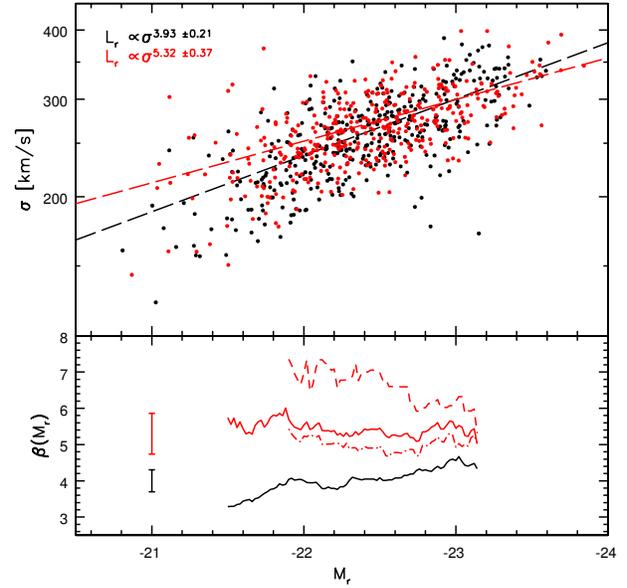}
%}
\caption{~Top panel: the Faber-Jackson relation for BCGs (red) and for the
  comparison sample CS1s. The dashed lines show the best linear
  fits to the relations. Bottom panel: the 
  variation of $\beta$ with $M_r$, i.e. $\beta$ determined from galaxies
  within $M_r \pm 1.0$ for BCGs meeting the early-type criteria (solid red
  line) and the corresponding 
  CS1s sample (solid black line). The dashed line shows the subset of these
  BCGs located in clusters with $\sigma_{\rm v, cl} > 400 \,{\rm
    km/s}\,$, the dash-dotted line the subset of BCGs in clusters with
  $\sigma_{\rm v, cl} < 500 \,{\rm km/s}\,$.
The typical error bars are displayed on the left.}   
\label{fig:faber_jackson}
\end{center}
\end{figure}

In Fig.~\ref{fig:faber_jackson}, we show the Faber-Jackson relation for the
BCGs and CS1s. Symmetric linear fits to each sample yield:
\begin{eqnarray*}
L_{\rm BCGs} &\propto& \sigma^{\;5.32 \pm 0.37} \; , \\
L_{\rm CS1s} &\propto& \sigma^{\;3.93 \pm 0.21} \; .
\end{eqnarray*}
We find a slope that is compatible with the standard $L \propto \sigma^4$
relation for non-BCGs and we confirm that $\sigma$  rises less steeply with
luminosity for BCGs.
\footnote{We find a similar change in
slope of the Faber--Jackson relation when using the $K$-band luminosities of
those BCGs with 2MASS photometry and a set of comparison galaxies. This is
contrary to recent 
claims by 
\citet{bmm06}, who use a much smaller sample of BCGs for their analysis.}

In the bottom panel of Fig.~\ref{fig:faber_jackson}, we investigate how
$\beta$ changes with luminosity. 
We find that for BCGs, $\beta$ is approximately
constant (within the typical error bars) and has a value $\sim 5.5$. For
non--BCGs, $\beta$ varies from values $\sim 3$ at the low-luminosity end to
values $\sim 4.5$ at high luminosities \citep[a similar range of $\beta$,
albeit over a larger luminosity interval, as was found by][]{dqm06}.

\citet{bmq06} find that in simulations of dissipationless mergers, $\beta$
increases with the eccentricity of the merger orbit. They also argue that
BCGs are expected to form through anisotropic merging due to the filamentary
structure surrounding galaxy clusters.

We test whether $\beta$ depends on cluster mass by
splitting the BCG sample according to cluster velocity dispersion, and fitting
$\beta$ separately for the two samples. 
We allow the sample to overlap in $\sigma_{\rm v, cl}$ to gain higher
statistical significance. We obtain the following results:
\begin{eqnarray*}
L_{\rm BCGs}\; (\,\sigma_{\rm v, cl} < 500 \,{\rm km/s}\,) 
&\propto& \sigma^{\;5.22 \pm 0.46} \\
L_{\rm BCGs}\; (\,\sigma_{\rm v, cl} > 400 \,{\rm km/s}\,) 
&\propto& \sigma^{\;5.91 \pm 0.69}
\end{eqnarray*}

The two values of $\beta$ are just consistent with each other within the
errors, and thus we cannot draw strong conclusions.  Our results indicate
that $\beta$ is larger for BCGs in more massive clusters. If the scenario
put forward by \citet{bmq06} is correct, this might imply that the merger
orbit eccentricity increases with cluster mass. Another explanation might be
that the number of (dissipationless) mergers is larger for BCGs in more
massive clusters.

\subsection{Stellar Populations}
\label{sect:stellarpop}

The availability of measurements of spectral indices for galaxies in
the spectroscopic SDSS catalog allows us to investigate the stellar
populations in BCGs and non--BCGs.
Both the distributions of the 4000\AA-break
\citep[Fig.~\ref{fig:histos_sample2_spec}, panel (d), measured as
$D_n(4000)$,][]{bmy99} and the $H\delta_A$ index
\citep[Fig.~\ref{fig:histos_sample2_spec}, panel (e); ][]{woo97} demonstrate
that the stellar populations of the BCGs and the comparison galaxies are
old, as is generally found for galaxies in this mass range \citep{khw03b}.
The metallicity, measured using the index [MgFe]'
\citep[Fig.~\ref{fig:histos_sample2_spec}, panel (f); ][]{tmb03} is also
typical for giant elliptical galaxies. There is a slight indication that the
stellar populations of BCGs are slightly older (larger $D_n(4000)$, lower
$H\delta_A$, and higher metallicity), but this is only significant for the
$H\delta_A$ index.

It should be noted that these measurements apply only to the galaxy
light contained within the fiber, i.e. the inner $3\arcsec$, whereas the
samples are matched in $g-r$ color within the
$23\,\mbox{mag}/\square\arcsec$ isophote. In panel (l) of
Fig.~\ref{fig:histos_sample2} we find that the color gradient between the
fiber aperture and the iso23 aperture is more prominent in the
non--BCGs. This is a confirmation of previous results that color gradients
in BCGs are weak or absent \citep{gsa97}, while non-BCG elliptical galaxies 
are generally redder in the center \citep{bcg05}. The presence of color
gradients is typically attributed to metallicity gradients \citep{jsd06},
however, we do not find evidence for
different metallicities in BCGs.

%On the other hand, the color gradient between the fiber aperture and the
%iso23 aperture is more prominent in the non--BCGs (Fig.~\ref{fig:histos_sample2}, panel [l]). Since the samples are
%matched in $(g-r)_{\rm iso23}$, this implies that non--BCGs are slightly
%redder in the center. This result might imply a difference in population age
%-- alternatively, it could also imply that non--BCGs are dustier in their
%centers than BCGs.

\begin{figure}
\begin{center}
%\setlength{\fboxsep}{-\fboxrule}
%\fbox{
%\includegraphics*[bb=3.7cm 2cm 16.8cm 25.5cm,width=0.6\hsize]
\includegraphics[width=1.0\hsize]
{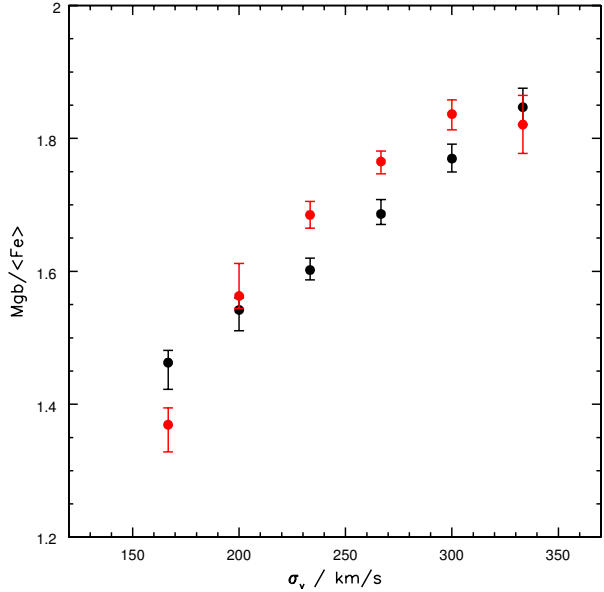}
%}
\caption{~The median value of Mgb/$\ave{\rm Fe}$ as a function of velocity
  dispersion for the BCGs (red) and the comparison sample CS1s (black).}  
\label{fig:vdisp_mgofe}
\end{center}
\end{figure}

We use the index Mgb/$\ave{\rm Fe}$ as an indicator of  
the $\alpha$/Fe ratio \citep{tmb03} and we find that BCGs have a systematically
higher Mgb/$\ave{\rm Fe}$ value than 
non-BCGs (Fig.~\ref{fig:histos_sample2_spec}, panel [g]). However, this index is known to correlate strongly  with
velocity dispersion, so this result is not independent of our previous result  that
BCGs have systematically larger velocity dispersions. In Fig.~\ref{fig:vdisp_mgofe}, we
plot Mgb/$\ave{\rm Fe}$ as a function of velocity dispersion for the BCGs
and the comparison sample CS1s. Except in the outermost bins, we do find
systematically higher $\alpha$/Fe ratios in the BCGs. Higher $\alpha$/Fe
ratios can be interpreted as an indication that star formation in the galaxy occurred over a
shorter time-scale \citep{gzs04}. The enhanced radio-AGN activity we find in BCGs
(Sect.~\ref{sect:radio}) may explain why  star formation has been shut off on 
shorter timescales in the BCGs.

\subsection{Emission line properties}

We also investigate the strengths of the four emission lines H$\alpha$,
H$\beta$, {\sc [Oiii]}$\lambda$5007, and {\sc [Nii]}$\lambda$6584 that are
commonly used to classify galaxies according to whether their emission line
luminosity is driven by star formation or AGN activity \citep{bpt81}.  We
limit our sample to galaxies with a signal-to-noise SNR $>3$ in the
respective line measurement(s). For the individual line measurements
H$\alpha$ / H$\beta$ / {\sc [Oiii]} / {\sc [Nii]} , this holds for 56\% /
42\% / 64\% / 50\% of the BCGs and 66\% / 47\% / 77\% / 63\% of the
comparison sample (note that this applies to the CS3s sample). Requiring
that SNR $>3$ in all four bands simultaneously leaves only 30\% of the BCGs
and 40\% of the comparison sample. These numbers already indicate that the
emission lines in BCGs are in general weaker than in non-BCGs, a result
which is further confirmed by the distributions of H$\alpha$ and {\sc
  [Oiii]} line luminosities shown in Fig.~\ref{fig:histos_sample2_spec},
panels (h) and (k). We note that it is particularly the high--mass BCGs in
which the emission line strength is suppressed compared to the comparison
sample.

\begin{figure}
\begin{center}
%\setlength{\fboxsep}{-\fboxrule}
%\fbox{
%\includegraphics*[bb=3.7cm 2cm 16.8cm 25.5cm,width=0.6\hsize]
\includegraphics[width=1.0\hsize]
{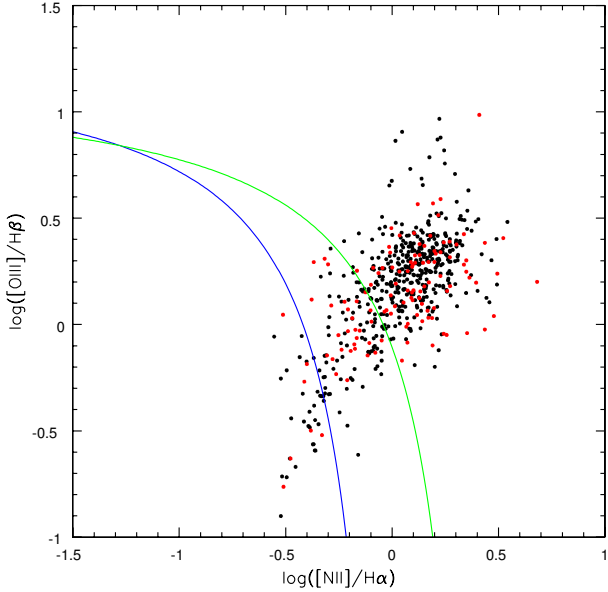}
%}
\caption{~BCGs (red) and the comparison sample CS3s (black) placed into the
  BPT diagram. Galaxies to the left of the blue line are classified as
  purely star-forming \citep{kht03}, galaxies to the right of the green line
  as purely AGN \citep{kds01}, and galaxies inbetween the lines as composite.
  Only galaxies with SNR $>3$ in all four line measurements are
  shown.}  
\label{fig:bpt}
\end{center}
\end{figure}

Fig.~\ref{fig:bpt} shows the BPT diagram of BCGs and the comparison sample
for those galaxies which satisfy SNR $>3$ in all four bands, i.e. 119 (out
of 391) BCGs and 472 (out of 1173) non-BCGs. Of these BCGs, 7 (i.e. 6\% of
the line--emitting sample / 2\% of the complete sample) are
classified as star-forming, 83 (70\% / 21\%) as AGN, and 29 (24\% / 7\%) as composite.
Of the non-BCGs, 29 (6\% / 2\%) are star-forming, 364 (77\% / 31\%) AGN, and
79 (16\% / 6\%) composite.  Our
sample is too small to draw detailed conclusions from these numbers, except
that for both samples, the emission line flux is dominated by AGN--like
emission. 
Their low {\sc [Oiii]} luminosities place the BCGs somewhat lower
in the BPT diagram than the non-BCGs, i.e. the BCGs are almost exclusively
classified as LINERs, whereas a few galaxies in the comparison sample could
be classified as low-luminosity Seyferts. 

%It should be noted that the line
%emission from BCGs is sometimes attributed to cluster cooling flows instead
%of optical AGN activity \citep[e.g.][]{vod97}. If so, this
%strengthens our result that BCGs appear to have less ongoing star
%formation and optical AGN activity than non--BCGs. This seems
%somewhat unexpected, as some cluster cores are known to have extended
%H$\alpha$ 
%emission structures \citep[e.g.][]{csf05}. However, the SDSS fiber only
%probes the inner galaxy, and is thus insensitive to such surrounding
%filaments. Furthermore, these structures are observed predominantly in
%cooling core clusters. \citet{ehb07} find that only BCGs at the centers of
%cooling 
%core clusters are more likely to display emission lines than other massive
%(cluster) galaxies. They also confirm that in optically selected cluster
%samples (such as our sample), the BCGs are not more likely to display
%emission lines than massive comparison galaxies.

For case B recombination, the unattenuated value of the Balmer decrement is
$\sim 3$ \citep[more specifically, it is H$\alpha$/H$\beta=2.86$ in
star-forming galaxies and H$\alpha$/H$\beta=3.1$ in AGN, ][]{ost89}.
Remarkably, a considerable fraction of the BCGs in our sample have Balmer
decrements below this value (Fig.~\ref{fig:histos_sample2_spec}, panel [j]).
It has been noted by \citet{kgk06} that 33\% of LINERs in the SDSS sample
have H$\alpha$/H$\beta<2.86$. A possible explanation is that the fits to the
stellar continuum are not entirely reliable for the most massive
galaxies, which tend to have very strong metallic absorption lines in their
spectra.

\subsection{Star formation in BCGs}
\label{sect:star_formation}

The results of the previous section suggest that BCGs do not have increased
amounts of star formation with respect to the comparison sample. This is
somewhat surprising, as there are several BCGs known with signs of recent
star formation \citep[blue colors, H$\alpha$ emission; see ][]{cae99}, and
some cluster cores are known to have extended H$\alpha$ emission structures
\citep[e.g.][]{csf05}. The latter are found to occur exclusively in cooling
core clusters \citep{cae99}. However, since the SDSS fiber only probes
the inner galaxy, it is insensitive to such surrounding filaments.  As
for nuclear line emission, \citet{ehb07} find that only BCGs at the centers
of cooling core clusters are more likely to display emission lines than
other massive (cluster) galaxies. They also confirm that in optically
selected cluster samples (such as our sample), the BCGs are not more likely
to display emission lines than massive comparison galaxies.

\begin{figure}
\begin{center}
%\setlength{\fboxsep}{-\fboxrule}
%\fbox{
%\includegraphics*[bb=3.7cm 2cm 16.8cm 25.5cm,width=0.6\hsize]
\includegraphics[width=1.0\hsize]
{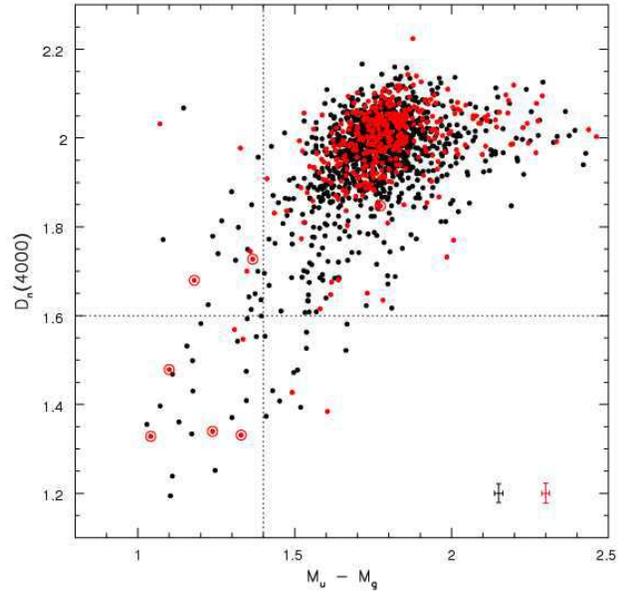}
%}
\caption{~The Balmer break $D_n(4000)$ vs. $M_u - M_g$ for the BCGs (red)
  and the comparison sample CS3s (black). Typical error bars are shown in
  the lower right corner. The BCGs which were identified as
  star--forming from the BPT diagram are marked by circles. Note that only
  one of these is an early-type galaxy and is also in the CS1s sample. In
  the latter sample, there is no BCG and only one comparison galaxy with
  $D_n(4000) < 1.6$.
}  
\label{fig:grd4n}
\end{center}
\end{figure}

Some of the line emission could be
attributed to cluster cooling flows instead of optical AGN activity
\citep[e.g.][]{vod97}.
In addition, only half of the galaxies have detectable emission lines, so
we would prefer a stellar age indicator that can be measured
for all galaxies. 
The strength of the Balmer break $D_n(4000)$ is measurable with a high SNR
in all the galaxy spectra and is an indication of the age of the stellar
population.  \citet{khw03} find that galaxies separate into two distinct
populations, with young, star--forming galaxies having $D_n(4000) \lesssim
1.6$ . We find that the number of galaxies with $D_n(4000) < 1.6$
is very similar to the number of star--forming galaxies identified from the
BPT diagram, for both the BCGs and non--BCGs (see Fig.~\ref{fig:grd4n}).
In Fig.~\ref{fig:grd4n}, we plot $D_n(4000)$ against $M_u - M_g$ within the
{\textit iso23} aperture. $M_u - M_g$ also straddles the Balmer break and can
thus serve to probe the average stellar population at radii larger than the
fiber aperture. The percentage of
galaxies with blue $M_u - M_g$ overall colors is compatible for BCGs and
non--BCGs, i.e. again there is no indication for enhanced star formation in
BCGs. Since we have matched in $M_g - M_r$ color and have argued
that this is essentially a match in stellar mass-to-light ratio, we do not
expect a systematic difference in $M_u - M_g$ (and thus this exercise may
serve to confirm this argumentation). Without the match in $M_g - M_r$, the
comparison sample contains more spiral galaxies, i.e. more star--forming
galaxies than the BCG sample. 

To conclude, we do not find evidence for increased amounts of star formation
in BCGs. However, we would like to caution that our sample is not well
suited for such an investigation: we probe only the very center of the
galaxies (and thus cannot detect line emission on larger scales), and we do
not have X-ray data for our sample (previous studies suggest there is a
strong connection between the cluster X-ray properties and star formation).

\section{Summary and discussion}
\label{sect:summary}

We have developed a refined algorithm for selecting
brightest cluster galaxies (BCGs)  in the C4 cluster
catalog, and we have improved the determination of velocity dispersion and
cluster membership. This refined cluster sample consists of 625 galaxy
clusters at $z<0.1$, and spans a wide range in velocity dispersion,
from galaxy groups to rich clusters of galaxies. This, along with the
detailed information available from SDSS for member galaxies, makes it a
suitable local comparison sample for optically--selected, high--redshift
cluster samples.

Since the original SDSS magnitude measurements of BCGs are affected by
excessive sky subtraction, we have developed a procedure to recover more
accurate magnitudes by adding a fraction of the difference between the {\tt
  local} and {\tt global} {\tt sky} background estimates to the radial
surface brightness profiles of the SDSS galaxies, and by determining new
magnitudes from these corrected profiles. We show that this procedure removes
the systematic bias in $z-J$ color as a function of angular size for
elliptical galaxies with photometry from both the SDSS and the 2MASS
surveys. We also show that our reconstructed surface brightness profiles of
BCGs agree well with previously published aperture photometry, at least to
the isophotal limit of
$r=23\mbox{mag}/\square\arcsec$ within which we measure the flux.\\

The properties of BCGs are governed  by two main factors: their large
(stellar) masses and their locations at the bottom of the  potential well of
their host cluster.
Our large sample of BCGs enables us to disentangle the influence of these two factors
and to assess  the extent to which  BCGs differ from
`ordinary' massive galaxies.

We investigate the occurence of radio--loud AGNs in the BCG population and we 
find that BCGs are more likely to be radio--loud than other galaxies of
the same stellar mass. This enhancement ranges from a factor 10 at stellar
masses of $5 \times 10^{10} M_{\odot}$ to less than a factor of two at masses larger than  $4 \times
10^{11} M_{\odot}$. This difference is
arguably the most striking difference between BCGs and non-BCGs, 
and is
likely due to the additional hot gas available in groups and clusters to
fuel the radio AGN.

The influence of the cluster environment is also evident as an increase in the
fraction of  dark matter in BCGs. The main observational signature of this
excess dark matter is that BCGs have larger radii and higher velocity 
dispersions than non-BCGs. However, since the
conversion of these quantities to dynamical mass depends on the shape of the
mass and light profiles of the galaxy, this result could also be mimicked by
non-homology between the BCGs and non-BCGs. Either case 
leads to a different slope of the
fundamental plane for BCGs, one that is much closer to the virial plane than
the observed fundamental plane of normal ellipticals.

It is interesting to note that the differences from `normal' galaxies are
particularly evident in lower-mass BCGs: these are the galaxies  that have 
a factor 10 higher probability of being radio--loud
and have a significantly larger dark matter mass to light ratio   
(or a larger degree of non-homology) when compared to the non-BCGs. 
The low mass BCGs also deviate most strongly  
from the generic fundamental plane.

We find that the slope of the Faber--Jackson relation is different for BCGs,
in that their velocity dispersion rises less steeply for a given increase in
luminosity than for non-BCGs. We also find evidence that this effect is stronger 
for BCGs in massive clusters. Such a change in the Faber--Jackson relation
is predicted if these systems  form in dissipationless
mergers along elliptical orbits. 
Our results thus support the scenario where BCGs form
mainly via dissipationless mergers, and imply that the merger orbits are preferentially
radial in the most massive clusters.

A difference in the Faber--Jackson relation also implies that BCGs can
follow at most one of the power--law relations often used to estimate the
mass of the supermassive black hole at their center, i.e. either the $M_{\rm
BH} - \sigma_{\rm v}$ or the $M_{\rm BH} - L$ relation. In fact, this is already obvious
from the parameter distributions shown in
Figs.~\ref{fig:histos_sample2} and \ref{fig:histos_sample2_spec}: for the
same distributions in stellar mass and in luminosity, we find a
systematically different distribution in velocity dispersion for our BCGs. 
The standard relations between black hole mass and bulge velocity dispersion
\citep{tgb02} have likely been derived for
non-BCGs and may lead to systematically wrong black hole estimates for
the brightest group and cluster galaxies.\\

We find that BCGs have very similar mean stellar ages and metallicities
to non--BCGs. They have slightly higher $\alpha$/Fe ratios, indicating that
their stars may have formed over a shorter time interval.  
Finally, BCGs display weaker optical emission lines than non-BCGs of the same
stellar mass.  In both BCGs and non-BCGs, the detected emission lines stem
predominantly from low-luminosity optical AGNs. In the accompanying paper
\citep{bes06b}, we further investigate the occurence of AGN activity in
BCGs, and argue that the radio--loud and the emission--line AGN activity are
independent, unrelated phenomena.

\section*{Acknowledgments}

We thank Stefano Zibetti, Anna Gallazzi, Brent Groves, Tim Heckman,
Gabriella De Lucia, and Mateusz Ruszkowski for helpful
discussions and comments, and the people behind CAS, particularly Alex
Szalay, Ani 
Thakar, and Eric Neilsen for setting up the SkyServer website, whose
functionalities eased a large part of this work. We thank the
  anonoymous 
referee for detailed comments which helped to clarify and strengthen the
paper.

PNB would like to thank the
     Royal Society for generous financial support through its University
     Research Fellowship scheme.

Funding for the Sloan Digital Sky Survey (SDSS) has been provided by the Alfred P. Sloan Foundation, the Participating Institutions, the National Aeronautics and Space Administration, the National Science Foundation, the U.S. Department of Energy, the Japanese Monbukagakusho, and the Max Planck Society. The SDSS Web site is http://www.sdss.org/.

The SDSS is managed by the Astrophysical Research Consortium (ARC) for the Participating Institutions. The Participating Institutions are The University of Chicago, Fermilab, the Institute for Advanced Study, the Japan Participation Group, The Johns Hopkins University, the Korean Scientist Group, Los Alamos National Laboratory, the Max-Planck-Institute for Astronomy (MPIA), the Max-Planck-Institute for Astrophysics (MPA), New Mexico State University, University of Pittsburgh, University of Portsmouth, Princeton University, the United States Naval Observatory, and the University of Washington.

\bibliography{refs.bib}

\appendix

\section{Improving SDSS Magnitudes}
\label{sect:magnitudes}

As described in Sect. \ref{sect:masses}, we need to correct the photometry of the
BCGs and around  200,000 galaxies at $z\lesssim0.1$  for the fact that the
standard SDSS photometric pipeline overestimates the sky background.

The SDSS photometric pipeline {\sc photo} estimates the {\tt global} {\tt
  sky} within a frame ($2048 \times 1498$ pixels; $13\arcminf5 \times 
9\arcminf8$) from the median value of the pixels in that frame, clipped at
$2.3 \sigma$. The {\tt local} {\tt sky} background is then determined with
the same sigma--clipping within
a box of $256 \times 256$ pixels ($1\arcminf7 \times 1\arcminf7$) on a grid
every 128 pixels, and interpolated between these positions. This sky
estimate is then subtracted from the image, and the photometry is performed
on the sky-subtracted image.

If a large fraction of the pixels in a $256 \times 256$ pixels box is part
of an object rather than  blank sky, this procedure causes the {\tt
  local} {\tt sky} to be an overestimate of the true sky background. This
may happen for  large galaxies, in crowded fields, and  also around stars, since
the wings of the PSF result in   a considerable stellar halo.

An overestimation of the sky background results in an underestimate of the
surface brightness of the object's pixels  -- thus, the total effect this has
on the flux of an object scales with the  square of
its radius. This effect is therefore particularly
severe for large galaxies. Indeed, \citet{lfr06} analyze the photometry of
BCGs analyzed in \citet{pol95} that lie in the SDSS DR4
and show that the discrepancy in the estimated luminosity
is a function of BCG radius (see also below).

It has been suggested that the true flux of an object may be recovered by
using the {\tt global} {\tt sky} estimate instead of the {\tt local} {\tt
  sky} estimate (Masataka Fukugita; SDSS mailing list). This approach should 
allow a new flux measurement without  performing a new,
independent photometric analysis of the raw images. For each object, the 1D
radial surface brightness profile, measured in 15 radial bins, is
available. By adding the difference between {\tt local} {\tt sky} and {\tt
  global} {\tt sky} to these surface brightnesses, the flux of an object,
assuming the {\tt global} {\tt sky} level in the respective field, can be
measured.

\subsection{Neighboring objects}

For isolated objects, the above argument implies that the {\tt global} {\tt
  sky} is generally a better sky estimate than the {\tt local} one.
 However, for blended objects, we find that the {\tt local} {\tt sky}
accounts for a large fraction of the flux from the respective neighbors; 
in these cases, therefore, using the {\tt global} {\tt sky} estimates would include
flux from the neighbors and thus lead to an overestimate of the luminosity of the object.
This is particularly true if the neighbor is a star of 
similar brightness to the object in question.

We thus need a `trigger' to determine for which objects the {\tt local} {\tt
sky} should be kept, and for which it should be replaced by the {\tt
global} {\tt sky}.

We make the trigger for each galaxy $i$ a function of the ratio ${\rm LR}_i$
of its 
luminosity $L_i$ to the luminosity $L_{\rm nb}$ of its neighbors (within
$1\arcminf6$), defined in the following way:
\begin{eqnarray*}
{\rm LR}_i &=& -2.5\, \log\,(L_i\, /\, L_{\rm nb}\,) \\
L_{\rm nb} &=& 
\frac{\sum\limits_{j\,\in\,\{\rm galaxies\}} L_j w_j}
{\sum\limits_{j\,\in\,\{\rm galaxies\}} w_j}
\;+\; 10\;\;
\frac{\sum\limits_{j\,\in\,\{\rm stars\}} L_j w_j}
{\sum\limits_{j\,\in\,\{\rm stars\}} w_j}\\
w_j &=& {\rm e}^{-\frac{d_{i,j}^2}{2 \left(2 r_{p,i}\right)^2}}
\end{eqnarray*}
where $r_{p,i}$ is the petrosian radius of the galaxy $i$, and $d_{i,j}$ is the
distance between galaxy $i$ and its  neighbor $j$. Thus, the contribution of a
neighbor to $L_{\rm nb}$ is weighted by a Gaussian of width equal to twice the
petrosian radius of the galaxy $i$ -- this is the aperture within which the
SDSS petrosian flux is measured. Since we find that the presence of a star  close
to the object results in a substantial overestimate of the galaxy luminosity if the 
{\tt global} {\tt sky} is used,
stars are weighted with an additional factor of
10 (the exact value of this factor makes only little difference, but we do
find slightly better results using a factor of 10 rather than  5). We
suspect that the stellar halo due to the broad PSF wings is accounted for
primarily as {\tt local} {\tt sky} background.

We find that for ${\rm LR}_i < -2.5$, the flux of neighboring objects is
negligible enough for the {\tt global} {\tt sky} to be the superior sky
estimate. For ${\rm LR}_i > -1$, the flux of neighboring objects contributes
a substantial fraction of  the {\tt local} {\tt sky} estimate, so that it cannot be
substituted by the {\tt global} {\tt sky}.

\subsection{The method}
\label{sec:sub-rules}

Rather than simply  substituting the {\tt global} {\tt sky} for the {\tt
  local} {\tt sky} (or not), we calculate         
the fraction $f_{\rm sky}$ of the difference between {\tt local}
and {\tt global} {\tt sky} to be added to the surface brightness
profile of each galaxy according to the following criteria:                               
\setlength{\leftmargini}{0.6cm}
\begin{enumerate}
\setlength{\itemindent}{0cm}
\setlength{\labelsep}{0.1cm}
\setlength{\labelwidth}{0.5cm}
\item
If $\Delta_{\tt sky} = {\tt sky}_{\tt local} - {\tt sky}_{\tt global} < 0$ in
any one of the 
five bands, then $f_{\rm sky}=0$ . In these cases, the two sky measurements are
essentially equivalent, and subtracting flux from the surface brightness
profile will add noise to the flux measurement.
\item Values of $\Delta_{\tt sky}$ are constrained to be  $ \le 10^{-9}
  \mbox{maggies}/\square\arcsec$.
\item The maximum value for $f_{\rm sky}$ is 0.7 .  In
  Sect.~\ref{sect:2mass_compare} we  
  show that this is superior to using $f_{\rm sky}^{\rm max}=1$ . This value
  is assigned to objects with ${\rm LR}_i \le -2.5$.
\item The minimum (non-zero) value for $f_{\rm sky}^{\rm min}$ is 0.1 . This value
  is assigned to objects with ${\rm LR}_i \ge -1$.
\item For objects with $-2.5 < {\rm LR}_i < -1$, $f_{\rm sky}$ is a linear
  function of ${\rm LR}_i$, being continuous at the endpoints with $f_{\rm
    sky}^{\rm max}$ and $f_{\rm sky}^{\rm min}$ .
\end{enumerate}

\subsection{Comparison for the BCGs of \citet{pol95}}

In order to assess the performance of our method to correct the SDSS
magnitudes, we need (an) external dataset(s) with accurate photometry. We
rely on two such datasets: the aperture photometry of BCGs published in
\citet[referred to as PL95 hereafter]{pol95}, and the 2MASS survey
\citep{scs06}. The method is 
considered successful if it can simultaneously reproduce the
curve-of-growth of a large fraction of the PL95 BCGs, and if galaxies do not
show any systematic bias in their median SDSS--2MASS  colors. 
\\

\begin{figure*}
\begin{center}
\includegraphics[width=0.48\hsize]{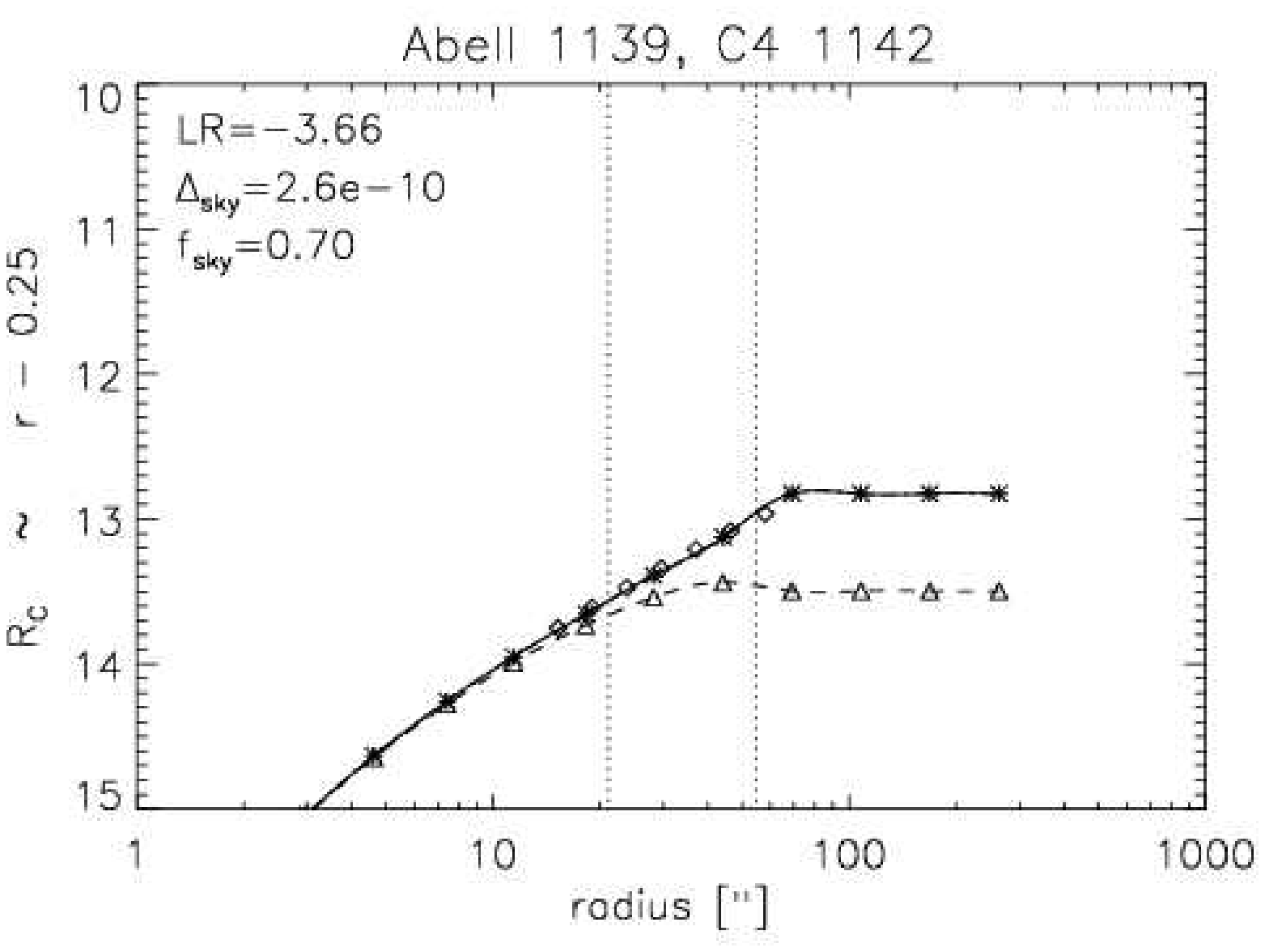}
\includegraphics[width=0.48\hsize]{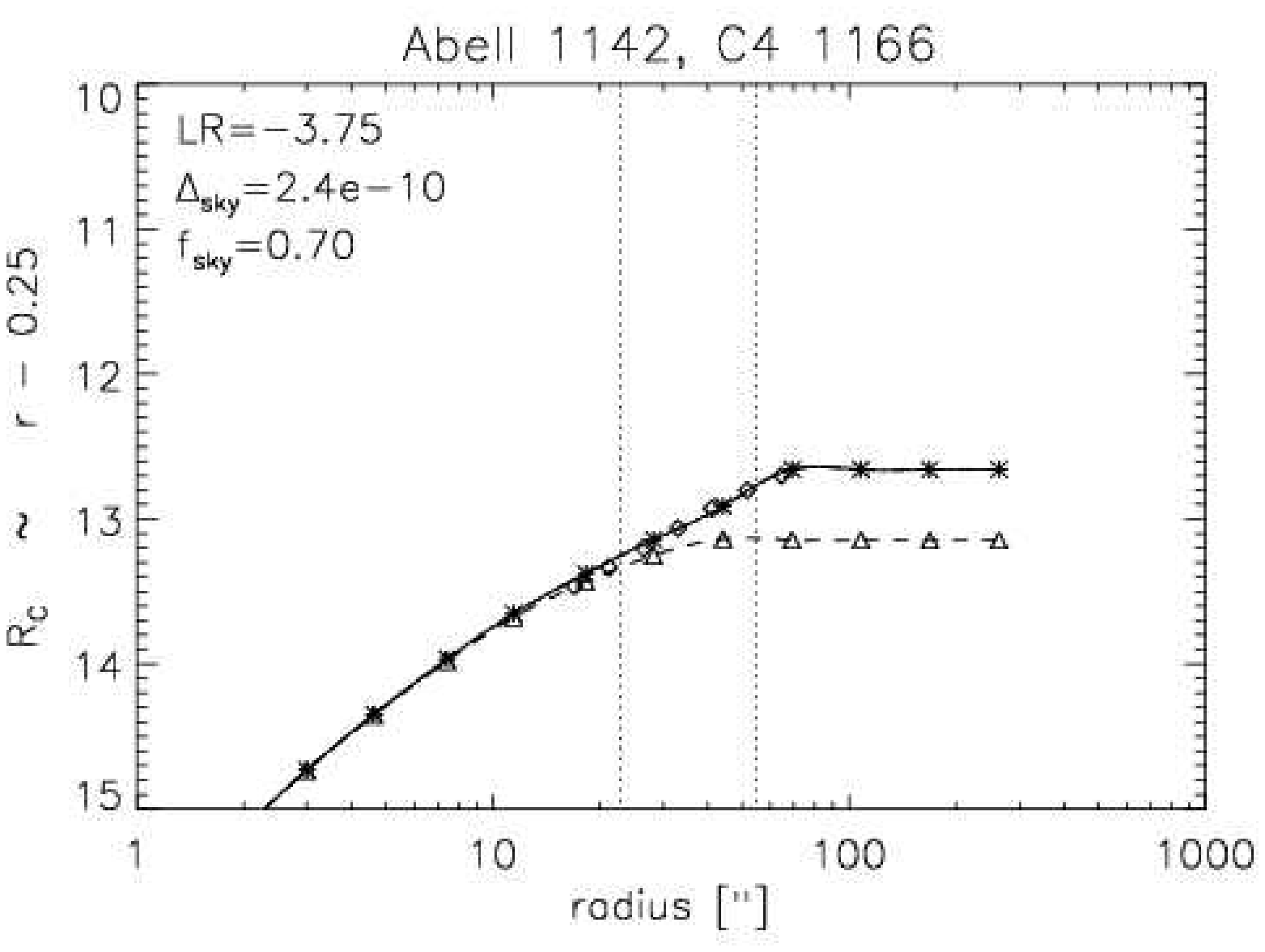}
\includegraphics[width=0.48\hsize]{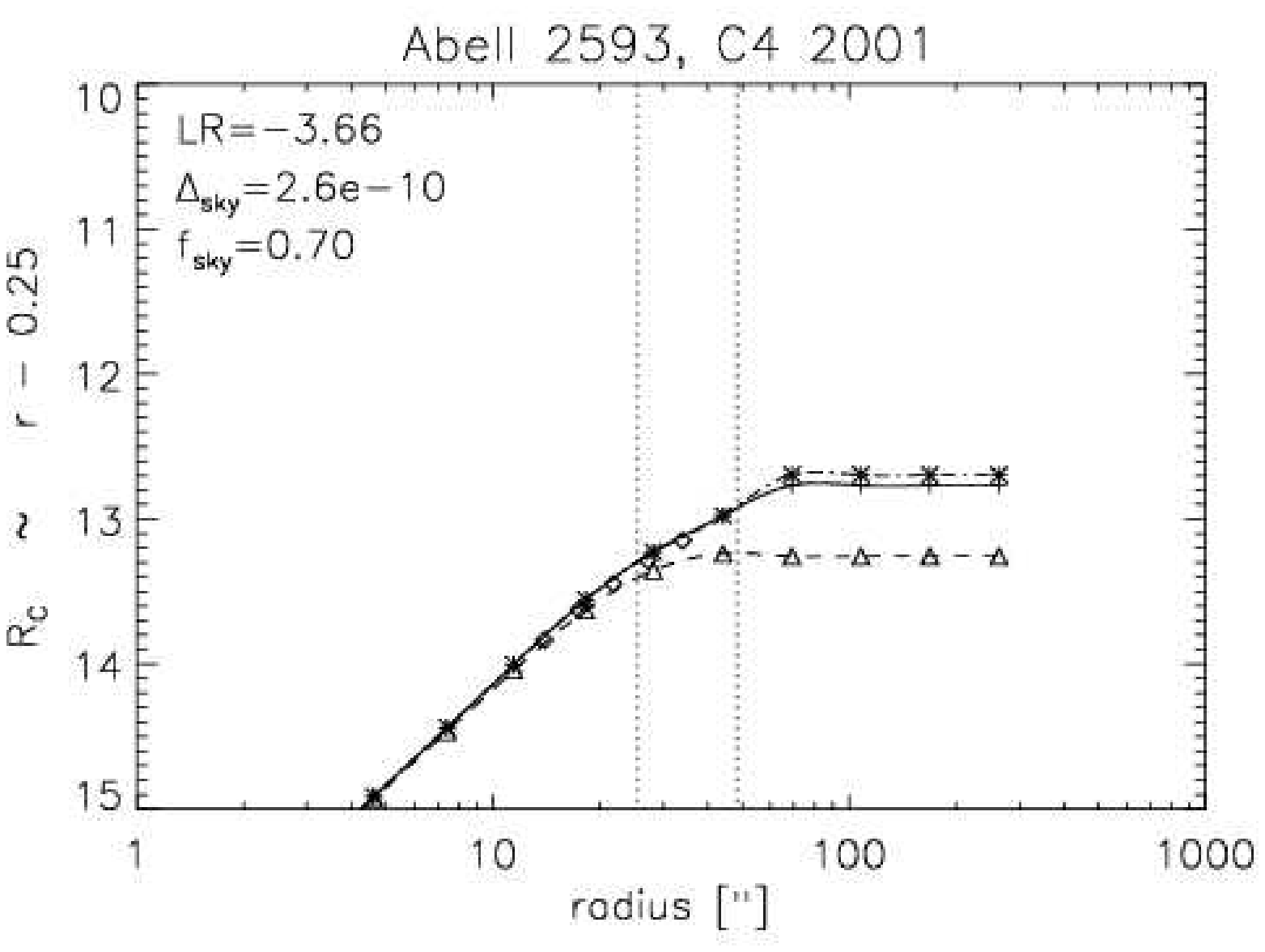}
\includegraphics[width=0.48\hsize]{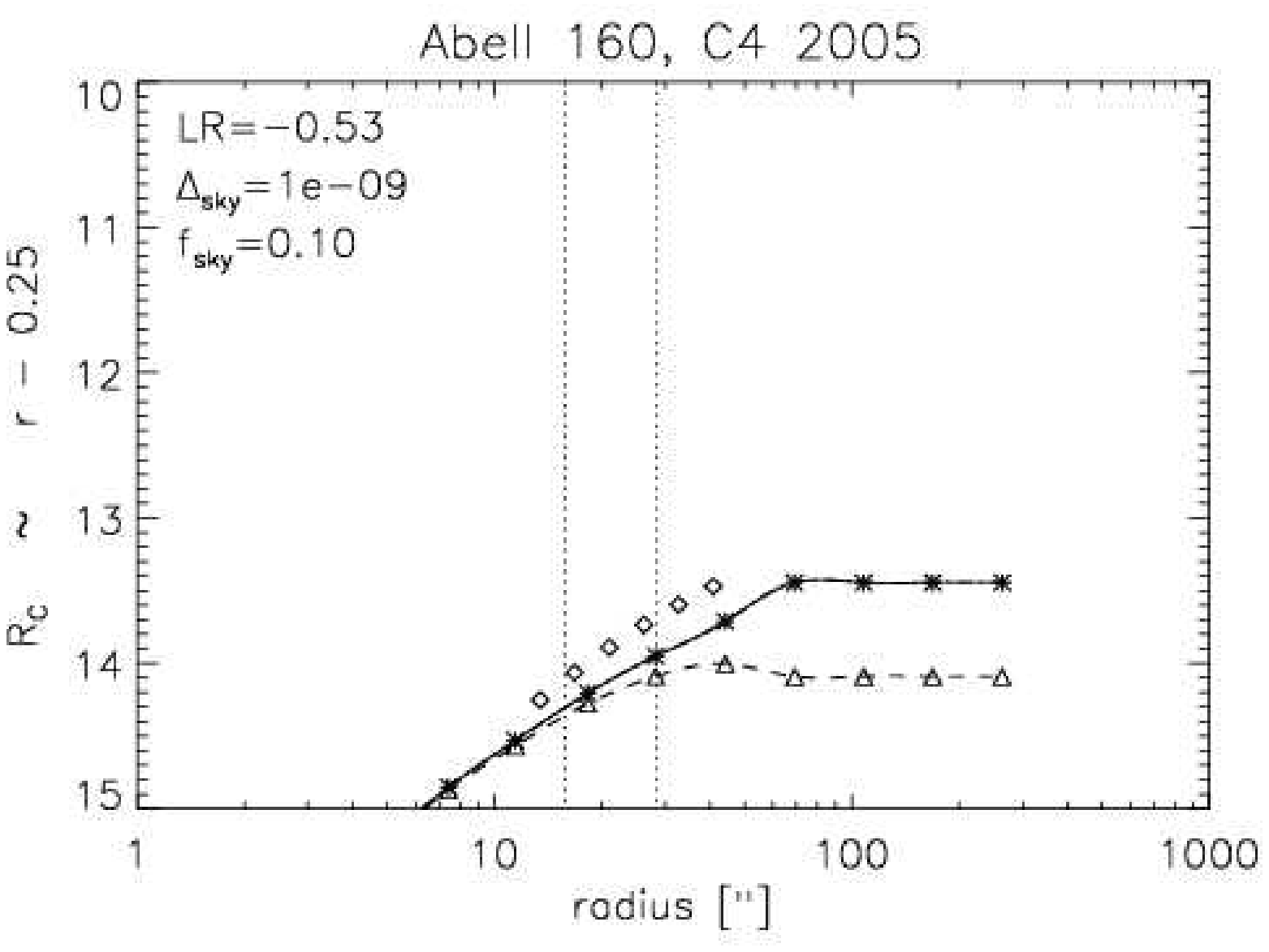}
\includegraphics[width=0.48\hsize]{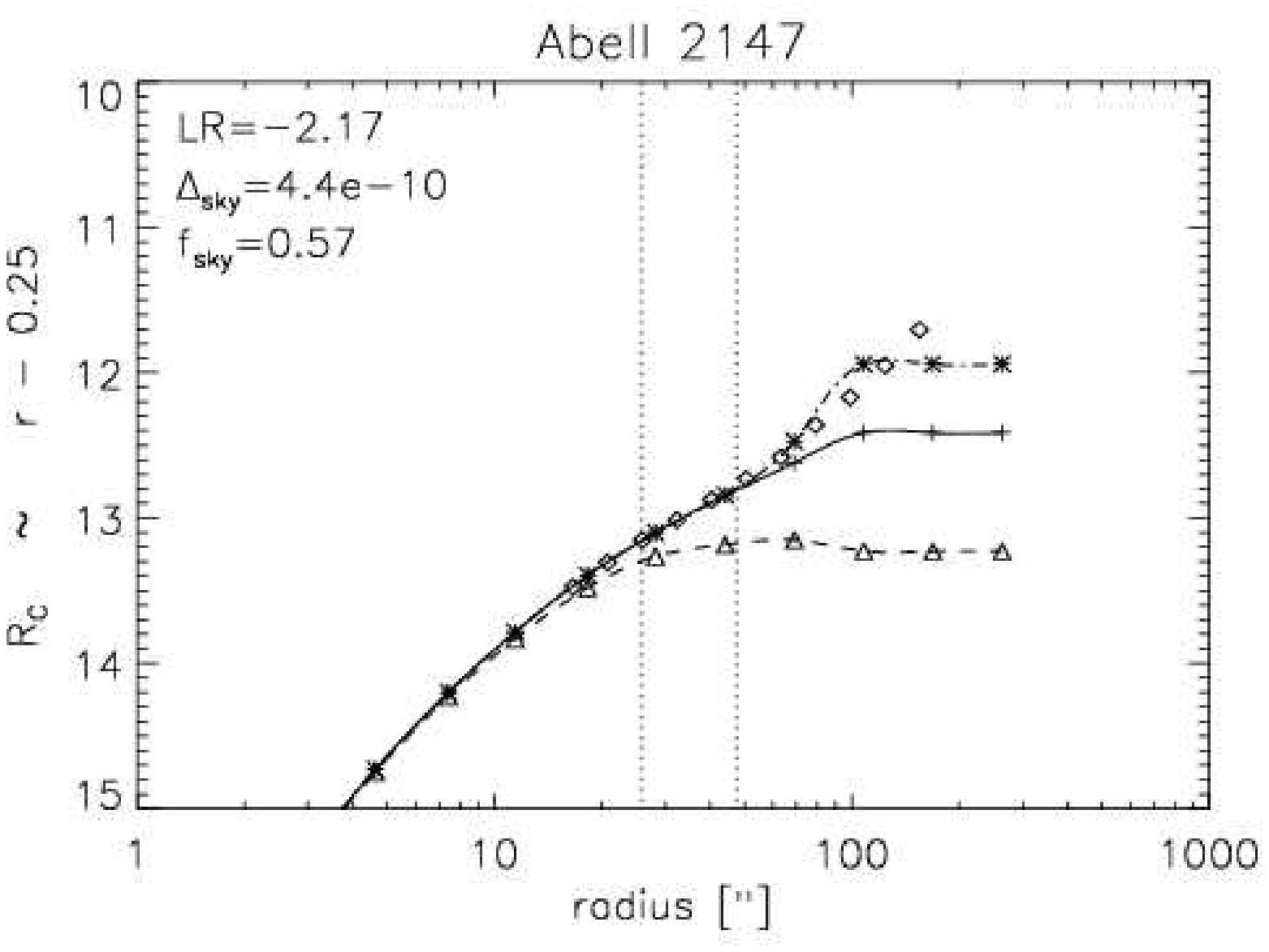}
\includegraphics[width=0.48\hsize]{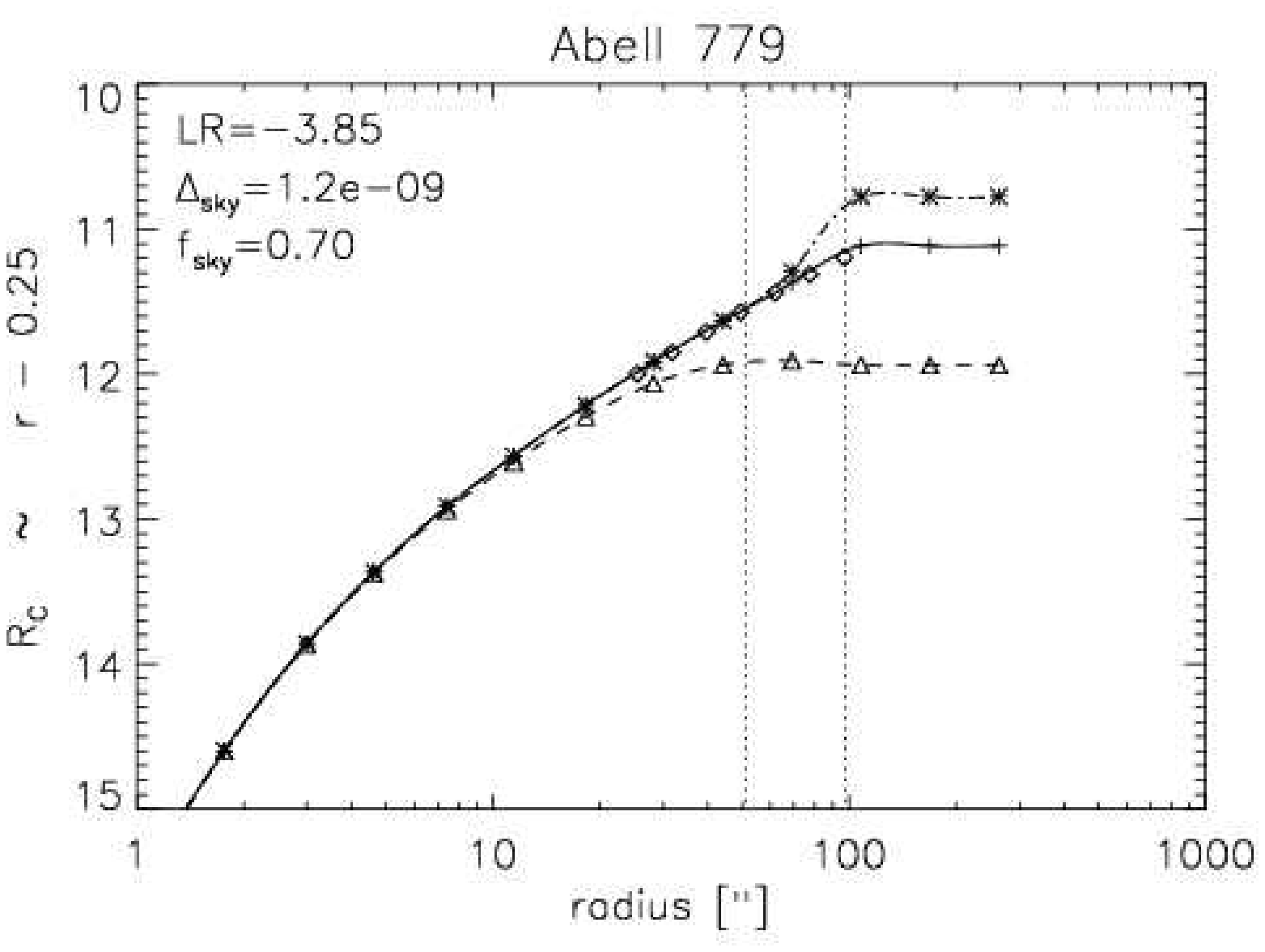}
\caption{~Comparison of the aperture photometry of \citet{pol95} (shown as
  diamonds) and SDSS photometry for 6 BCGs. The uncorrected SDSS magnitudes
  are shown as triangles, with the respective cubic spline fit as a dashed
  line. By adding flux of $f_{\rm sky} \Delta_{\rm sky}$, we obtain the
  curve-of-growth shown by $+$ symbols and as dash-dotted line. Noisy
  photometry at large radii can cause jumps in the curve-of-growth; this
  can be alleviated by the method described in the text, and is shown as
  $\times$ symbols, and as solid line. The dotted lines indicate the radii
  at which the surface brightness profile reaches a magnitude of
  $r=23\,\mbox{mag}/\square\arcsec$ (left line) and
  $r=24\,\mbox{mag}/\square\arcsec$ (right line).}  
\label{fig:mags_bcgs_pol95}
\end{center}
\end{figure*}

In Fig. \ref{fig:mags_bcgs_pol95} we demonstrate that our method can
reproduce the aperture photometry of the PL95 BCG sample. Of the
119 BCGs in that sample, 35 have imaging data from 
DR5, and 12 are also in our BCG sample (which was based on DR3).
%(\textit{why do 19 show up in the image list?}) 
We show the curve-of-growth
of six representative BCGs in Fig. \ref{fig:mags_bcgs_pol95}.
We assume $r' - R_{C} = 0.25$ as the typical color of an elliptical at $z=0$
\citep{fsi95} in order to  
compare measurements taken in different bands.

The examples show how the original SDSS photometry breaks down at radii
larger than $\sim 20\arcsec$. Our improved photometry is able to reproduce
the PL95 aperture photometry much better, out to $\gtrsim 80\arcsec$.  Of
the 12 BCGs in common with our sample, our photometry fails to accurately
reproduce PL95 only for the BCG of Abell~160 (C4\_DR3\_2025): this is because {\tt
  PHOTO} attributes a large fraction of the BCG luminosity to a secondary
nucleus; hence LR is rather large, and only little sky background is added
to the brightness profile. Of the 35 BCGs in DR5, this is also the case for
the BCG of Abell~1185 (not shown). The crowding in the fields around the
BCGs of Abell~2040 and Abell~2052 also leads to an underestimation of the
flux by about 0.2~mag (not shown).

In addition to bad sky subtraction, noise in the surface brightness profile plays an
important role particularly at large radii, as can be seen in the
curve-of-growth of the BCG of Abell~779 (Fig.~\ref{fig:mags_bcgs_pol95}).
The last non-zero bin of the SDSS surface brightness  profile causes a jump in total magnitude of
more than 0.5~mag, which is not seen in the PL95 photometry. To avoid such
jumps, we compare the cumulative flux for each bin of radius $\ge 18\arcsec$ to
the flux predicted for that bin from the slope of the curve-of-growth from
the two bins prior in radius. If the flux is more than 30\% larger than
predicted (or if the 
surface brightness in the respective bin is negative), the measured
flux is substituted by the predicted flux. For several BCGs this procedure
improves the agreement between  the SDSS and PL95 brightness profiles.  For
one (the BCG of Abell~2147), this causes disagreement at large
radii. However, this occurs at a significantly larger radius than the one
within which we measure the magnitudes.

\subsection{Comparison to 2MASS magnitudes}
\label{sect:2mass_compare}

Of the 200,000 unique SDSS galaxies at $z<0.1$, about half have a counterpart
in the 2MASS survey's Extended Source Catalog (XSC); we made use of the list
of 2MASS objects matched to SDSS galaxies provided via the NYU VAGC
\citep{bss05} to identify these galaxies and obtain their 2MASS properties. 
\begin{figure}
\begin{center}
\includegraphics[width=\hsize]{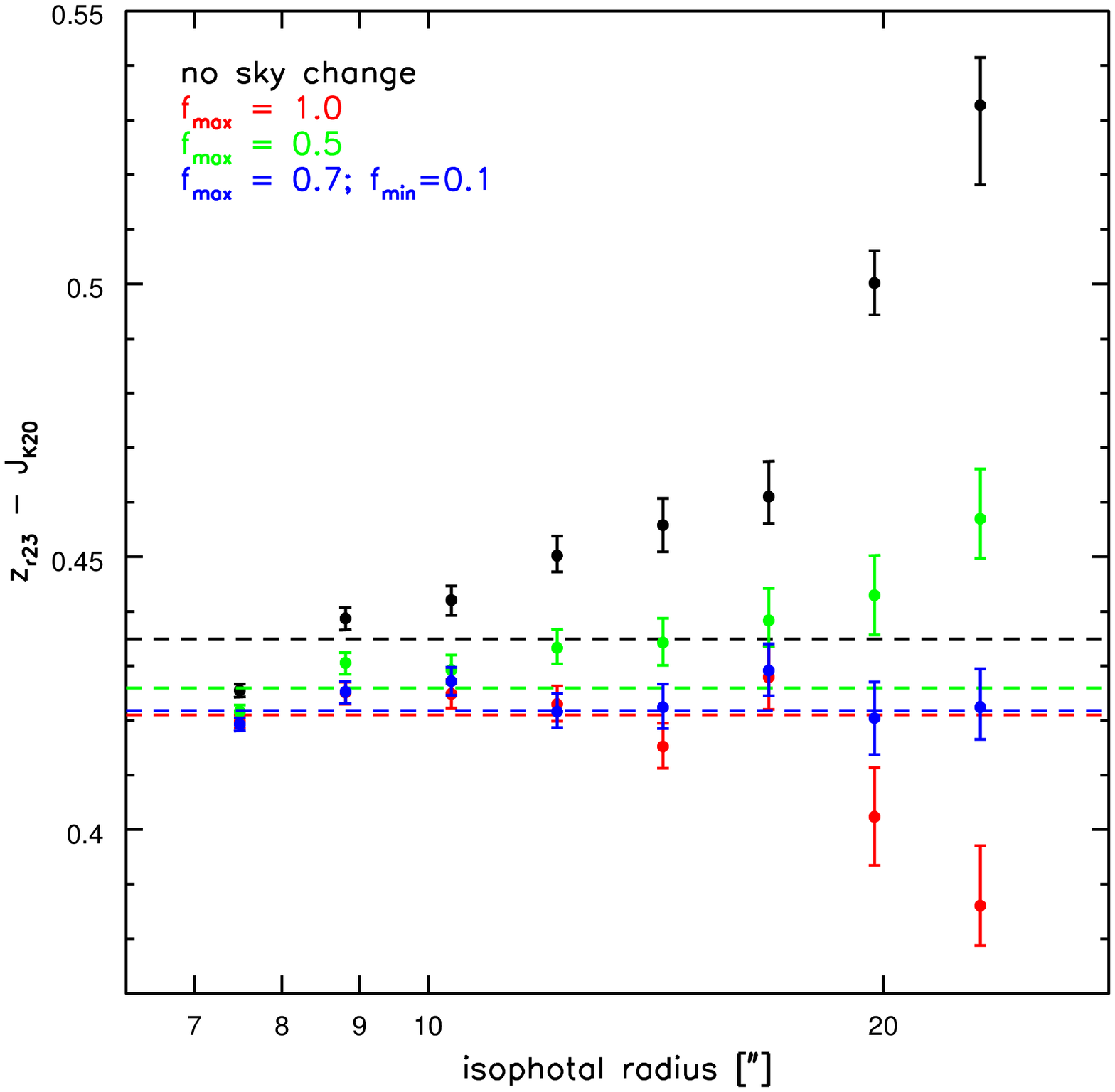}
\caption{~The difference between SDSS $z$ magnitude (measured within the
  $r=23\,\mbox{mag}/\square\arcsec$ radius) and 2MASS $J$ magnitude (measured
  within the $K=20\,\mbox{mag}/\square\arcsec$ radius) as a function of galaxy
size (given by the $r=23\mbox{mag}/\square\arcsec$ isophotal
radius). We measure the median difference in bins of radius; the error bars
denote the 68\% levels of the distribution in each bin, divided by the
square root of the number of galaxies in that bin. For the black symbols,
the sky subtraction has not been changed from the original SDSS value. For
the red symbols, $f_{\rm sky}^{\rm max} = 1.0$, i.e. if an object meets the
criteria to replace the {\tt local} by the {\tt global} {\tt sky}, the full
difference is added (see Section~\ref{sec:sub-rules} for details). For the
green symbols, $f_{\rm sky}^{\rm max} = 0.5$, and for the blue symbols,
$f_{\rm sky}^{\rm max} = 0.7$ and $f_{\rm sky}^{\rm min} = 0.1$. The dashed
lines show the median values for all galaxies.} 
\label{fig:mags_2mass}
\end{center}
\end{figure}

In order to be able to directly compare the SDSS and XSC data,
we work with isophotal magnitudes. One of the magnitude measurements
provided by 2MASS is measured within the $K = 20\,\mbox{mag}/\square\arcsec$
isophote. $K=20$ corresponds approximately to $r=23$, hence we measure SDSS
magnitudes within the radius where the surface brightness profile drops to
$r=23\,\mbox{mag}/\square\arcsec$. We compare these magnitudes for
elliptical galaxies, selected using standard cuts that have
been used in previous SDSS studies (i.e. $c_{\rm SDSS} = R_{90}^{\rm
  SDSS}/R_{50}^{\rm SDSS} > 2.86$ and $g_{\rm petro}^{\rm SDSS} - r_{\rm
  petro}^{\rm SDSS}$). We also limit the sample to galaxies where the SDSS
and 2MASS radii agree to within 30\% (this criterion changes for each
magnitude measurement). We consider only galaxies with
a radius larger than $7\arcsec$, the minimum radius necessary to
avoid PSF effects for 2MASS data \citep{jcc00}.
After correcting for galactic extinction (and converting the $J$-band
magnitude to an AB magnitude), we compare
the SDSS $z$-band magnitudes to the 2MASS $J$-band magnitudes, since these
bands are adjacent in wavelength and thus color differences from different
stellar populations can be expected to be minimal.

In Fig.~\ref{fig:mags_2mass}, we investigate the color $(z_{r23} - J_{K20})$ as a
function of galaxy size (given by the $r=23\,\mbox{mag}/\square\arcsec$
isophotal radius in the SDSS) for four different sky subtractions. Clearly,
for the original SDSS sky subtraction, the color term is a strong function of
galaxy size, indicating that the sky is systematically oversubtracted
in the optical band. This
accounts for a systematic difference of the order of $0.1$ mag for galaxies
larger than 20$\arcsec$ 
\citep[for individual galaxies, this may be much more, as
demonstrated by ][]{lfr06}. However, substituting the {\tt local} {\tt
  sky} estimate by the {\tt global} one (i.e. $f_{\rm sky}^{\rm max} = 1.0$)
leads to an underestimation of the sky, and thus an overestimation of the
luminosity.  We find that with $f_{\rm sky}^{\rm max} = 0.7$, there is 
little variation of the median color term with galaxy size, and no
systematic trend (note that setting $f_{\rm sky}^{\rm min} = 0.1$ makes only
a small difference).

\subsection{Final magnitudes}

The isophotal magnitudes are not
redshift independent because of  cosmological surface brightness dimming. 
We thus modify the algorithm to correct both 
for $(1+z)^4$ surface brightness dimming and for
galactic
extinction (i.e. these corrections applied directly to the radial profile
before the magnitudes are measured).

We choose to use isophotal magnitudes corresponding to $r=23$ at $z=0$
(i.e. $r=23.41$ at $z=0.1$). This is a rather bright cut-off, but is less
sensitive to residual errors in the sky subtraction and/or surface
brightness measurements (cf. Fig.~\ref{fig:mags_bcgs_pol95}).

\subsection{Influence of the sky subtraction on our results}

\begin{table*}
\begin{center}
\caption{Comparison of the results of Kolmogorov--Smirnov tests
(expressed as the decimal logarithm of $1$ minus the
confidence
level at which the null hypothesis, that the distributions are drawn from an
identical parent population, can be rejected) on BCG and non--BCG
distributions of various parameters, for different versions of sky
subtraction applied to the SDSS photometry.
}
\label{tab:skysub_tests}
\renewcommand{\arraystretch}{1.3}
\begin{tabular}{l c c c c c c c c c c c c c c}
sky subtraction & 
\multicolumn{1}{c}{$R_{50}$} & 
\multicolumn{1}{c}{$R_{\rm iso23}$} & 
\multicolumn{1}{c}{$c^{\prime}$} & 
\multicolumn{1}{c}{$\mu_{50}$} & 
\multicolumn{1}{c}{$\mu_{\rm iso23}$} & 
\multicolumn{1}{c}{$b/a$} & 
\multicolumn{1}{c}{$\Delta(g-r)$} & 
\multicolumn{1}{c}{$\sigma_{\rm v}$} &
\multicolumn{1}{c}{$M_{\rm dyn, 50}/M_{\star}$} &
\multicolumn{1}{c}{$D_n(4000)$} &
\multicolumn{1}{c}{H$\delta_{\rm A}$} &
\multicolumn{1}{c}{L(H$\alpha$)} &
\multicolumn{1}{c}{L(H$\beta$)}
\\
\hline
\hline
$f_{\rm sky}^{\rm max} = 0.7$ & -5.44 & -3.60 & -4.58 & -8.78 & -12.43 & -7.28 & -3.13 & -2.06 & -13.33 & -0.88 & -2.67 & -1.60 & -0.01 \\
\hline
$f_{\rm sky}^{\rm max} = 0.5$ & -6.13 & -3.37 & -4.84 & -10.38 & -14.46 & -7.71 & -2.61 & -2.27 & -14.00 & -0.57 & -2.71 & -1.04 & -0.04 \\
$f_{\rm sky}^{\rm max} = 1.0$ & -4.91 & -3.29 & -3.48 & -6.44 & -12.76 & -6.44 & -2.16 & -2.19 & -12.57 & -0.61 & -2.74 & -1.75 & -0.01 \\
\end{tabular}
\end{center}
\end{table*}

Our analysis of BCG properties (Sect.~\ref{sect:optical})
as a function of stellar mass  depends critically
on the correct determination of the luminosities of the objects in our sample.
If the luminosities we assign to BCGs are
underestimated, then the comparison galaxies (which are matched in stellar
mass) would be systematically less massive than the BCG, leading
to possibly spurious differences in their physical properties.

To estimate the effect that sky subtraction has on our results, we repeat the
analysis presented in Sect.~\ref{sect:optical} with photometry derived with
values of $f_{\rm sky}^{\rm max} = 0.5$ and $f_{\rm sky}^{\rm max} =
1.0$. To evaluate whether there could be significant quantitative
effects on our  results, we compare the results of
the Kolmogorov--Smirnov test on the BCG and non--BCG distributions for
several parameters (as shown also in
Figures~\ref{fig:histos_sample2} and \ref{fig:histos_sample2_spec}) in
Table~\ref{tab:skysub_tests}. We find only minor differences to our initial
results; none of our conclusions are altered as a result of using a different value of
$f_{\rm sky}^{\rm max}$.

\section{Cluster Examples}
\label{sect:clusters}

In the following we present finding charts and redshift histograms for
clusters referred to in the text. The finding charts are centered on the BCG,
and spectroscopically confirmed cluster members are marked by triangles. In
the redshift histograms, a short-dashed line indicates the
cluster redshift, the long-dashed line the redshift of the BCG, and the
dotted lines the $\pm3\sigma_{\rm v, cl}$ limits. Also, the redshift (both the C4 and
our measurement) as well as the velocity dispersion measurements (given in
km/s) are listed in the redshift histograms.

\begin{figure*}
%[htbp]
\begin{center}
\begin{minipage}{0.48\hsize}
\includegraphics[width=1\hsize]{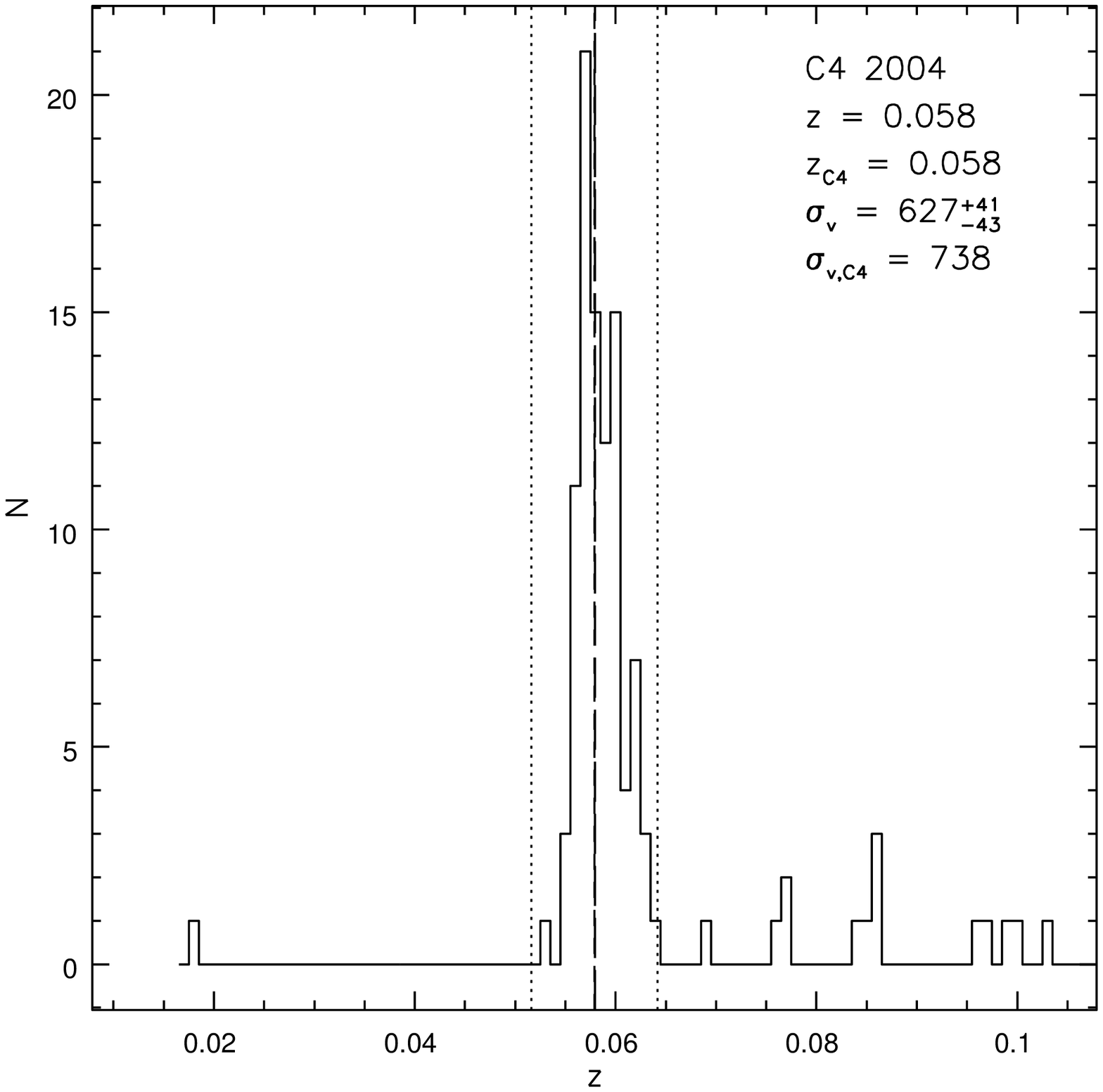}
\end{minipage}
\begin{minipage}[][0.45\hsize][t]{0.45\hsize}
\caption{~\textbf{C4\_2003.}
Since most
galaxies seem to cluster around 
the galaxy shown at the center of this image, this is the galaxy we identify
as the BCG, even though the galaxy about 6\arcmin west-south-west of the
center (marked by two lines) is brighter by a third of a magnitude. The
$R_{200}$ of this cluster is 1.5~Mpc, which translates to $\sim 20\arcmin$.
}
\label{fig:c4_2004}
\end{minipage}

\begin{minipage}{0.87\hsize}
\includegraphics[width=1\hsize,]{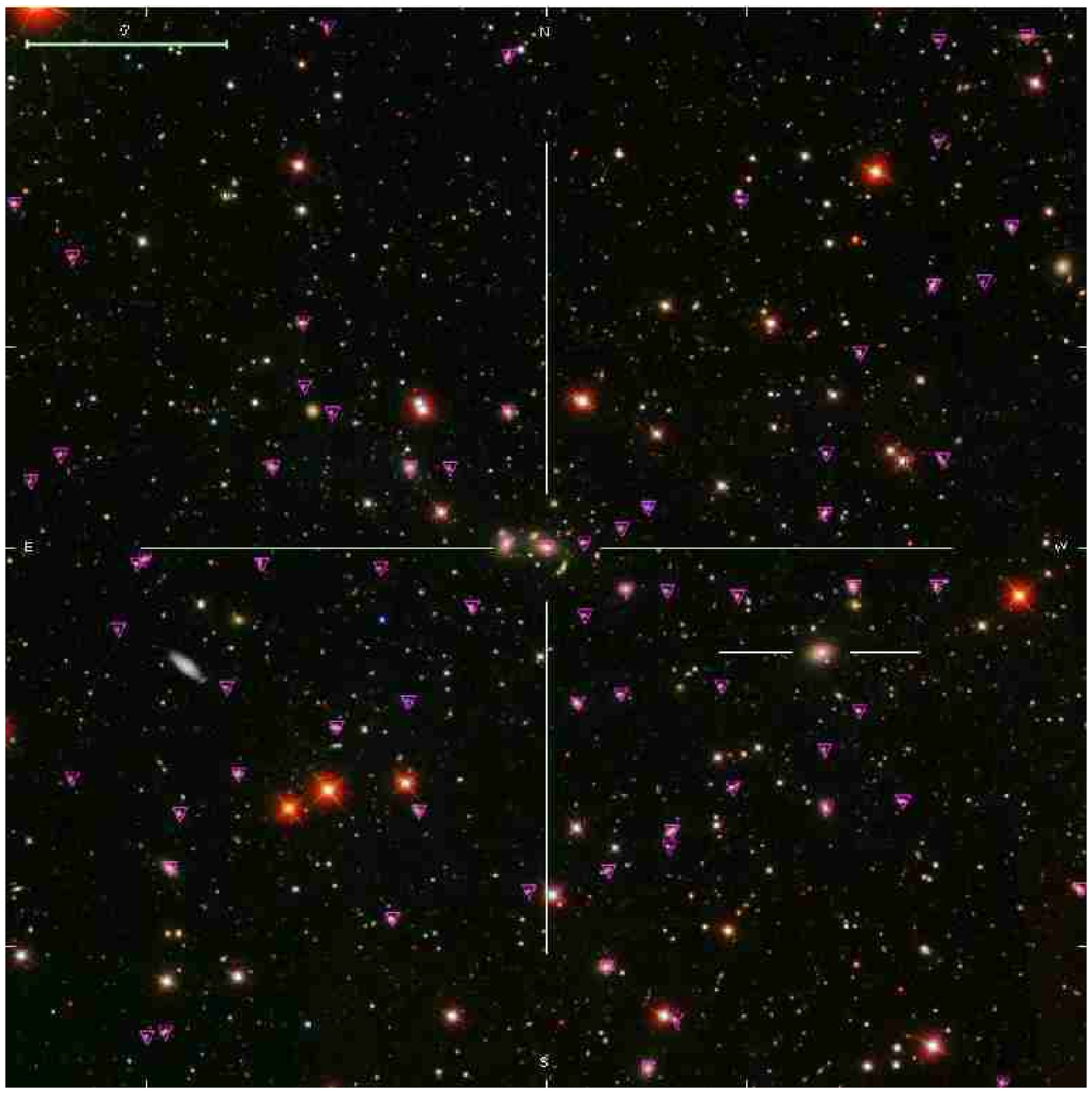}
\end{minipage}
\end{center}
\end{figure*}

\begin{figure*}
%[htbp]
\begin{center}
\begin{minipage}{0.48\hsize}
\includegraphics[width=1\hsize]{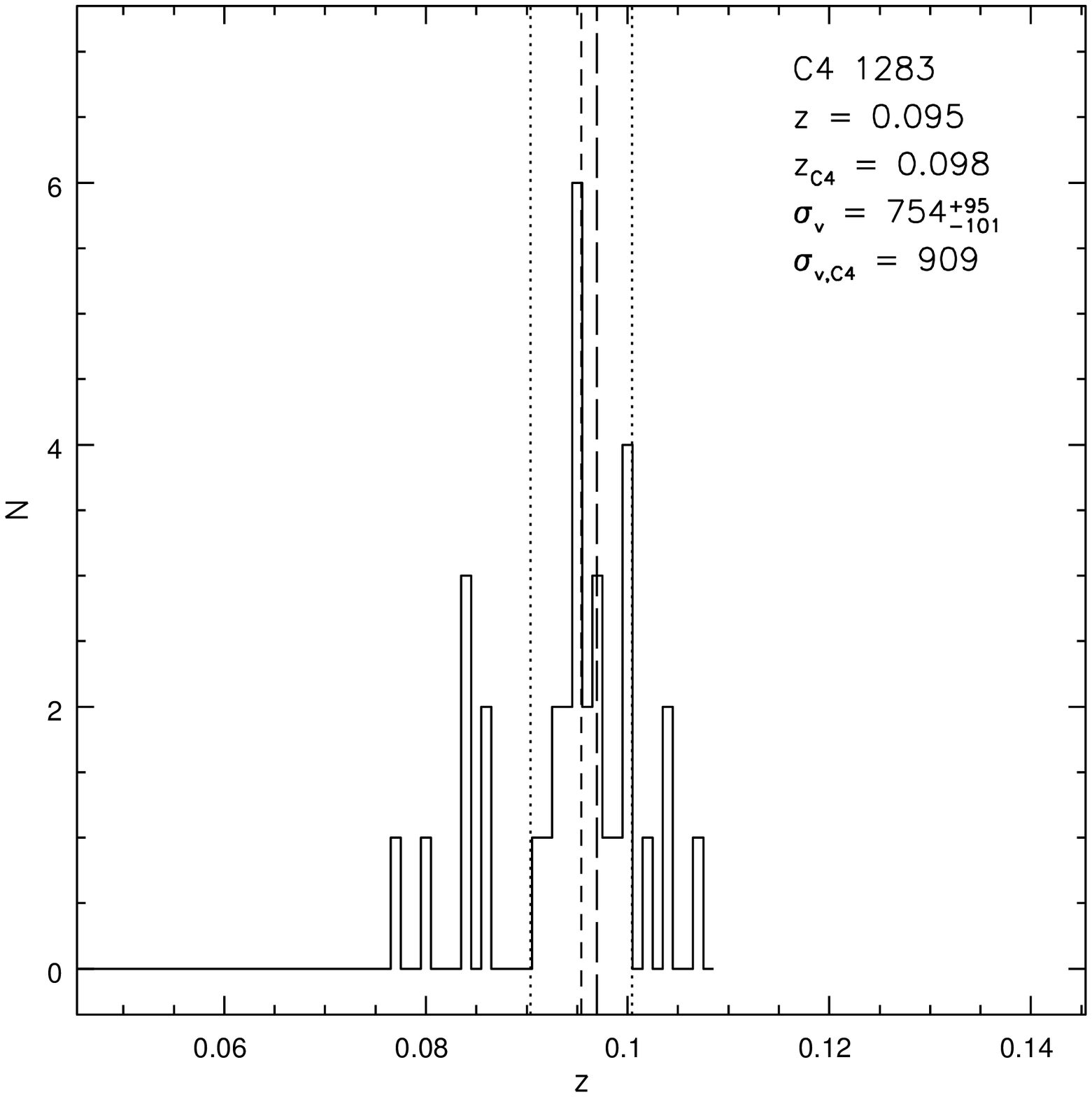}
\end{minipage}
\begin{minipage}[][0.45\hsize][t]{0.45\hsize}
\caption{~\textbf{C4\_DR3\_1283.} For this cluster, the C4 mean galaxy (marked by
  diagonal lines in the finding chart) lies 2.8~Mpc from 
  the BCG which we identify. The large-scale distribution of galaxies at the
  cluster redshift suggests that C4 selected part of the infall region of
  this cluster, but fails to pick up the cluster itself. Note that the
  brightest galaxy in the field is IC 0504 at $z=0.013$, whereas many of the
  other  bright galaxies in the foreground belong to C4\_DR3\_1356 at
  $z=0.03$. Below, we show thumbnail images of the C4 mean galaxy (left) and
  our BCG (right).
}
\label{fig:c4_1283}
\vspace{0.2cm}
\includegraphics[width=0.48\hsize]{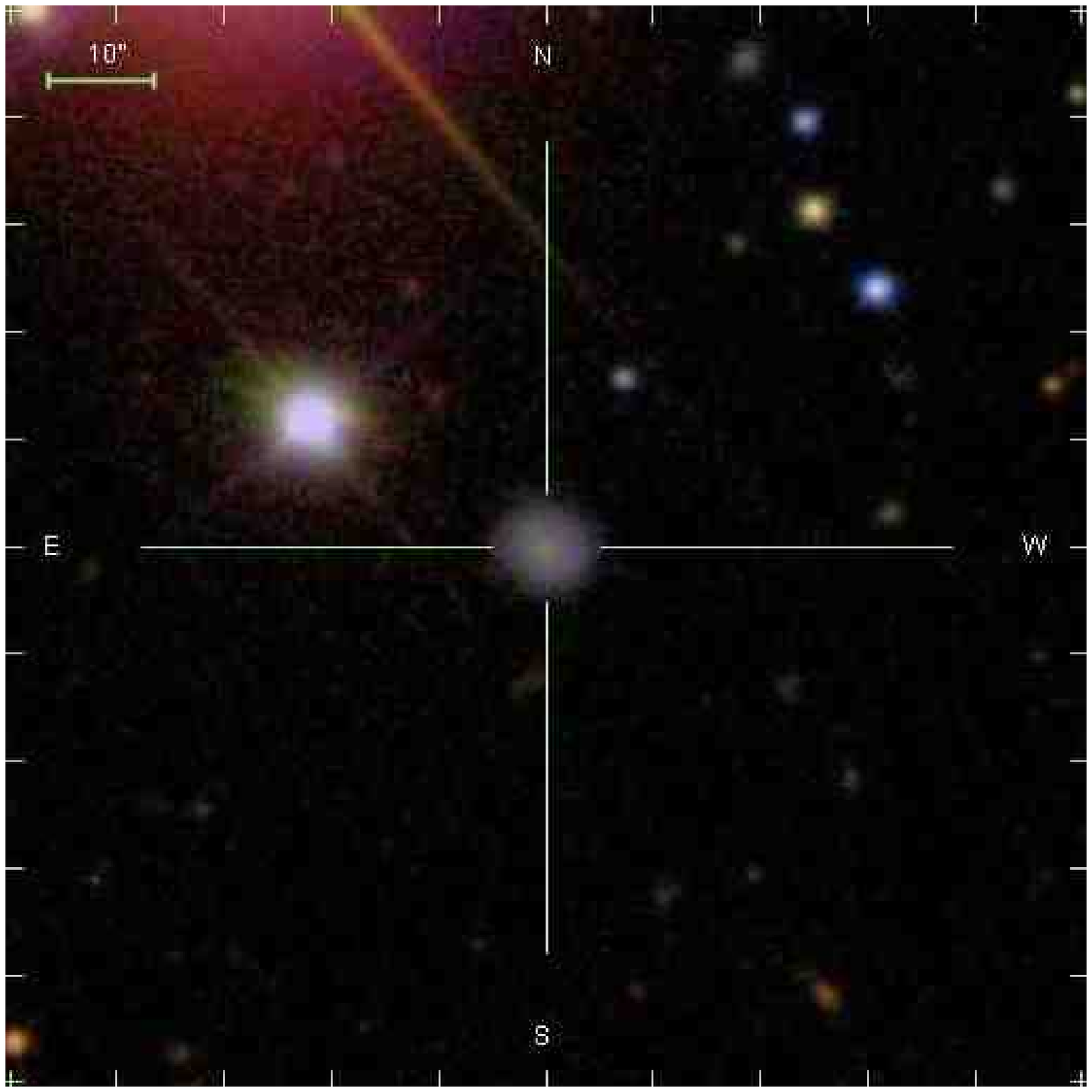}
\includegraphics[width=0.48\hsize]{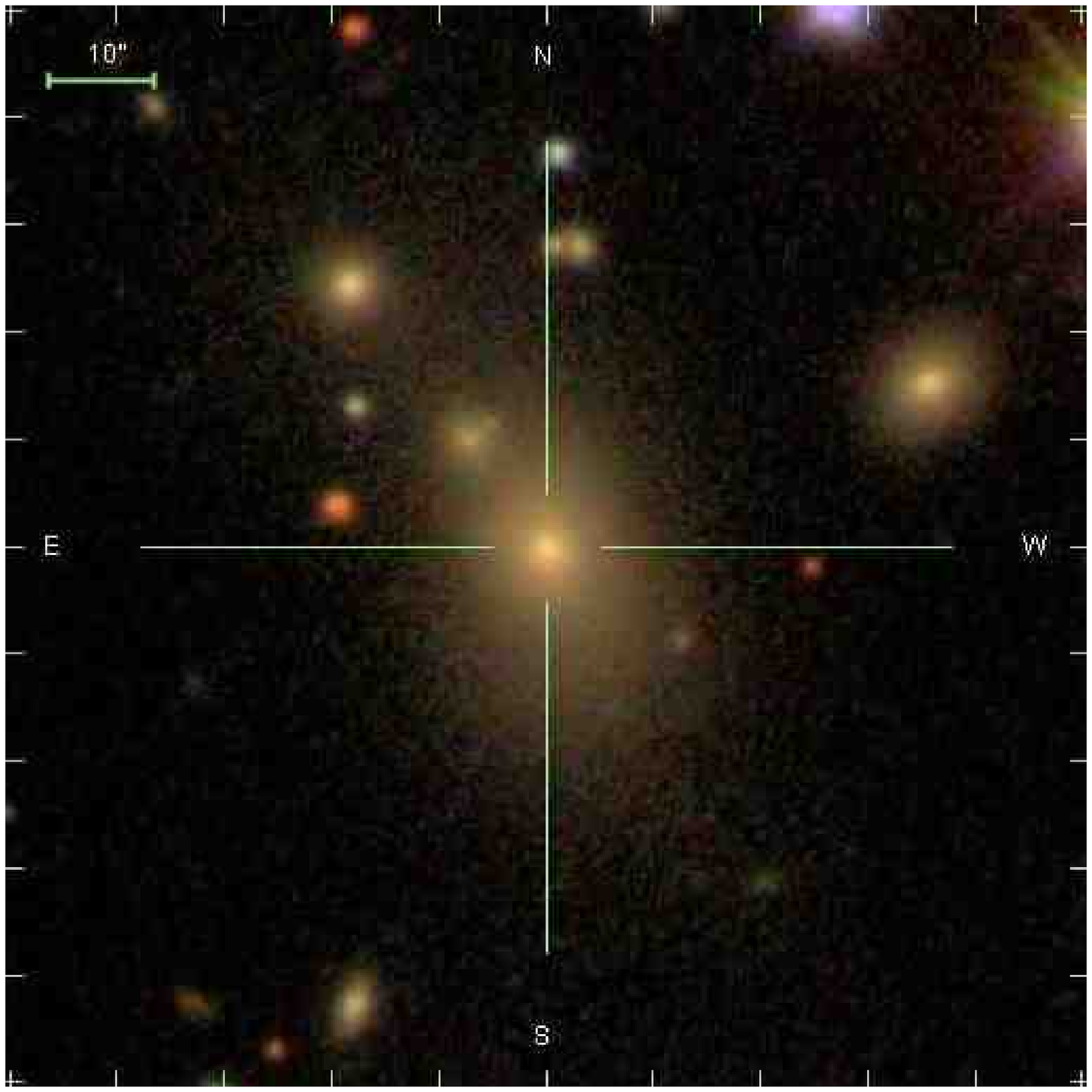}
\end{minipage}

\begin{minipage}{0.87\hsize}
\includegraphics[width=1\hsize,]{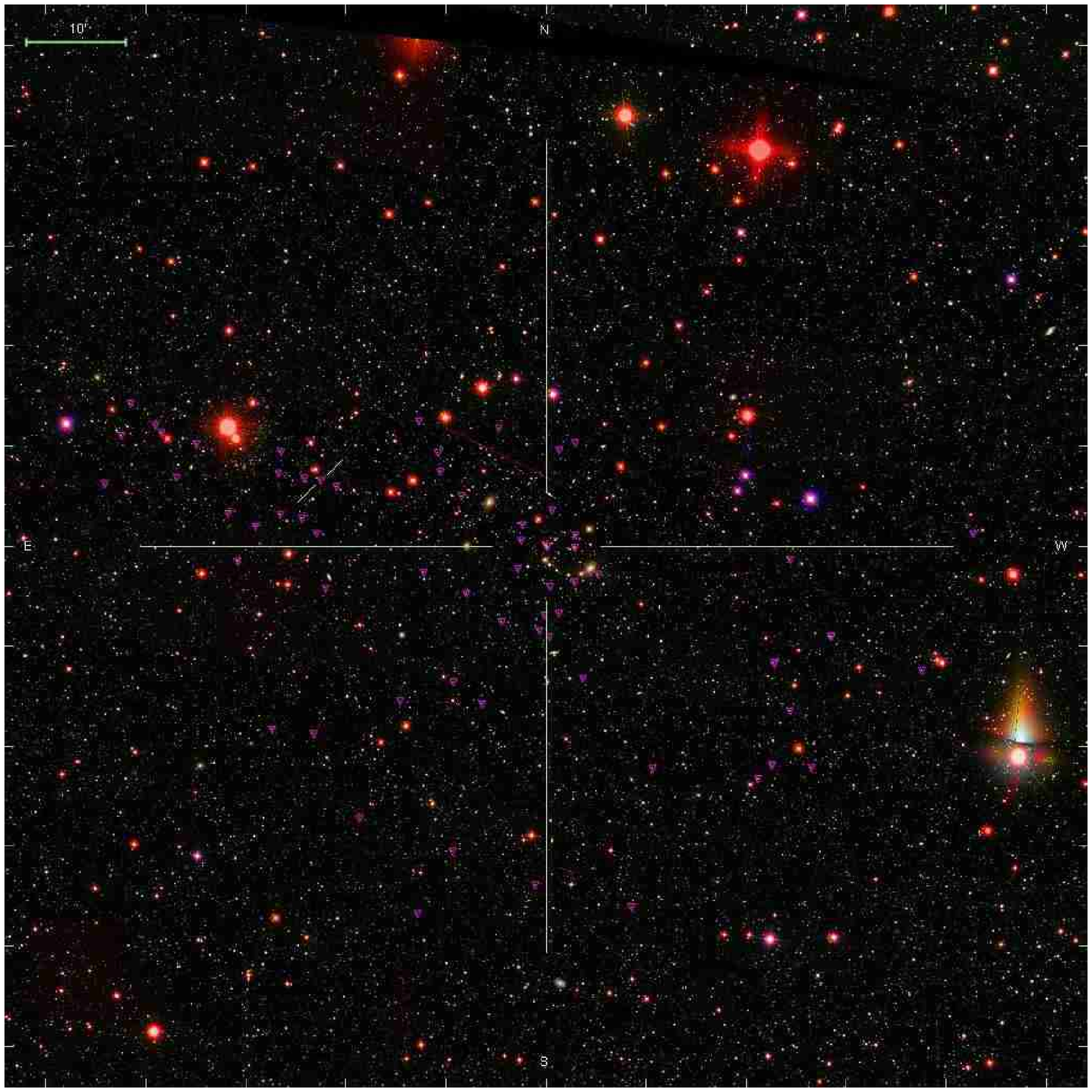}
\end{minipage}
\end{center}
\end{figure*}

\begin{figure*}
%[htbp]
\begin{center}
\begin{minipage}{0.48\hsize}
\includegraphics[width=1\hsize]{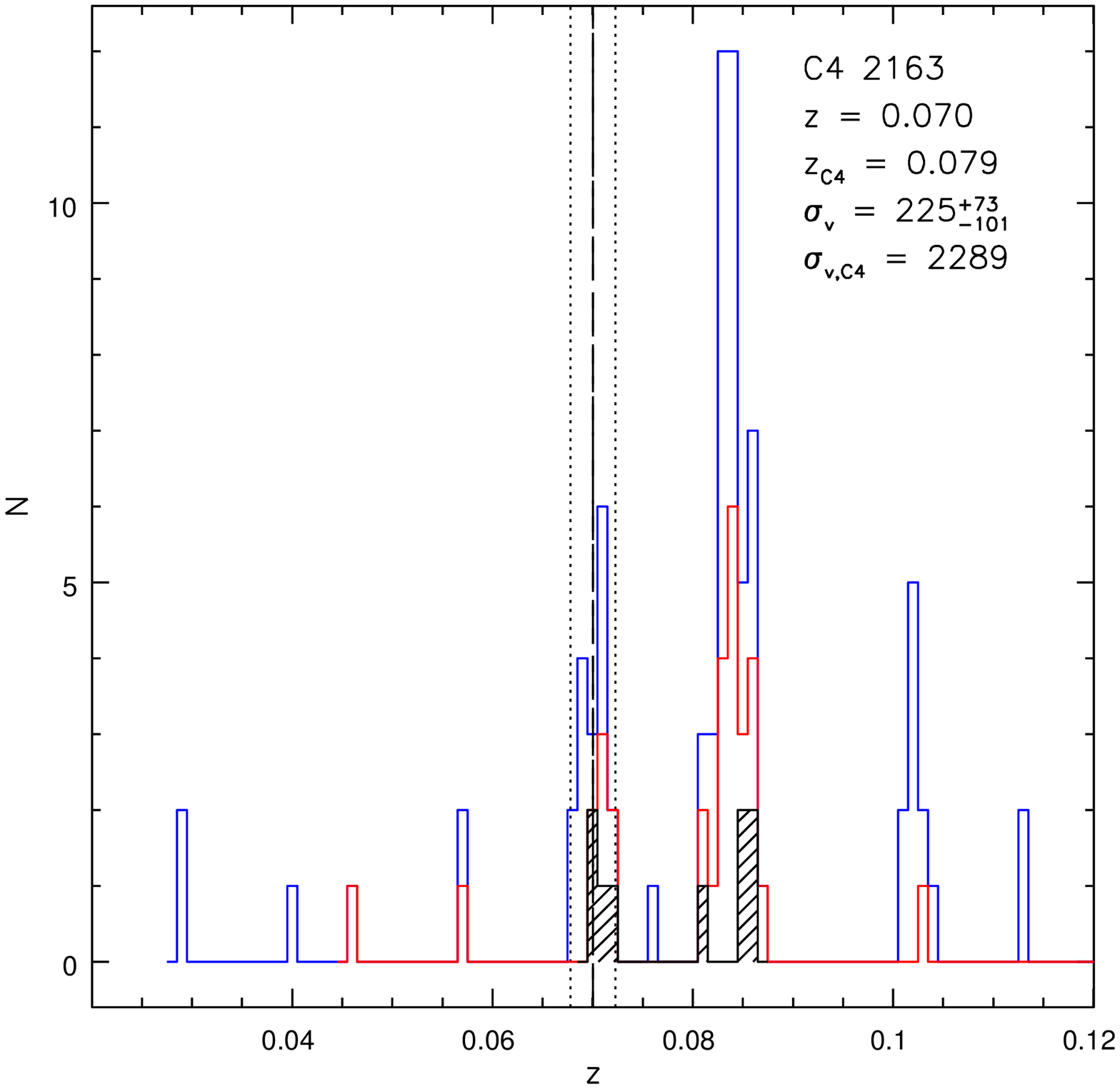}
\end{minipage}
\begin{minipage}[][0.45\hsize][t]{0.45\hsize}
\caption{~\textbf{C4\_DR3\_2163.} In the C4 catalog, this cluster is listed as a
  structure at $z=0.079$, with $\sigma_{\rm v, cl}>2000 \mbox{ km/s}$. The mean galaxy
  is marked in the finding chart below and is part of the group of
  galaxies at $z=0.07$, which our algorithm identifies. In the
  redshift histogram, we show the distribution of galaxies within
  projections of $1R_{200}$
  (black, shaded), $3R_{200}$ (red), and $5R_{200}$ (blue).
  Obviously, the difference in the redshift measurements as well as the
  large C4 velocity dispersion are due to another
  structure at $z\sim0.082$. In Fig. \ref{fig:c4_2101}, we show that this
  background structure is associated with C4\_2124.
}
\label{fig:c4_2163}
\end{minipage}

\begin{minipage}{0.87\hsize}
\includegraphics[width=1\hsize,]{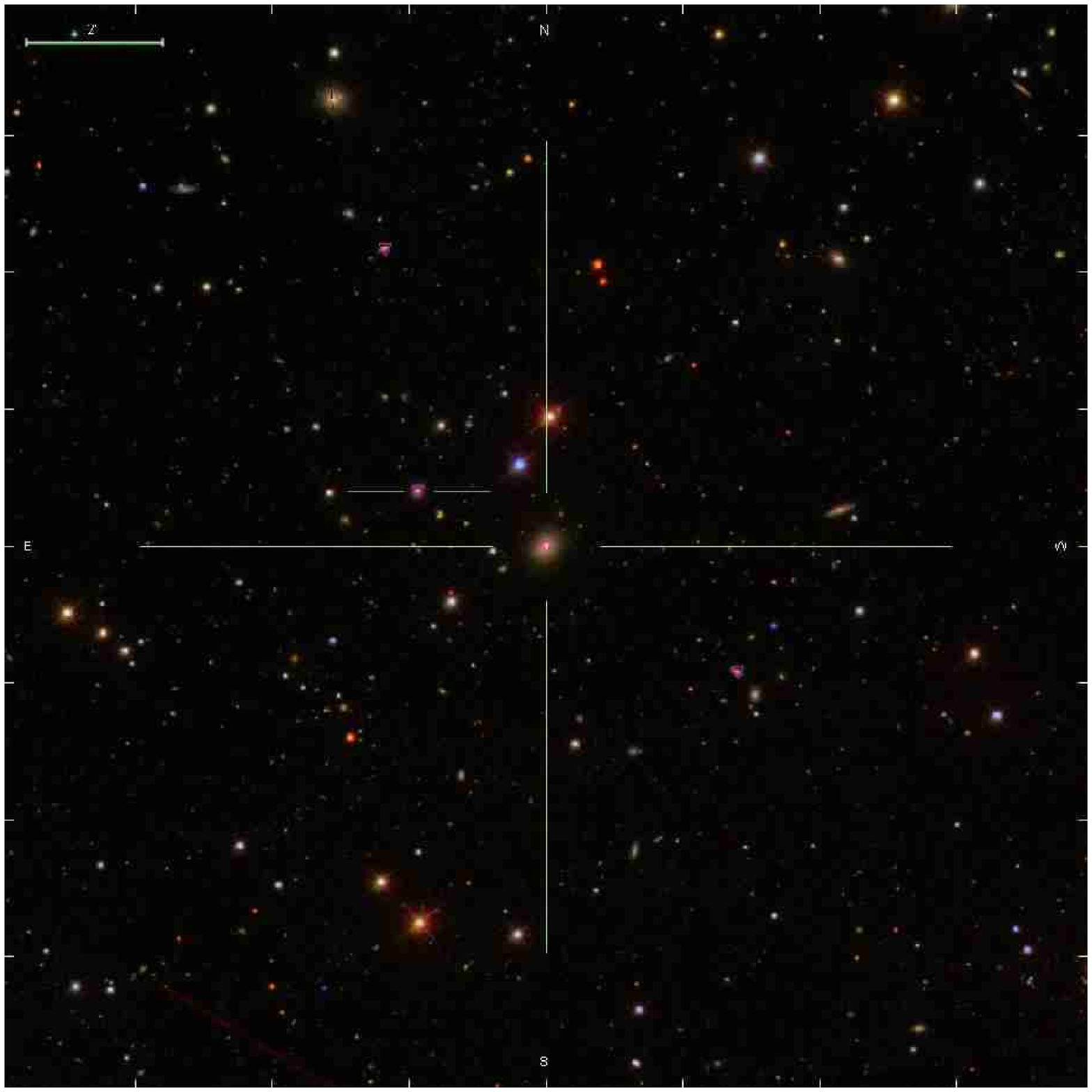}
\end{minipage}
\end{center}
\end{figure*}

\begin{figure*}
%[htbp]
\begin{center}
\begin{minipage}{0.48\hsize}
\includegraphics[width=1\hsize]{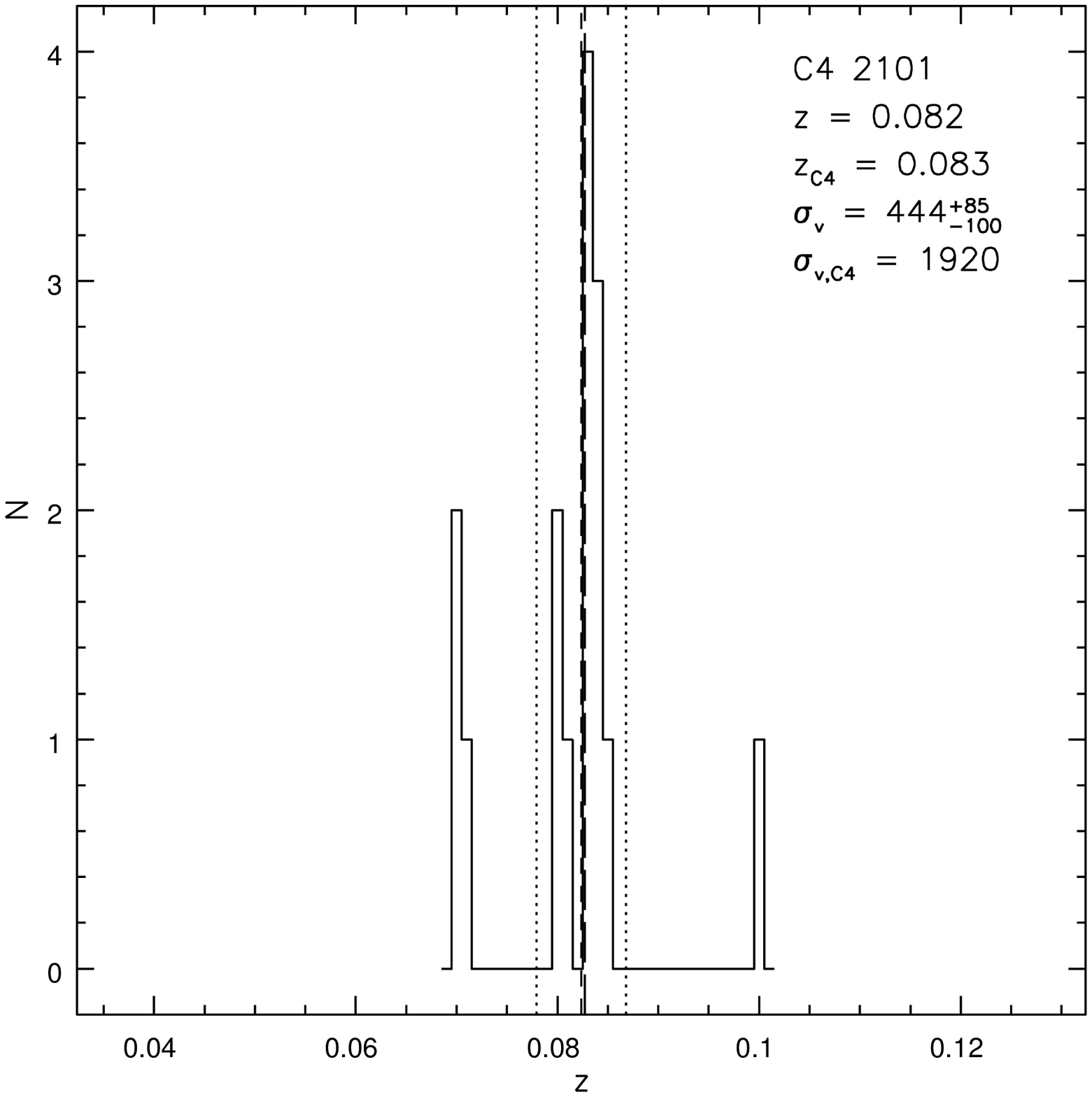}
\end{minipage}
\begin{minipage}[][0.45\hsize][t]{0.45\hsize}
\caption{~\textbf{C4\_2124.} The finding chart below is not centered on the
  BCG of C4\_2124, but on the BCG of C4\_DR3\_2163, in order to illustrate the
  sheet-like structure at $z\sim0.082$ which led to the deviating redshift
  measurement for C4\_DR3\_2163. The BCG of C4\_2124 is marked in the finding
  chart, and its $R_{200}$ (equivalent to 1~Mpc) is indicated by a dashed circle.
}
\label{fig:c4_2101}
\end{minipage}

\begin{minipage}{0.87\hsize}
\includegraphics[width=1\hsize,]{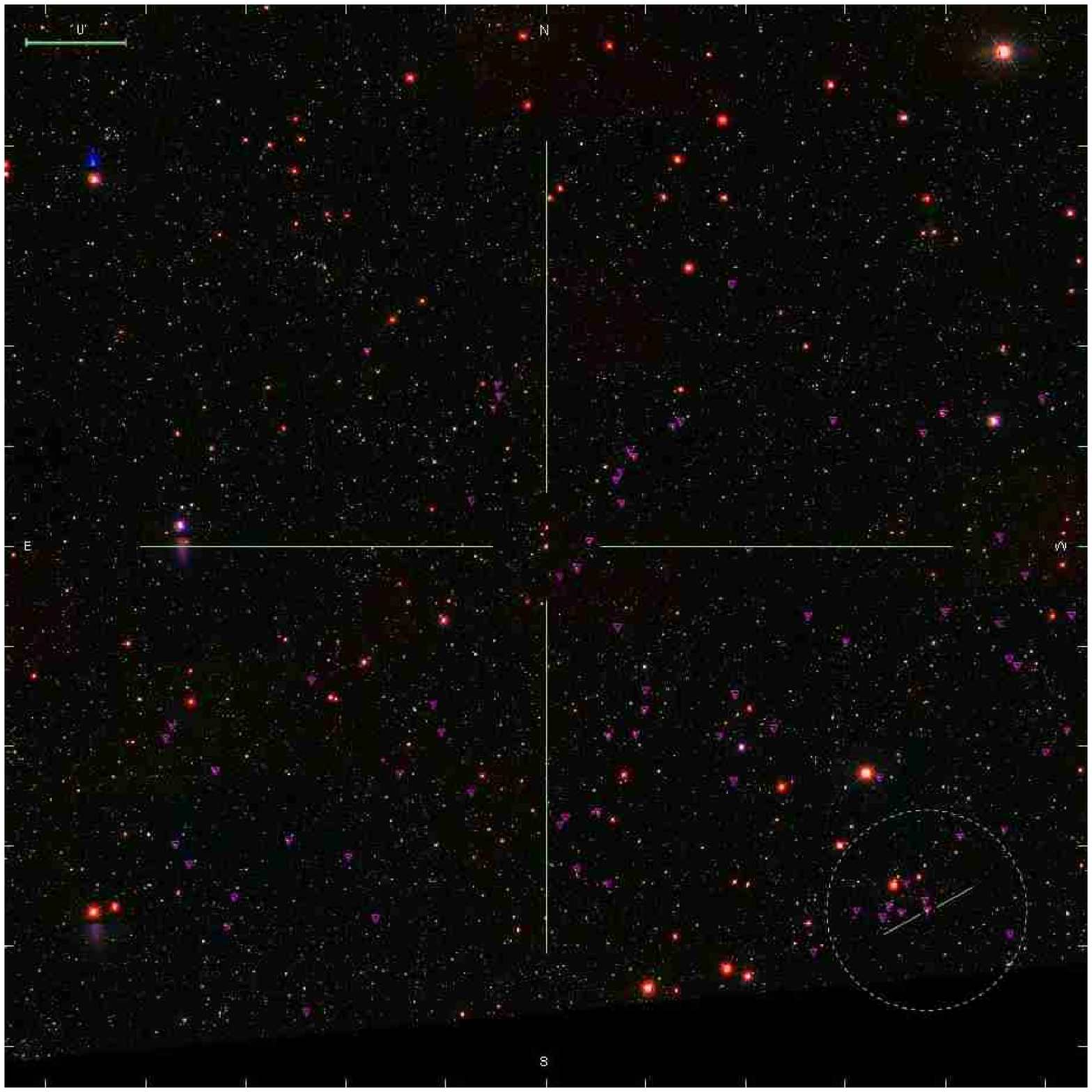}
\end{minipage}
\end{center}
\end{figure*}

\label{lastpage}

\end{document}